\documentclass[letterpaper,aps,prd,twocolumn,tightenlines,preprintnumbers,nofootinbib,showkeys,superscriptaddress,longbibliography
]{revtex4-1}
\pdfoutput=1

\makeatletter
\renewenvironment{widetext@grid}{%
  \par\ignorespaces
  \onecolumngrid
  \vskip16\p@
  \prep@math@patch
}{%
  \par
  \vskip14.5\p@
  \twocolumngrid\global\@ignoretrue
  \@endpetrue
}
\AtBeginDocument{\let@environment{widetext}{widetext@grid}}
\makeatother

\usepackage{XCharter}
\usepackage[T1]{fontenc}
\usepackage{blkarray}
\usepackage{booktabs}
\usepackage{array}
\usepackage{fullpage}
\usepackage{amsfonts}
\usepackage{amsmath}
\usepackage{slashed}
\usepackage{amssymb}
\usepackage{mathptmx}    
\usepackage{graphicx}
\usepackage{epic}
\usepackage{eepic}
\usepackage{epsfig}
\usepackage{latexsym}
\usepackage[dvipsnames]{xcolor}
\usepackage[export]{adjustbox}
\usepackage{float}
\usepackage{multirow}
\usepackage{hyperref}
\usepackage{enumitem}
\usepackage{lipsum}
\hypersetup{colorlinks=true,citecolor=red,linkcolor=NavyBlue,urlcolor=NavyBlue}
\usepackage[caption=false,labelformat=simple]{subfig}
\renewcommand\thesubfigure{(\alph{subfigure})} 
\usepackage{natbib}
\usepackage{relsize}
\usepackage[left=1.6cm,right=1.6cm,top=2.0cm,bottom=2.0cm]{geometry}
\usepackage{comment}
\usepackage{shorthand}
\usepackage{dcolumn}
\usepackage{longtable}
\usepackage{siunitx}

\usepackage{pifont}
\begin{document}


\title{Magnetized Dense Matter in the Nambu-Jona-Lasinio Model\\
and Its Applications to Compact Stars}

\author{Arijit Das}
\email{arijit21@iisertvm.ac.in}
\affiliation{Indian Institute of Science Education and Research Thiruvananthapuram, Vithura, Kerala, 695 551, India}

\author{Prashanth Jaikumar}
\email{prashanth.jaikumar@csulb.edu}
\affiliation{California State University Long Beach,1250 Bellflower Blvd, Long Beach, CA 90840, USA}

\author{Tanumoy Mandal}
\email{tanumoy@iisertvm.ac.in}
\affiliation{Indian Institute of Science Education and Research Thiruvananthapuram, Vithura, Kerala, 695 551, India}

\begin{abstract}
\noindent
We review the properties of strongly interacting matter under the influence of an external magnetic field, with emphasis on results obtained using the Nambu–Jona-Lasinio model and its variants. Strong magnetic fields, relevant for heavy-ion collisions and magnetars, introduce modifications to the phase structure of quantum chromodynamics. In the first half of the review, we discuss the Nambu-Jona-Lasinio model and its major variants, including Polyakov loop extended model, Kobayashi-Masakawa-'tHooft extended model, entangled model, and nonlocal formulations, highlighting their successes and limitations. Particular attention is given to magnetic catalysis and inverse magnetic catalysis and the tension between model predictions and lattice simulation results. We further review the emergence of phases at high density and low temperature such as the magnetic color-flavor-locked phase, the magnetic two-flavor superconducting phase, the crystalline color superconductor and magnetized quarkyonic matter, along with possible mesonic $\pi$ and $\rho$ meson condensation channels. Implications for strongly magnetized neutron stars are discussed wherever relevant. This review aims to provide a focused overview of the capabilities and limitations of approaches within the Nambu-Jona-Lasinio model and to outline open challenges in understanding strongly interacting magnetized matter at intermediate densities.\end{abstract}
\maketitle 


\section{Introduction}

Strongly interacting matter (SIM) under extreme conditions exhibits a rich phase structure, and intense effort has been invested in understanding the details of the phase diagram of quantum chromodynamics (QCD). The phase structure of QCD bears direct relevance to heavy-ion collisions, the early universe, and compact astrophysical objects such as magnetars where intense magnetic fields are naturally involved. The vast literature on this subject suggests that at low temperature and sufficiently high baryon density (equivalently, large baryon chemical potential), quarks can form color superconducting phases \citep{Barrois:1977xd,Bailin:1983bm} with many variants, e.g., a two-flavor color superconductor (2SC) \citep{Bailin:1983bm}, a crystalline color superconductor (CCS) or LOFF phase\citep{Larkin:1964wok,Fulde:1964zz,Alford:2000ze}, a color-flavor locked phase (CFL) \citep{Bailin:1983bm} or a color-spin-locked phase \citep{Marhauser:2006hy}.
Another interesting suggestion at intermediate baryon density is that of quarkyonic matter \citep{McLerran:2007qj}. 

\par Among the various external conditions that can significantly influence the phase structure of SIM, strong magnetic fields play a particularly important role. There are numerous compelling reasons to expect that very strong magnetic fields existed in the early universe \citep{Grasso:2000wj} and we may expect similar environments in relativistic heavy ion colliders \citep{Skokov:2009qp}, making these studies relevant across both cosmological and terrestrial laboratory settings. A clear motivation for studying magnetized SIM comes from Astrophysics. Neutron stars serve as natural astrophysical testing grounds that harbor extreme magnetic fields, and even more so in case of magnetars, whose surface magnetic fields can reach up to $10^{14}\text{--}10^{15}\,\mathrm{G}$ while internal fields can potentially reach up to $10^{16}\text{--}10^{18}\,\mathrm{G}$. If the core density of these systems is sufficiently high to support superconducting quark phases, it can potentially have macroscopic influence, meaning that compact stars can serve as natural laboratories for dense matter in the presence of strong magnetic fields. 
Additional motivation arises from ultra relativistic heavy-ion collisions (RHIC), where extremely intense magnetic fields are created during collisions of ions for a very short time. Calculations of such spontaneously generated magnetic fields \citep{Kharzeev:2007jp,Skokov:2009qp,Bzdak:2011yy,Shen:2025unr,Sun:2023rhh} estimate the magnitude of these magnetic fields to be as large as $10^{19}~\text{G}$. \citep{Lasky:2013bpa}.

\par Moreover, magnetic fields introduce an additional scale in the system that can significantly modify the nature of chiral symmetry breaking. This is exemplified by the phenomenon of magnetic catalysis (MC) \citep{Gusynin:1995nb}, where the chiral condensate is enhanced due to dimensional reduction \citep{Gusynin:1995nb} of quark dynamics. Many studies based on effective models of SIM have consistently predicted MC. However, lattice QCD (LQCD) results at finite temperature reveal a qualitatively different behavior of inverse magnetic catalysis (IMC)~\citep{Bali:2011qj}. 
Reconciling these two seemingly contradictory phenomena is essential for constructing a consistent picture of the QCD phase diagram in extreme magnetic fields, and also for interpreting the imprints of such phases in magnetars and heavy-ion collisions. To investigate this rich phase structure, it is pertinent to ask which theoretical frameworks may be chosen that can reliably describe these phenomena.

\par Below a density $\rho \lesssim 2\rho_0$, where $\rho_0$ is the nuclear saturation density, such systems can be described fairly accurately by chiral perturbation theory \citep{Andersen:2012zc}. Beyond this density regime, predictions from the chiral expansion scheme come with large uncertainties. On the other extreme $\rho \gtrsim 40\rho_0$, perturbative QCD \citep{Fraga:2025juh} gives accurate results but densities reached in RHIC or inside the core of magnetars fall in the regime of ``intermediate densities'' where neither of these formalisms are reliable. 
Effective models of QCD have emerged as a versatile framework to investigate these effects across a wide range of temperatures, densities, and magnetic field strengths. These models include the magnetized versions of Nambu-Jona-Lasinio (NJL) model \citep{Nambu:1961tp,Nambu:1961fr}, quark-meson model \citep{Berges:1998rc} and their Polyakov loop corrected versions~\citep{Schaefer:2007pw,Fukushima:2003fw} and Kobayashi–Maskawa–’t Hooft (KMT) extensions \citep{Moreira:2020wau,Rai:2018ufz}, nonlocal NJL model \citep{Bowler:1994ir} as well as approaches based on Dyson–Schwinger formalism \citep{Maskawa:1974vs} and functional renormalization group techniques \cite{Guan:2025mce}. 

\par Given the multitude of competing and often seemingly contradictory effects induced by strong magnetic fields, a completely exhaustive review of recent developments based on existing literature is a daunting task. Therefore, the scope of this review is rather modest. We focus on summarizing the principal results obtained using the NJL model and its extensions that include magnetic fields. Wherever relevant, implications for strongly magnetized compact objects are also highlighted. This review is broadly organized into two parts. The first part presents the theoretical framework of NJL model and its extensions in the presence of a constant and uniform magnetic field. These include the Polyakov loop extended NJL (PNJL) model (Sec.~\ref{PNJL_section}), the entangled Polyakov loop extended NJL (EPNJL) model (Sec.~\ref{EPNJL_Section}), the NJL model with the KMT interaction (Sec.~\ref{NJL_KMT_Section}), the nonlocal NJL (NNJL) model (Sec.~\ref{NNJL_Section}), its Polyakov loop extension (Sec.~\ref{NNJL_Polyakov_Section}), and finally its KMT extension (Sec.~\ref{Polyakov_KMT_NNJL_Section}).

\par The second part focuses on physical phenomena and phases of QCD in strong magnetic fields. We first discuss MC and IMC in Sec.~\ref{MC_section}, respectively. This is followed by a review of magnetized phases of QCD, including the magnetized CFL (MCFL) phase (Sec.~\ref{MCFL_Section}), the magnetized 2SC phase, CCS/LOFF phase, and the associated chromomagnetic instability (Sec.~\ref{2SC_Section}), as well as magnetized quarkyonic matter (Sec.~\ref{Quarkyonic_Matter_Section}). We then examine the effects of magnetic fields on mesonic masses and condensates e.g. pion and $\rho$-meson condensates (Secs.~\ref{Pion_condensate_Section} and \ref{rho_condensate_section}). Finally, in Sec.~\ref{Discuss_Section}, we present concluding remarks and highlight some avenues to add to the existing literature on the subject of this review.

\section{Notations and Conventions}
\label{MAGNETIZED EFFECTIVE MODELS}
An inherent challenge in any wide survey of evolving research is the wide variety of notational conventions used across different works. In this review, we adhere to a single set of conventions wherever possible unless explicitly specified. We use the natural units with $\hbar$ = $c$ = $k_\text{B}$ = 1. All effective field-theoretic models are treated within the relativistic mean-field (RMF) approximation. 
A generic fermion multiplet is denoted by $\psi$. We denote by $N_C$, $N_f$, and $N_S$ the number of colors, flavors, and spin degeneracies, respectively. The fields $\sigma$, $\omega$, $\rho$, and $\pi$ represent the scalar, vector, isovector, and pseudoscalar mesons. The chiral and diquark condensates are denoted by $\overline{\sigma}$ and $\overline{\Delta}$, respectively. For fermions of flavor $f$, $m_f$ and $M_f$ ($m$ and $M$ in flavor degenerate case) denote the current and dynamical masses, respectively. In the absence of a magnetic field, the single-fermion dispersion relation will be written as $\epsilon_f(\vec{p}) = \sqrt{\vec{p}^{\,2} + M_f^2}$ and in the flavor-degenerate case, $\epsilon(\vec{p}) = \sqrt{\vec{p}^{\,2} + M^2}$ will be used.
\par In the presence of a uniform and static magnetic field of strength $B$ (which will be assumed to be in the $z$-direction), the energy spectrum of charged fermions is quantized into Landau levels, and the corresponding dispersion relation should be changed as $\epsilon_{f\text{k}} = \sqrt{p^2_z + M^2_f + 2\text{k}|\widetilde{Q}_fB|}$ in terms of rotated charge $\widetilde{Q}_f$ of quarks and the Landau level index $\text{k} \in \mathbb{Z}_{\geq 0}$. However, wherever possible, we will use $\epsilon_{f \text{k}, a} = \sqrt{p^2_\perp + p^2_z + M_f^2}$ where $a$ are the possible rotated $\widetilde{Q}_f$ charges of quarks in units of $\widetilde{e}$. The quantity $p^2_\perp$, the momentum transverse to $\vec{B}$, is defined as $p^2_\perp = p^2_x + p^2_y$ if $a = 0$ (in which case the index \text{k} will be dropped and will be written as $\epsilon_{f,0}$) and $p^2_\perp = 2\text{k}|a|\widetilde{e}B$ if $a = \pm1/2$. The Landau level degeneracy factor is defined as $\alpha^{f}_\text{k} = 2 - \delta_{\text{k}0}$ for a fermion of flavor $f$. Moreover, we will use $\epsilon^{\pm}(p) = \epsilon(p) \pm \mu$, $\epsilon_{f}^{\pm}(p) = \epsilon_f(p) \pm \mu_f$ and $\epsilon_{f\text{k},a}^{\pm}(p) = \epsilon_{f\text{k},a}(p) \pm \mu_f$ where $\mu_f$ will denote the chemical potential of a fermion flavor $f$. The symbol $\nu_{\text{max}} = \text{Int}\!\left[\Lambda^2/2|a| \widetilde{e} B \right]$ will denote the highest occupied Landau level, where a momentum cut-off scale $\Lambda\approx~0.6-0.7~\text{GeV}$ is introduced due to non-renormalizable interactions present in the NJL model. The inverse temperature will be written as $\beta = 1/T$, $T$ being the temperature of the system. The quantities $\Phi$ and $\overline{\Phi}$ will represent the Polyakov loop operator and its conjugate, respectively, while $\mathcal{U}(\Phi, \overline{\Phi}, T)$ will denote the associated Polyakov loop potential. The thermodynamic grand potential will be denoted by $\Omega$, with or without subscripts and superscripts, depending on context. For magnetized systems, we will define $x_f = m_f^2/2|Q_f B|$. The Hurwitz zeta function will be denoted by $\zeta(a,x)$, and its derivative with respect to the first argument by $\zeta'(a,x) = \partial \zeta(a,x)/\partial a$.

\section{Nambu-Jona-Lasinio model and its variants}
\label{MAGNETIZED EFFECTIVE MODELS}

\par A wide range of effective models have been proposed to capture the essential features of QCD in regimes relevant to nuclear and quark matter. These models, while not derived directly from first principles, incorporate key symmetries of QCD and often provide valuable insights into the phase structure, the thermodynamic properties of SIM, and possible astrophysical consequences. While no single model should be expected to capture all the symmetries of QCD, the NJL model and its variants have been used for decades and have been reasonably successful in understanding the properties of SIM under such circumstances. In the following subsections, we briefly describe such variations of the NJL model, leading up to the inclusion of magnetic fields. 

\subsection{The Nambu Jona-Lasinio model}
\label{NJL_NoB}

The NJL model~\citep{Nambu:1961tp,Nambu:1961fr} was proposed in the 1960s, well before the advent of QCD, in order to explain the origin of the nucleon mass using spontaneous chiral symmetry breaking.
The NJL model offers a schematic picture of the basic dynamics behind the spontaneous breaking of chiral symmetry, with pions as Goldstone bosons emerging from a local four-point interaction between quarks.
Given the wide range of extensions and variants, we start with the simplest $SU(2)_L \times SU(2)_R$ NJL model, also called the $SU(2)_f$ NJL model with only the scalar, pseudoscalar and diquark channels. The Lagrangian in the mass degenerate case ($m_u$ = $m_d$ = $m$) is given as~\citep{Huang:2001yw}
\begin{widetext}
\begin{equation}
\mathcal{L} \supset \overline{\psi}\left(i\slashed{\partial} - m\right)\psi
 + G_S \left[ \left(\overline{\psi}\psi\right)^2
    + \left(\overline{\psi}i\tau^a\gamma^5\psi\right)^2 \right] +
G_D\bigg[(i\overline{\psi}^C\epsilon\varepsilon^b\gamma^5\psi)(i\overline{\psi}\epsilon\varepsilon^b\gamma^5\psi^C)\bigg],
\end{equation}
\end{widetext}
where $G_S$ and $G_D$ are the effective couplings in the (pseudo)scalar and diquark channels, respectively. The gap equations dynamically generate the quark mass even if the current quark mass $m = 0$, breaking chiral symmetry spontaneously, $SU(2)_L \times SU(2)_R \rightarrow SU(2)_V$ and one obtains the pseudoscalar pions ($\pi^a: a=1,2,3$) in the theory. After a Hubbard-Stratonovich transformation, the Lagrangian takes a more tractable form,
\begin{widetext}
\begin{align}
\mathcal{L} \supset ~\overline{\psi}\left(i\slashed{\partial} - M + i\gamma^5 \tau^a \pi^a\right)\psi - \frac{\sigma^2 + \vec{\pi}^2}{2G_S} - \frac{\Delta^{*b}\Delta^{b}}{2G_D} -\dfrac{1}{2}\Delta^{*b}(i\overline{\psi}^C\epsilon\varepsilon^b\gamma^5\psi) -\frac{1}{2}\Delta^{b}(i\overline{\psi}\epsilon\varepsilon^b\gamma^5\psi^C).
\label{NJL_B}
\end{align}
\end{widetext}
where $\sigma$ and $\pi^a$ are auxiliary scalar and pseudoscalar meson fields, respectively and $M = m + \sigma$ in the mean field approximation. The grand potential $\Omega$ in the mass degenerate case ($m_u$ = $m_d = m$) is given by~\citep{Huang:2001yw}

\begin{widetext}
\begin{align}
\Omega ={}& \frac{\sigma^2}{4G_S} + \frac{\Delta^2}{4G_D}
- 2\mathlarger{\sum}_{f}\int^{\Lambda} \frac{d^3\vec{p}}{(2\pi)^3}
\bigg[\epsilon_{f}(p) + T\ln\!\left(1 + e^{-\beta \epsilon^{+}_{f}}\right)
+ T\ln\!\left(1 + e^{-\beta \epsilon^{-}_{f}}\right) \nonumber\\
&+ \epsilon^{+}_{f,\Delta} + T\ln\!\left(1 + e^{-\beta \epsilon^{+}_{f,\Delta}}\right)
+ \epsilon^{-}_{f,\Delta} + T\ln\!\left(1 + e^{-\beta \epsilon^{-}_{f,\Delta}}\right)\bigg].
\end{align}
\end{widetext}
Here, $\epsilon^{\pm 2}_{f,\Delta} = \epsilon^{\pm 2}_{f} + \Delta^2$. In the presence of an electromagnetic field $A_\mu$, the bosonized Lagranian can be obtained from 
Eq.~\eqref{NJL_B} by replacing the ordinary derivative by the covariant derivative, $D_\mu = \partial_\mu - i|a|\widetilde{e}A_{\mu}$ where $\widetilde{e}$ and $\widetilde{Q}$ are rotated charge and rotated charge matrix, respectively. Upon activation of magnetic fields, the momentum integral measure is discretized as \citep{Noronha:2007wg,Mandal:2009uk,Mandal:2012fq,Mandal:2016dzg,Mandal:2017ihr,Avancini:2019wed,Aguirre:2020tiy,Mei:2024rjg}
\begin{align}
\label{Int_Disc_Scale}
\int_0^\Lambda \frac{d^3 \mathbf{p}}{(2\pi)^3} \rightarrow \frac{|a|\widetilde{e}B}{8\pi^2}\mathlarger{\sum}_{n=0}^{\nu_{max}}\int^{\Lambda^{\prime}}_{-\Lambda^{\prime}}dp_z
\end{align}
where $\Lambda^{\prime} = \sqrt{\Lambda^2 - 2|a|\widetilde{e}B\text{k}}$. Notice that this replacement of the momentum integral measure has reduced the fermion dynamics from $3 + 1$ dimensions to $1 + 1$ dimensions. This is termed as dimensional reduction, which plays a decisive roles in the phenomenon of magnetic catalysis, to be discussed in later sections. The grand potential changes as follows:
\begin{widetext}
\begin{align}
\Omega_{B \neq 0} ={}&
\frac{\sigma^2}{4G_S} + \frac{\Delta^2}{4G_D}
- \mathlarger{\sum}_{f=u,d}\int_0^\infty \frac{d^3 \vec{p}}{(2\pi)^3}
\Big[ f(\epsilon_{f,0}^+) + f(\epsilon_{f,0}^-) \Big] - \frac{\widetilde{e}B}{8\pi^2}
\mathlarger{\sum}_{f=u,d}~ \mathlarger{\sum}_{\text{k}=0}^\infty \alpha^f_\text{k}
\int_{-\infty}^{\infty} dp_z \,
\Big[ f(\epsilon_{f\text{k},1}^+) + f(\epsilon_{f\text{k},1}^-) + 2 f(\epsilon_{f\text{k}\Delta,1/2}^+) \\ \nonumber
&+ 2 f(\epsilon_{f\text{k}\Delta,1/2}^-) \Big].
\end{align}
\end{widetext}
Here, $\epsilon^{\pm}_{f\text{k}\Delta,a} = \sqrt{\epsilon_{f,a}^{\pm 2} + M^2_f}$. The function $f$ is defined as $f(\epsilon) = \epsilon + 2T\text{ln}(1 + e^{-\beta \epsilon})$.
\subsection{Regularization Schemes}
As the NJL model is fundamentally non-renormalizable, it is only ready for application only after it is coupled with a regularization scheme. Many such schemes are used, with their own merits and limitations. In the present section, we present the basic philosophy of such regularization schemes and present examples wherever necessary.
\subsubsection{3-Momentum Sharp Cut-Off, Dimensional Regularization and 3-Momentum Sharp Cut-Off}
The most straightforward such schemes are dimensional regularization and three momentum cutoff scheme \cite{Klevansky:1992qe,Buballa:2003qv,Fukushima:2003fw,Ratti:2005jh,Inagaki:2012re,Garibli:2019mzv}. In the former scheme, one first analytically continues $4\rightarrow D$ dimensional phase space integral
\begin{align}
    \int\frac{d^4\vec{p}}{(2\pi)^4}\longrightarrow \int\frac{d^D\vec{p}}{(2\pi)^D}\ = \frac{2(4\pi)^{\frac{1-D}{2}}}{\Gamma[\frac{D - 1}{2}]}M^{4-D}_{0}\int_{0}^{\infty}dp~p^{D-2}
\end{align}
($M_0$ is the renormalization scale) while in the latter case, one replaces \cite{Kohyama:2015hix,Inagaki:2012re}
\begin{align}
    \int\frac{d^4\vec{p}}{(2\pi)^4} \longrightarrow \int\frac{dp^0}{(2\pi)}\int_{0}^\Lambda \frac{p^2~dp}{(2\pi)^3}
\end{align}
$\Lambda$ being the regularization scale applied so that it removes $|\vec{p}| > \Lambda$ states and vacuum contributions are finite.
\par These two schemes are simple to implement, and hence, widely used. However, there are considerable problems with this scheme. There is no first principle way of justifying the widely used choice of $\Lambda = 0.693~$GeV which is obtained from fitting vacuum observables. The 4-point fermion coupling $G$ depends strongly on the $\Lambda$, $G\Lambda^2$ being a much more meaningful quantity compared to $G$. Lorentz invariance is explicitly broken and physical quantities contain regulator artifacts e.g. unphysical oscillations and artifact of large critical temperatures for small chemical potentials \cite{Buballa:2003qv}. Moreover, at very high densities and temperatures, the quark Fermi momentum may exceed $\Lambda$ and occupied quark states would be thrown out. To preserve Lorentz invariance, one uses a four momentum cut-off scheme in the Euclidean space where one uses $p^2_{\text{E}} = p^2_{\text{4}} + p^2_0 < \Lambda^2$ and one integrates over a 4-sphere of radius $\Lambda$. In the Wick rotated framework, one replaces

\begin{align}
    \int \frac{d^4p_{\mathrm{E}}}{(2\pi)^4}
    \;\longrightarrow\;
    \int^{\Lambda}\frac{p^3_{\text{E}}~dp_{\mathrm{E}}}{(2\pi)^3}
    \int d\Omega_{3}
\end{align}
and similar replacements can be made for finite temperatures \cite{Kohyama:2015hix}.

\noindent However, that does not deal with the fundamental limitation that NJL model is non-renormalizable and physical results still remain dependent on the cutoff $\Lambda$.

\subsubsection{Smooth Cut-Off Scheme}
Soft cut-off scheme \cite{Liu:2024nbf,Chaudhuri:2022oru} addresses the issue and gradually suppresses contributions from the large momenta using decaying weight functions $f_{\Lambda}(k)$ as \cite{Bowler:1994ir,Xue:2021ldz,Liu:2024nbf,Buballa:2003qv} as follows

\begin{align}
    \int_{0}^{\Lambda}dp~g(p) \longrightarrow \int_{0}^{\infty}f_{\Lambda}(p)g(p)
\end{align} 
The removal of sharp cut-off artifacts is necessary for analytic continuation of functions given on imaginary Matsubara frequencies. However, the choice of $f_{\Lambda}(p)$ is not dictated by any first principles arguments. Moreover, apart from $\Lambda$, usually there are other parameters involved in $f_{\Lambda}(p)$ and usage of such $f_{\Lambda}(p)$ requires justification of the choice of these parameters. As changing $f_\Lambda(p)$ changes physical quantities e.g. condensates and mesons masses, the parameters e.g. parameters e.g. $G$, $m_C$ and $\Lambda$ must be refitted.

\subsubsection{Pauli-Villars Regularization}
The Pauli-Villars regularization scheme introduces fictitious heavy regulators $\{ c_i, M_i \}$ to cancel out the UV divergences and replaces
\begin{align}
    \int_{0}^{\infty}~dp~f(p,M) \longrightarrow~\int_{0}^{\infty}dp~\mathlarger{\sum}_{i}c_i~f(p,M_i)
\end{align}

As a example, consider again the unregularized part of the vacuum contribution to the ferrmionic condensate in the $SU(2)_f$ NJL model in the presence of a magnetic field. By the above prescription, one defines a modified version

\begin{align}
\langle \overline{\psi}\psi \rangle^{\text{PV}}_{\text{vac}}
={}& -\frac{N_C M_f |q_f|B}{\pi^2}\mathlarger{\sum}_{\text{k}}\alpha_{\text{k}}\int_{0}^{\infty}dp_z \nonumber\\
&\times\Bigg[\frac{1}{\sqrt{p_z^2 + \eta^2_{f,\text{k}}}}
- \mathlarger{\sum}_{i}\frac{c_i}{\sqrt{p_z^2 + \eta^2_{f,\text{k},i}}}\Bigg]
\end{align}
where $\eta_{f,\text{k}} = M^2_f + 2|q_f|B\text{k}$ and $\eta_{f,\text{k}, i} = M^2_f + \Lambda^2_i + 2|q_f|B\text{k}$. The divergent pieces are
\begin{align}
    \text{ln}(p_z + \sqrt{p^2_z + \eta_{f,\text{k}(, i)}})\xrightarrow[p_z\to\infty]{}\text{ln}(2p_z) + \frac{\eta^2_{f,\text{k}(, i)}}{4p^2_z}
\end{align}
and cancellation criteria e.g $\sum_i c_i = 1$ and $\sum_i c_i \Lambda^2_i = 0$ conditions lead to vanishing of the divergent terms. However, this scheme does not preserve gauge invariance in chiral and non-abelian gauge theories. Although generalizations of the scheme for such field theories exist \cite{Okuyama:1996fy}, they are fairly complicated compared to the original scheme. Moreover, depending on the divergence structure, loop calculations require  several regulator scales $\{\Lambda_i\}$ and carefully chosen coefficients $\{c_i\}$, which can lead to very computationally complicated structures. Moreover, there is no all-agreed-upon choice of regulator masses, coefficients, number of regulators.
\subsubsection{Proper Time Regularization}
The basic idea behind this proper time regularization is the following manipulation of the Gamma function
\begin{align}
    \frac{1}{A(p^2)^n} \longrightarrow \frac{1}{\Gamma(n)}\int_{\frac{1}{\Lambda^2}}^{\infty}d\tau~\tau^{n-1}e^{-A\tau}
\end{align}
As an example \cite{Wang:2019jze}, one can show that the vacuum part of the 2-fermion condensate in the $SU(2)_f$ NJL model can be written as

\begin{align}
    \langle \overline{\psi}\psi \rangle_{\text{vac}} = \Lambda^2 e^{-\frac{M^2_f}{\Lambda^2}} - M^2_f~\Gamma\bigg(0, \frac{m^2_f}{\Lambda}\bigg)
\end{align}
$\Gamma(s, x)$ being the upper incomplete gamma function. In this scheme, ultraviolet contributions are damped by the factor $e^{-M^2_f/\Lambda^2}$ and the cutoff happens smoothly. However, in the presence of a magnetic field, an important subtlety arises. One can show that $\langle \overline{\psi}\psi \rangle_{\text{vac}}(B)$, when expanded around $B = 0$, leads to 
\begin{align}
\langle \overline{\psi}\psi \rangle_{\text{vac}}(B)
={}& -\frac{N_C M_f}{2\pi^2}\int_{0}^{\infty}d\tau\, e^{-\tau M_f^2} \nonumber\\
&\times\bigg[\frac{1}{\tau^2} + \frac{(q_fB)^2}{3}
- \frac{(q_fB)^4}{45}\,\tau^2 + \cdots\bigg]
\end{align}
showing that only the first term needs to be regularized and the B-dependent part is finite. Therefore, to prevent artificial regularization of B-dependent parts, it is essential to first separate the $B$-independent and $B$-dependent terms in all physical quantities. This is one of the principal motivators of MFIR and zMFIR schemes, which we discuss in the next subsection.
\subsubsection{MFIR and zMFIR Schemes}
Choice of a proper regularization in the presence of strong magnetic fields scheme is of particular importance. We have already demonstrated in the context of proper time regularization procedure that the magnetic field dependent contributions are artificially regularized unless separation of physical quantities in B-independent and B-dependent parts are made, which actually turns out to be unique, evident from identical results in completely different regularization schemes used in \cite{Ebert:2003yk} and \cite{Avancini:2015ady,Menezes:2008qt}. Therefore, it is important to devise regularization schemes that can separate magnetic and nonmagnetic effects in a clean way. This is achieved in MFIR and zMFIR schemes.
\par The MFIR scheme \cite{Avancini:2018svs,Menezes:2008qt} separates all physical quantities in three parts: a vacuum part, a B-dependent part and a medium part. For instance, the pressure contributed by a quark of flavor $f$ \cite{Menezes:2008qt} can be obtained as
\begin{align}
\mathcal{P} ={}& -\frac{(M - m_c)^2}{4G} + \mathcal{P}^{\text{med}} \nonumber\\
&+ \frac{N_C}{2\pi}\mathlarger{\sum}_{f=u,d}\mathlarger{\sum}_{\text{k}=0}^{\infty}\alpha_{\text{k}}\,|q_f|B
\int_{-\infty}^{\infty}\frac{dp_z}{2\pi}\,\epsilon_{f\text{k}}(B)
\end{align}
where $\mathcal{P}_{f}^{\text{med}}$ is the finite medium dependent part. In order to regularize the last term, one adds and subtracts the lowest Landau level (LLL) term $\text{k} = 0$. Afterwards, one analytically continues the integral to $d$ dimensions, sets $d = 1 - \epsilon$, $x = M^2_f/2qB$, uses the definition of Hurwitz-Zeta function, expands around $\epsilon = 0$ and adds and subtracts the vacuum contribution to the pressure, $\mathcal{P}^{\text{vac}}$, to get rid of the $\mathcal{O}(1/\epsilon)$ divergence to finally obtain the total pressure $\mathcal{P}_f$ as
\begin{align}
    \mathcal{P}_f = -\frac{(M - m_c)^2}{4G} + \mathcal{P}_{f}^{\text{med}} + \mathcal{P}^{\text{vac}} + \mathcal{P}_f^{\text{mag}}
\end{align}
where the finite magnetic contribution

\begin{align}
\mathcal{P}^{\text{mag}} &= \frac{N_C}{2\pi^2}\mathlarger{\sum}_{f=u,d}(q_fB)^2\,\mathcal{F}(x_f), \\
\mathcal{F}(x) &= \zeta'(-1,x) - \tfrac{1}{2}\left(x^2 - x\right)\ln x + \tfrac{x^2}{4}.
\end{align}

In this expansion, only the vacuum contribution $\mathcal{P}^{\text{vac}}_{f}$ need regularization, usually done using sharp cut-off scheme. However, the decomposition of physical quantities into vacuum and magnetic contributions is not always straightforward for models beyond the mean-field approximation. Furthermore, its implementation can become involved for certain observables e.g. for the calculation of meson pole masses in a magnetized medium which require calculation of mesonic polarization loops \cite{Ebert:2003yk,Menezes:2008qt}. To overcome these difficulties, Ref.~\cite{Avancini:2018svs} introduced the zMFIR scheme, a framework where summations over Landau levels are absorbed inside the definitions of the Hurwitz-Zeta functions. In a large class of effective models in the RMF approximation, one can write the grand canonical potential and the thermodynamical properties in the following general form 
\begin{align}
    I = \mathlarger{\sum}_{f}\mathlarger{\sum}_{s = \pm~1}\int\frac{d^3\vec{p}}{(2\pi)^3}~f(\epsilon_f(\vec{p})) = \mathlarger{\sum}_{s = \pm~1}I_f.
\end{align}
The integral $I$ usually contains divergent parts that need to be regularized. In the presence of magnetic field, $I_f$ is changed to 
\begin{align}
    I_f(B) = \frac{|q_f|B}{2\pi}\mathlarger{\sum}_{k = 0}^\infty\alpha_\text{k}\int_{-\infty}^{\infty}\frac{dp_3}{(2\pi)}f(\epsilon_{f\text{k}})
\end{align}
One introduces a non-normalized density of states $g_f(\epsilon, B)$ and writes $I_f$ as

\begin{align}
    I_f(B) = \int_{M_f}^{\infty}d\epsilon~g_f(\epsilon, B)f(\epsilon)
\end{align}
In order to dissect $g_f(\epsilon, B)$ into non-magnetic contribution $g_f(\epsilon)$ and purely magnetic contribution $\tilde{g}_f(\epsilon, B)$ one takes the limit $B\rightarrow~0$ and uses the asymptotic expansion of $\zeta(1/2, q_E\xrightarrow[]{} 0)$ \cite{Dib:2000hd}. In the end, $I_f$ can also be dissected as $I_f(B) = I_f(0) + \Tilde{I}_f(B)$. Applying these results to thermodynamic potential $SU(2)_f$ NJL model model one can show that the magnetic contribution to the vacuum part of the grand potential \cite{Menezes:2008qt} is
\begin{widetext}
\begin{align}
\widetilde{\Omega}_{\mathrm{vac}}(B)
={}& -N_c\mathlarger{\sum}_{f=u,d}\frac{(2\beta_f)^2}{(2\pi)^2}
\bigg[\frac{1}{24}\ln x_f + \frac{1}{12}
- \int_{0}^{1} dq_E\,\zeta\big(-\tfrac{1}{2},q_E\big)\,\zeta\big(\tfrac{1}{2},x_f+q_E\big)
\nonumber\\
&+ \int_{0}^{\infty} dq_E\,
\frac{\zeta\big(-\tfrac{1}{2},q_E\big) + \tfrac{2}{3}q_E^{3/2} - \tfrac{1}{2}q_E^{1/2} + \tfrac{1}{24}q_E^{-1/2}}{\sqrt{x_f+q_E}}\bigg].
\end{align}
\end{widetext}

Just as MFIR, zMFIR scheme isolates all ultraviolet divergences into the pure vacuum contribution, B-dependent contribution is finite and spurious oscillations are eliminated. Instead of writing the magnetic contribution using summation over Landau levels, it is written in terms of Hurwitz zeta functions which are much more convenient to tackle numerically and also provides closed form expressions. This is particularly powerful for calculating meson masses, polarization functions and
screening masses, where the zeta-function representation considerably simplifies intermediate expressions.
\par However, one needs to keep in mind that zMFIR is completely equivalent to the MFIR scheme Ref. \cite{Avancini:2018svs}. It only provides a different mathematical representation useful in calculating certain quantities described above. The vacuum contribution in the zMFIR result is still cut-off dependent. The closed-form expressions in terms of Hurwitz-Zeta functions are valid only in the RMF approximation and in the presence of constant magnetic field. For more involved extensions of the NJL model it can not be used so directly. 
\subsection{Polyakov-loop corrected NJL (PNJL) model}
\label{PNJL_section}

\noindent One of the most serious limitations of the local NJL model, namely the lack of a confinement mechanism, can be addressed by including the Polyakov loop potential $\mathcal{U}(\Phi,\overline{\Phi},T)$ \citep{Dexheimer:2009hi} induced by quark one-loop vacuum fluctuation \citep{Skokov:2010sf}. The operator describing such fluctuations is related to the trace over the Polyakov loop operator $\Phi$ \citep{Quandt:2022lhe,McLerran:1981pb} which serves as an order parameter for color confinement-deconfinement phase transition \citep{Fukushima:2017csk} in the pure gauge sector. Various functional forms of $\mathcal{U}(\Phi,\overline{\Phi},T)$ have been used, with their own successes \citep{Lo:2013hla} and shortcomings \citep{Mueller:2014tea}. Incorporating the Polyakov loop potential in the NJL model accounts for introducing confinement effects phenomenologically. However, this simple addition shows excellent agreement with LQCD calculations \citep{Borsanyi:2012cr,Miyahara:2016din,Bhattacharyya:2010wp}. 
\noindent In the flavor symmetric case, the thermodynamic potential in the mean field approximation for the $SU(2)_f$ PNJL model becomes \citep{Sakai:2008um}

\begin{widetext}
\begin{align}
\Omega ={}& V + \mathcal{U}(\Phi,\overline{\Phi},T)
- 2N_f \int^{\Lambda} \frac{d^3\vec{p}}{(2\pi)^3}\,
\Big[N_c\,\epsilon(p)
+ T \ln\!\Big(1 + 3\big(\Phi + \overline{\Phi}\, e^{-\beta \epsilon^-}\big) e^{-\beta \epsilon^-} + e^{-3\beta \epsilon^-}\Big) \nonumber\\
&+ T \ln\!\Big(1 + 3\big(\overline{\Phi} + \Phi\, e^{-\beta \epsilon^+}\big) e^{-\beta \epsilon^+} + e^{-3\beta \epsilon^+}\Big)\Big]
= \Omega_{\text{quark}} + \mathcal{U}(\Phi,\overline{\Phi},T).
\label{PNJL_Omega}
\end{align}
\end{widetext}

where $V(\sigma)$ contains the quardratic contributions from the scalar (and other) condensate. In the presence of a magnetic field, the grand potential \citep{Mao:2024gox,Shao:2019hen} can be obtained using Eq.~\eqref{PNJL_Omega}, using the replacement Eq.~\eqref{Int_Disc_Scale} and replacing $\epsilon_f$ by $\epsilon_{f\text{k}}$. After proper regularization of the vacuum part, the quark part becomes~\citep{Andersen:2012jf}

\begin{widetext}
\begin{align}
\Omega_{\text{q}} ={}& \frac{(M - m)^2}{4G}
- \frac{N_c}{4\pi^2}\bigg[\Lambda\,(2\Lambda^2 + M^2)\sqrt{\Lambda^2 + M^2}
- M^4 \ln\frac{\Lambda + \sqrt{\Lambda^2 + M^2}}{M}\bigg] \nonumber\\
&- \frac{N_c}{2\pi^2}\mathlarger{\sum}_f (Q_fB)^2\Big[\zeta'(-1, x_f) - \tfrac{1}{2}\left(x_f^2 - x_f\right)\ln x_f + \tfrac{1}{4}x_f^2\Big] \nonumber\\
&- T\mathlarger{\sum}_{f,\text{k}} \alpha^f_{\text{k}}\,\frac{|Q_f B|}{2\pi}\int_{-\infty}^{\infty}\frac{dp_z}{2\pi}
\Big[\ln\!\big(1 + 3\Phi\, e^{-\beta \epsilon^{-}_{f\text{k}}} + 3\overline{\Phi}\, e^{-2\beta \epsilon^{-}_{f\text{k}}} + e^{-3\beta \epsilon^{-}_{f\text{k}}}\big) \nonumber\\
&+ \ln\!\big(1 + 3\overline{\Phi}\, e^{-\beta \epsilon^{+}_{f\text{k}}} + 3\Phi\, e^{-2\beta \epsilon^{+}_{f\text{k}}} + e^{-3\beta \epsilon^{+}_{f\text{k}}}\big)\Big]
\end{align}
\end{widetext}

\subsection{Entangled Poylakov Loop Extended NJL (EPNJL) Model}
\label{EPNJL_Section}

\noindent
Although chiral and deconfinement transitions take place (nearly) at the same temperature~\citep{Aoki:2006br}, it is not obvious that the two phenomena should be so connected. To understand whether this is coincidence or correlation, a conjecture of entanglement between
these two transitions \citep{Sakai:2010rp,Sugano:2014pxa}, implemented within the entangled PNJL (EPNJL) model \citep{Sakai:2010rp,Sakai:2011fa}, is proposed. The origin of the four-quark vertex $G_S$ is the one-gluon exchange diagram between two quarks and its higher-order diagrams in QCD. On the other hand, the relativistic mean field value of the temporal component of the gluons $\langle G_0 \rangle$ plays a crucial role as it couples to the quarks through the covariant derivative. Moreover, the Polyakov loop operator is defined in terms of $\langle G_0 \rangle$. However, as gluons are colored, gluon dynamics itself is affected by this uniform background temporal gluon field. Hence, $G_S$ (and all other effective couplings) can be elevated from the status of a constant parameter to an entanglement vertex, a function of the Polyakov loop operator $\Phi$, $G_S(\Phi, \overline{\Phi})$, chosen in such a way that the functional form is $\mathbb{Z}_3$ symmetric. For phenomenological studies, the most popular functional form \citep{Mattos:2021alf} of these couplings is
\begin{align}
    \widetilde{G}_S(\Phi, \overline{\Phi}) &= G_S[1 - \alpha_1(\overline{\Phi}\Phi) - \alpha_2(\Phi^3 + \overline{\Phi}^3)] \label{EPNJL_G}\\
    \widetilde{G}_V(\Phi, \overline{\Phi}) &= G_V[1 - \alpha_1(\overline{\Phi}\Phi) - \alpha_2(\Phi^3 + \overline{\Phi}^3)] \label{EPNJL_GV}
\end{align}
With these choices of $\widetilde{G}_S$ and $\widetilde{G}_V$, the model has 4 free parameters given that $\widetilde{G}_S/\widetilde{G}_V$ is independent of $\Phi$ and $\overline{\Phi}$. The EPNJL model with $T_0 \approx 190$ MeV and $\alpha_1 = \alpha_2 = 0.2$ reproduces well the full LQCD results for the deconfinement and chiral transition lines at zero and imaginary $\mu$ \citep{Sakai:2010rp}. The thermodynamic potential for the EPNJL model same as that of PNJL model with only the coupling $G_S$ replaced by the entangled vertex $G_S(\Phi, \overline{\Phi})$. The effective mas $M = m_0 - 2\widetilde{G}_S\sigma$ and because of the presence of vector interaction, the quark chemical potential $\mu_q$ is replaced by the effective chemical potential $\widetilde{\mu}_q = \mu_q - 2\widetilde{G}_V n_q$. The same logic of replacement $G_S \rightarrow \widetilde{G}_S(\Phi, \overline{\Phi})$ remains valid for evaluating the grand potential for EPNJL model in the presence of a magnetic field. The details of the calculations can be found in ref.\citep{Ferreira:2013oda}.
\subsection{NJL Model with Kobayashi Maskawa ’t Hoof (KMT) Interaction}
\label{NJL_KMT_Section}
The global $U(3)_L \times U(3)_R$ symmetry of the QCD Lagrangian (for massless quarks) is broken by
the $U(1)_A$ Adler–Bell–Jackiw anomaly of the $SU(3)$ singlet axial current $\overline{q}\gamma^\mu \gamma^5 q$. It has also been seen from the mesonic spectrum that the $U(1)_A$ symmetry of the QCD Lagrangian is severely broken. The SU(3) singlet pseudoscalar $\eta^{\prime}$ is too heavy to be the ninth Goldstone boson. Moreover, the $U(1)_A$ anomaly is responsible for the $\eta^{\prime}-\pi$, K, $\eta$ splitting \citep{Polyakov:1975rs}. In instanton studies \citep{Diakonov:1995ea}, it has been realized that this anomaly has physical effects with the result that the theory contains neither a conserved $U(1)$ quantum number, nor an extra Goldstone boson. Instead, effective quark interactions arise, which are known as Kobayashi-Masakawa-’t Hooft (KMT) interactions \citep{Moreira:2020wau},
\begin{align}
    \mathcal{L}_{\text{KMT}} =  K~\text{det}_f[\overline{\psi}(1 + \gamma^5)\psi] + K\text{det}_f[\overline{\psi}(1 - \gamma^5)\psi].
\end{align}
To break the $U(1)_A$ symmetry of the NJL model, the 't Hooft term is added by hand, mimicking the axial anomaly of QCD. In the mean field approximation \citep{Steiner:2005jm}, the quark masses are dynamically modified in terms of quark condensates and using $\langle \overline{u}_L u_R \rangle = \langle \overline{u}_R u_L \rangle = \sigma_u$, $\langle \overline{d}_L d_R \rangle = \langle \overline{d}_R d_L \rangle = \sigma_d$ and $\langle \overline{s}_L s_R \rangle = \langle \overline{s}_R s_L \rangle = \sigma_s$ one gets the dynamical mass $M_f$ of a single quark of flavor $f$ as $M_f = m_f - (2G_S/3)\sigma_f - K\prod_{i \neq f}\sigma_i$ and in the presence of the thermal bath, the vacuum expectation value $\langle \overline{f}f \rangle$ is replaced by the thermal average $\langle \langle \overline{f}f \rangle \rangle$: $\langle \langle \overline{u}u \rangle \rangle = \widetilde{\sigma}_u$, $\langle \langle \overline{d}d \rangle \rangle = \widetilde{\sigma}_d$, and $\langle \langle \overline{s}s \rangle \rangle = \widetilde{\sigma}_s$. 
The grand potential becomes \citep{Islam:2023zpl}

\begin{align}
&\Omega_{\text{tHooft}}(\widetilde{\sigma}_u, \widetilde{\sigma}_d, \widetilde{\sigma}_s)
= \mathcal{V}(\widetilde{\sigma}_u, \widetilde{\sigma}_d, \widetilde{\sigma}_s) \cdot V
- 2N_c T \mathlarger{\sum}_{f}\int\frac{d^3\vec{p}}{(2\pi)^3} \nonumber\\
&\times\Big[\ln\{1 + \exp(-\beta\epsilon^{-}_{f})\}
+ \ln\{1 + \exp(-\beta\epsilon^{+}_{f})\}\Big]
\end{align}

\noindent where \(V\) is the volume of the system. Furthermore,
\begin{align}
&\mathcal{V}(\widetilde{\sigma}_u, \widetilde{\sigma}_d, \widetilde{\sigma}_s) 
= \nonumber\\
&-2N_c \mathlarger{\sum}_{f} 
\int^{\Lambda} \frac{d^3\vec{p}}{(2\pi)^3} \epsilon_{f} 
+ \Big\{ G_S (\widetilde{\sigma}_u^2 + \widetilde{\sigma}_d^2 + \widetilde{\sigma}_s^2) 
+ 4 K \widetilde{\sigma}_u \widetilde{\sigma}_d \widetilde{\sigma}_s \Big\}.
\end{align}
is the vacuum energy term. In the presence of a constant, uniform magnetic field \citep{Chatterjee:2011ry},

\begin{widetext}
\begin{align}
\Omega_{\text{tHooft}}
={}& - \mathlarger{\sum}_{f}\mathlarger{\sum}_{\text{k}} \frac{N_c \alpha^f_\text{k} |Q_f B|}{(2\pi)^2} \int dp_z \, \epsilon_{f\text{k}}
+ 2G_S \mathlarger{\sum}_f I_f^2 + 4K I_u I_d I_s \nonumber\\
&- \mathlarger{\sum}_{f, \text{k}} \frac{N_c \alpha^f_\text{k} |Q_f B|T}{(2\pi)^2} \int dp_z\,
\bigg[\ln\left(1 + e^{-\beta\epsilon^{-}_{f\text{k}}}\right) + \ln\left(1 + e^{-\beta\epsilon^{+}_{f\text{k}}}\right)\bigg].
\end{align}
\end{widetext}

where the expressions of $I_f$ and $\sin^2 \theta_{\pm}^{f,n}$ can be found in \citep{Chatterjee:2011ry}.
\subsection{Nonlocal Nambu Jona-Lasinio (NNJL) Model}
\label{NNJL_Section}
Although the local NJL model is simple yet powerful, results of observables derived from this model strongly depend on the regularization procedure used \citep{Avancini:2020xqe}. Apart from not having any mechanism for confinement, the usual procedure of putting a sharp cutoff in the momentum space misses the running of the QCD coupling, and hence, the reduction in the coupling constant as a function of magnetic field needs to be put in by hand, which will be described in more detail in the context of IMC in a later section. Also, in the local version of the NJL model, it is impossible to describe radial excitations of mesons.
\par Such drawbacks of the standard NJL model can be rectified satisfactorily in the framework of nonlocal models \citep{Dumm:2021vop} which naturally arise in the context of several successful approaches to low-energy quark dynamics as, for example in the instanton liquid model \citep{Schafer:1996wv} and the Schwinger–Dyson resummation techniques \citep{Roberts:2000aa}. It captures the essence of the running of the QCD couplings using form factors which lead to IMC effect naturally without introducing an explicit magnetic field dependent coupling \citep{Pagura:2016pwr}. Moreover, appropriate choices of form factors regularizes the model in such a way that anomalies are preserved \citep{RuizArriola:1998zi}. It also addresses some issues associated with confinement as well \citep{Plant:1997jr}.
In terms of the form factors $f(x)$ and nonlocal quark currents
\begin{align}
    J_I(x) = \int\int d^4x_1 d^4x_2 f(x_1)f(x_2)\overline{\psi}(x - x_1)\Gamma_I \psi(x + x_2)
\end{align}
\noindent where the matrix $\Gamma_I$ is defined as $\Gamma_{\sigma} = 1, \Gamma^{a}_{\pi} = i\gamma^g\tau^a, \Gamma^{\mu a}_{\rho} = \gamma^{\mu}\tau^a$ and $\Gamma^{\mu a}_{\rho} = \gamma^5\gamma^{\mu}\tau^a$, the action for the approximate $SU(2)_L \times SU(2)_R$-symmetric nonlocal
NJL model motivated by the instanton theories \citep{Anikin:2000rq} is given by \citep{Volkov:2005kw}:
\begin{align}
    S_{\text{NNJL}} = \int d^4x \Bigg[ 
        \overline{\psi}(x)(i\slashed{\partial} - m)\psi(x) 
        + \frac{G_S}{2} \Big( J_{\sigma}~J_{\sigma} + J^{a}_{\pi}~J^{a}_{\pi} \Big)\Bigg]
\end{align}
\noindent The form factors $f(x)$ capture the nonlocal nature of the 4-point interaction. Different phenomenological functional forms of these form factors have been used for different purposes in the literature, such as, gaussian \citep{Bowler:1994ir}, Lorentzian \citep{Grigorian:2006qe}, Woods–Saxon \citep{Grigorian:2006qe} and sharp cutoff \citep{Grigorian:2006qe}. The thermodynamic free energy for nonlocal PNJL model is given by
\begin{align}
    \Omega_{NNJL} = -4T\mathlarger{\sum}_{n\in \mathbb{Z}}\int\frac{d^2\vec{p}_E}{(2\pi)^3}\text{ln}\bigg[\omega^2_{n} + \vec{p}_E^2 + M^2(\omega_{n}, \vec{p}_E)\bigg] + \frac{\sigma^2}{2G} 
\end{align}
where $\vec{p}_E = (p_z, p_4 = -ip_0)$ and $\omega_n = (2n + 1)\pi T$, $n\in \mathbb{Z}$ are the Matsubara frequencies. This entire formalism can be extended to cases with external magnetic field \citep{Pagura:2017rey}. We present here only the final important mathematical result obtained using the Ritus eigenfunction method \citep{Pagura:2016pwr}.
The mean field approximated bosonised action per unit 4-volume can be expressed as

\begin{align}
&\frac{S^{\text{MFA}}_{\text{BOS}}}{V^{(4)}}
= \frac{\sigma^2}{2G_S} - N_C\mathlarger{\sum}_{f = u,d}\frac{|Q_fB|}{2\pi}\int\frac{d^2\vec{p}_E}{(2\pi)^2} \nonumber\\
&\times\Big[\ln\big[\vec{p}^{\,2}_E + \big(M^{s_f, f}_{\vec{p}_E,0}\big)^2\big]
+ \mathlarger{\sum}\nolimits_{i=1}^{\infty}\ln\big[\big(2i|Q_fB| + \vec{p}^{\,2}_E \nonumber\\
&\qquad + M^{-1, f}_{\vec{p}_E,i}M^{+1, f}_{\vec{p}_E,i}\big)^2
+ \vec{p}^{\,2}_E \big(M^{+1, f}_{\vec{p}_E,i} - M^{-1, f}_{\vec{p}_E,i}\big)^2\big]\Big]
\end{align}

where
\begin{align}
&M^{\lambda,f}_{\vec{p}_E, i}
= (-1)^{i - (1 - \lambda s_f)/2}\int_{0}^{\infty}dr\, r\, e^{r^2/2} \nonumber\\
&\times\bigg[m_0 + \sigma\, g\bigg(\frac{|Q_fB|}{2}r^2 + \vec{p}^{\,2}_E\bigg)\bigg]
L_{i - (1 - \lambda s_f)/2}(r^2)
\end{align}

where $f(k^2)$ stands for the Fourier transform of $f(x)$ and $L_n(x)$ are Laguerre polynomials. $S^{\text{MFA}}_{\text{BOS}}/V^{(4)}$ can be used to obtain the grand canonical partition function $\mathcal{Z}$ and hence the grand potential as $\Omega = -T\text{ln}\mathcal{Z}$.

\subsection{Polyakov Loop Extended Nonlocal NJL Model}
\label{NNJL_Polyakov_Section}
\begin{figure*}[t]
\centering
\subfloat[\centering ]{\includegraphics[width=0.48\textwidth]{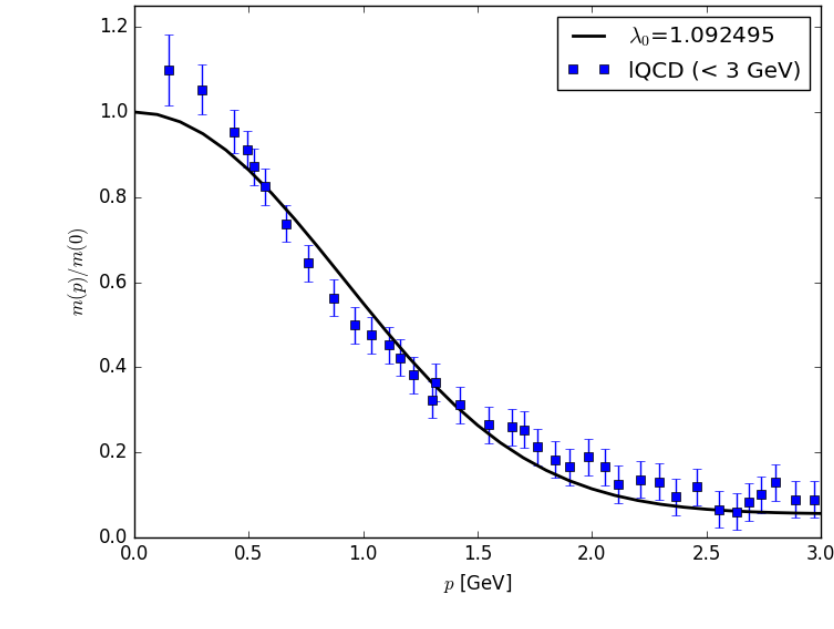}}
\hfill
\subfloat[\centering ]{\includegraphics[width=0.48\textwidth]{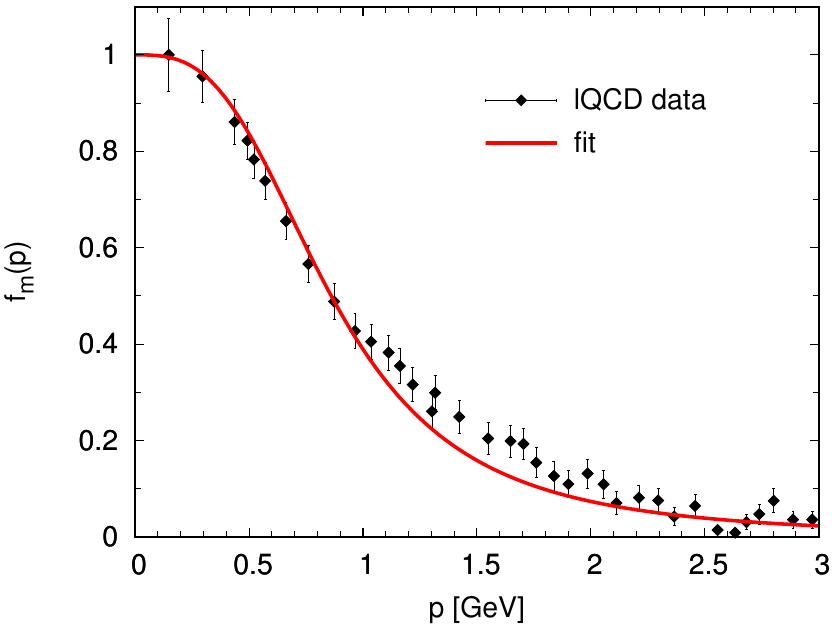}}
\caption{
    (a) Fit of lattice data from \citep{Parappilly:2005ei} for the momentum dependent non-perturbative meson mass function $m(p)$ in the nonlocal $SU(2)$ PNJL model \citep{Carlomagno:2019yvi}. 
    (b) Fit of lattice data from \citep{Parappilly:2005ei} for $f_m(p)$ in the ansatz $m_f + \alpha_m f_m(p)$ for the momentum dependent non-perturbative meson mass in the nonlocal $SU(3)$ PNJL model \citep{Carlomagno:2018tyk}.
    }
\label{fig:pnjl_combined}
\end{figure*}

\noindent One of the advantages of nonlocal PNJL model \citep{Carlomagno:2023clk} is that the critical temperature for deconfinement transition can be made naturally compatible with lattice QCD estimates, as shown in Fig.\ref{fig:pnjl_combined}, without spoiling the entanglement between deconfinement and chiral restoration transition temperatures even for imaginary chemical potential. Moreover, it shows robust consistency for non-perturbative meson masses with lattice QCD results.
\par As a simple example, the Euclidean action for the $SU(2)_L\times SU(2)_R$ model with the scalar/pseudoscalar interactions, is as follows
\begin{align}
S_{\text{PNNJL}} = \int d^4x\, \Big[&\overline{\psi}\left(-i\slashed{D} + \hat{m}\right)\psi \nonumber\\
&- \frac{G_S}{2}\left(j_{a} j_{a} - j_{P} j_{P}\right) + \mathcal{U}(\Phi, \overline{\Phi}, T)\Big]
\end{align}

where $\hat{m}$ is the quark current mass matrix. In the process of bosonisation one introduces two scalars $\sigma_{1,2}(x)$ and a pseudoscalar $\pi^a(x)$. The mean field thermodynamical potential at finite temperature and arbitrary (in general, complex) chemical potential in the Polyakov gauge is given by \citep{Contrera:2016rqj}
\begin{widetext}
\begin{equation}
\Omega^{\text{MFA}} = -4T\mathlarger{\sum}_{c = r, g, b}~\mathlarger{\sum}_{n=-\infty}^{\infty}\int\frac{d^3\vec{p}}{(2\pi)^3}\text{ln}\Bigg[\frac{(\rho^c_{n,\vec{p}})^2 + M^2(\rho^c_{n,\vec{p}})}{Z^2(\rho^c_{n,\vec{p}})}\Bigg] + \frac{\sigma^2_1 + \kappa^2_p\sigma_2^2}{2G_S} + \mathcal{U}(\Phi, \overline{\Phi}, T)
\end{equation}
\end{widetext}

Here $M(k) = Z(k)[m_f + \sigma_1g(k)]$ and $Z(k) = [1 - \sigma_2f(k)]^{-1}$ where $f(k)$ and $g(k)$ are fourier transforms of the form factors $\mathcal{F}(x)$ and $\mathcal{G}(x)$ \citep{Contrera:2016rqj} for the currents $j_a$ and $j_P$ respectively. Also $\rho^c_{n,\vec{p}} = \vec{p}^2 + [(2n + q)\pi T -i\mu + \phi_c]^2$, where the quantities $\phi^c$ are defined as $\phi = \text{diag}(\phi_r, \phi_g, \phi_b)$ and $\phi_r = \phi_3 + \phi_8/\sqrt{3}$, $\phi_g = -\phi_3 + \phi_8/\sqrt{3}$ and $\phi_b = -2\phi_8/\sqrt{3}$ in the Polyakov gauge where $\Phi$ can be written as $\Phi = \phi_3 T^3 + \phi_8 T^8$. Now formally $\Omega^{MFA}$ is divergent and has to be regularized. It is shown that the properly regularized potential \citep{Pagura:2011rt}
\begin{align}
    \Omega^{\text{MFA}}_{\text{NNJLP}} 
    = \Omega^{\text{MFA}} 
    - \Omega^{\text{MFA}}(\sigma_1 = 0 = \sigma_2) 
    + \Omega^{\text{free}}_{\text{reg}} 
    + \Omega_{0}
\end{align}
where the regularized expression of $\Omega^{\text{free}}_{\text{reg}}$ is given by
\begin{align}
&\Omega^{\text{free}}_{\text{reg}} = -4T\int\frac{d^3\vec{p}}{(2\pi)^3}\mathlarger{\sum}_{c = r, g, b}~\mathlarger{\sum}_{s = \pm 1} \nonumber\\
&\times\mathrm{Re}\,\ln\!\Big[1 + e^{-[\epsilon(p) + s(\mu + i\phi_c)]/T}\Big]
\end{align}

In the presence of an applied magnetic field, the grand potential in the mean field approximation becomes \citep{Carlomagno:2022kxx} 
\begin{align}
&\Omega^{\mathrm{MFA}}_{B,T,\mu}
= \frac{\overline{\sigma}^2}{2G}
- T \mathlarger{\sum}_{n=-\infty}^{\infty} ~\mathlarger{\sum}_{c=r,g,b} ~\mathlarger{\sum}_{f=u,d}
\frac{|Q_f B|}{2\pi} \int \frac{dp_3}{2\pi} \nonumber\\
&\times\Big[\ln\!\big(p_{\parallel}^2 + \big(M^{\lambda,f}_{0,p_{\parallel}}\big)^2\big)
+ \mathlarger{\sum}\nolimits_{k=1}^{\infty} \ln \Delta^{f}_{k,p_{\parallel}}\Big]
+ \mathcal{U}(\Phi,\overline{\Phi}, T)
\label{nonlocpnjlomega}
\end{align}
where the expressions of $\Delta^{f}_{k,p_{\parallel}}$ and $M^{\lambda,f}_{k,p_{\parallel}}$ can be found in \citep{Carlomagno:2022kxx}. The quantity $p_{\parallel} = (p_3, \omega_n - i\mu - \phi_c)$. The subscripts $c = r, g, b$ stand for color index. Using the techniques in \citep{GomezDumm:2017iex}, the regularized thermodynamic potential is given by \citep{Carlomagno:2022kxx} 

\begin{align}
\Omega^{\mathrm{MFA,reg}}_{B,T,\mu}
= \Omega^{\mathrm{MFA}}_{B,T,\mu}
- \Omega^{\mathrm{free}}_{B,T,\mu}
+ \Omega^{\mathrm{free,reg}}_{B,T,\mu}
\end{align}
\noindent Here, $\Omega^{\mathrm{free}}_{B,T,\mu}$ is evaluated at $\sigma = 0$, keeping the interaction with the magnetic field and the Polyakov loop. The term corresponding to $\Omega^{\mathrm{free,reg}}_{B,T,\mu}$ can be written as
\begin{align}
&\Omega^{\mathrm{free,reg}}_{B,T,\mu}
= - \frac{3}{2\pi^2} \mathlarger{\sum}_f (Q_f B)^2
\left[ \zeta'(-1, x_f) + F(x_f) \right] \nonumber\\
&- T \mathlarger{\sum}_{c,f,k} \frac{|Q_f B|}{2\pi}\, \alpha_k
\int \frac{dp}{2\pi} \, G^{f}_{k,p}(\phi_c, \mu, T)
\end{align}
with $F(x_f)$ and $G^{f}_{k,p}(\phi_c, \mu, T)$ can be found in \citep{Carlomagno:2022kxx}.
\subsection{KMT Interactions and Polyakov Loop Corrected Nonlocal NJL Model}
\label{Polyakov_KMT_NNJL_Section}
The KMT (to take care of the $U(1)_A$ anomaly of QCD) and Polyakov loop (to take care of confinement) corrected $SU(3)_L \times SU(3)_R$ NNJL model is given by the following Euclidean action
\begin{widetext}
\begin{equation}
S_{E} = \int d^4_E x \Bigg[\overline{\psi}(x)(-i\slashed{D} + \hat{m}_0)\psi(x)
-\frac{G}{2}\big(j^S_a~j^S_a - j^P_a~j^P_a\big) -\frac{H}{4}
T_{abc}\big(j^S_a~j^S_b~j^S_c-3j^S_a~j^P_b~j^P_c\big) + \mathcal{U}(\Phi[G(x)])\Bigg]
\end{equation}
\end{widetext}
where the term involving the coupling H is the nonlocal 'tHooft term and $T_{abc}$ are the structure constants of $SU(3)_C$. The grand potential for the model at finite temperature is as follows:
\begin{widetext}
\begin{align}
\Omega_{MFA} ={}& -2T\mathlarger{\sum}_{f,c}\int\frac{d^3\vec{p}}{(2\pi)^3}\mathlarger{\sum}_{n = -\infty}^{\infty}\text{Tr}\text{ln}[\omega^2_{nc} + \Sigma^2_{fc}(\omega^2_{nc})] - \frac{1}{2}\Bigg[\mathlarger{\sum}_{f}\bigg(\overline{\sigma}_f\overline{S}_f
+ \frac{G_S}{2}\overline{S}^2_f\bigg) + \frac{H}{2}\overline{S}_u\overline{S}_d\overline{S}_s\Bigg] + \mathcal{U}(\Phi, \overline{\Phi}, T)
\end{align}
\end{widetext}
Here, $\omega^2_{nc} = (\omega_n - \phi_c)^2 + \vec{p}^2$,  $\phi = \text{diag}(\phi_r, \phi_g, \phi_b )$ and $\overline{S}_f$ are auxiliary fields. The constituent masses $\Sigma_{fc}$ are defined as 
\begin{align}
    \Sigma_{fc}(\omega^2_{nc}) = m_f + r^2(\omega^2_{nc})\overline{\sigma}_f
\end{align}
As the first term of $\Omega_{MFA}$ is formally divergent, it needs to be regularized. Authors of \citep{Contrera:2007wu} have used the strategy used in \citep{GomezDumm:2004sr} to regularize the potential.
\section{Effects of Magnetic field on the EoS: Anisotropic Pressure}
Although many works have been done on the effects of magnetic fields on neutron stars \cite{Martinez:2003dz,Yue:2009jh,Chen:2005dx,Harding:2006qn,Wei:2005aga} and quark stars \cite{Rabhi:2009ih,Menezes:2009uc,Menezes:2008qt,Felipe:2007vb}, no consistent scheme was followed in these works and hence, such works have different opinions about what the correct magentic field contributions to the pressure and energy density should be. However, it is well known that in the presence of a magnetic field, the pressure splits into components transverse ($p_{\perp}$) and parallel ($p_{||}$) to the field direction. Many works, in fact, have ignored the pressure anisotropy even at very strong magnetic fields where such effect is most relevant. Ref.\cite{Ferrer:2010wz} has formalized a systematic approach to calculate the EoS of such magnetized system where the main quantity of interest is the Maxwell part of the energy-momentum tensor
\begin{align}
    \tau^{\mu\nu}_{M} = (\epsilon - p)u^\mu u^\nu + p(\eta^{\mu\nu}_{||} - \eta^{\mu\nu}_{\perp})
\end{align}
where $\epsilon = p = H^2/2$, $u_{\mu}$ is the four velocity vector of the medium which in the rest frame of the system becomes $u_{\mu} = (1, \vec{0})$, $\eta^{\mu\nu}_{||} = \text{diag}(1, 0, 0, 1)$ and $\eta^{\mu\nu}_{\perp} = \text{diag}(0, 1, 1, 0)$ are the longditudinal and transverse metrics. The energy-momentum tensor of the Dirac field is given by 
\begin{align}
    \tau^{\mu\nu}_\psi = \frac{1}{2}\overline{\psi}\big[\gamma^\mu \overrightarrow{D}^\nu + 
\gamma^\nu \overrightarrow{D}^\mu\big]\psi - \eta^{\mu\nu}\overline{\psi}\big[i\gamma^{\rho}D_{\rho} - m\big]\psi
\end{align}
The quantum statistical average of the energy momentum tensor is 
\begin{align}
\langle \tilde{\tau}^{\rho \lambda} \rangle
&= \frac{1}{\mathcal{Z}}\,\mathrm{Tr}\Big[\tilde{\tau}^{\rho\lambda}e^{-\beta(H - \mu N)}\Big] \nonumber\\
&= \frac{1}{\mathcal{Z}}\int [d\phi_0]\,\tilde{\tau}^{\rho\lambda}
\exp\!\bigg[\int_{0}^{\beta}d\tau\int d^3x\, \mathcal{L}(\tau, x)\bigg]
\end{align}
where 
\begin{align}
    \tau^{\rho\lambda} = \int_{0}^{\beta}d\tau\int d^3x \big[\tau^{\rho\lambda}_{M} + \tau^{\rho\lambda}_{\psi}\big]
\end{align}

\begin{figure}[hbt]
\centering

\begin{minipage}[t]{0.48\textwidth}
    \centering
    \includegraphics[
        angle=-90,
        origin=c,
        height=0.9\linewidth
    ]{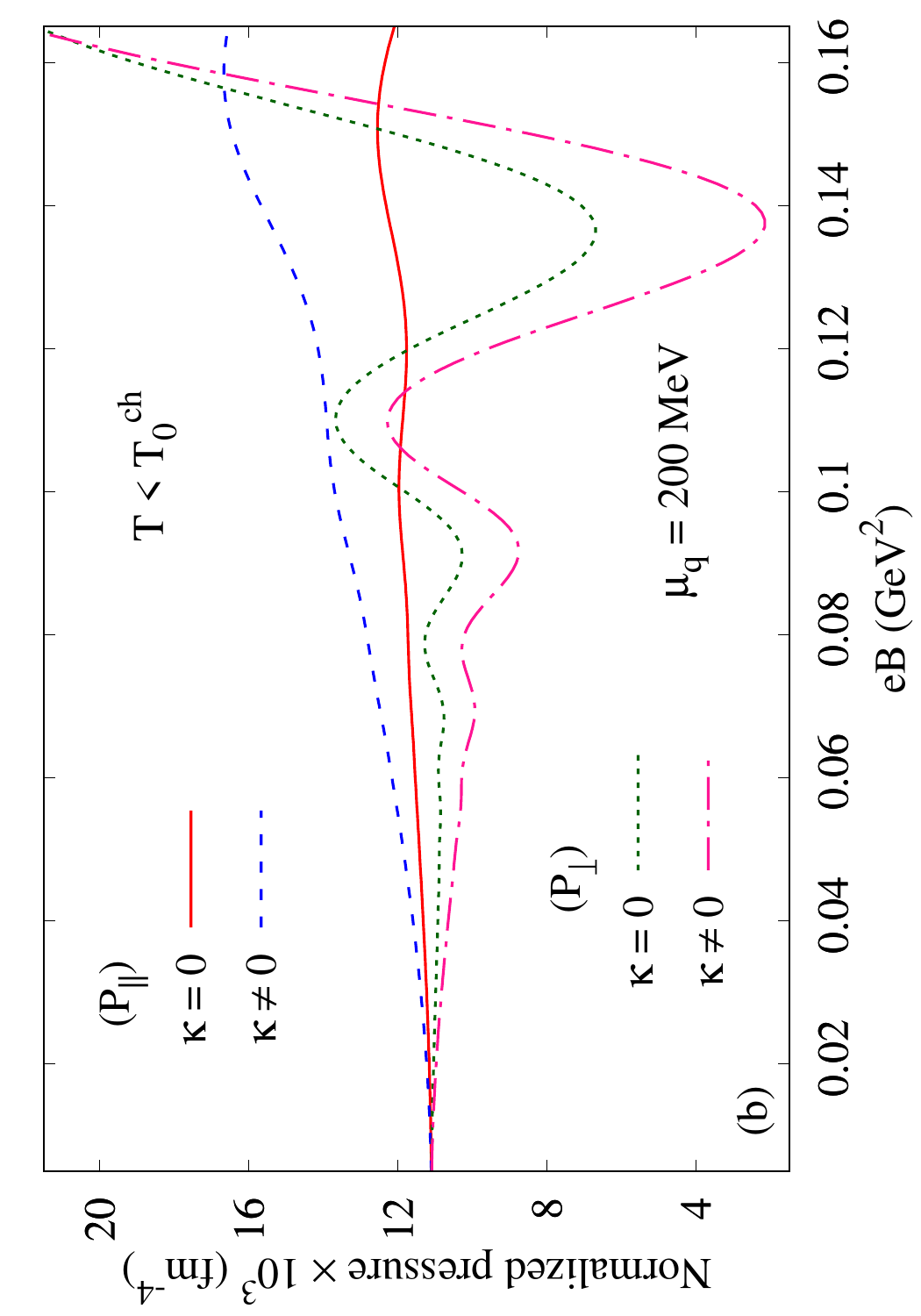}

    \vspace{-10mm}
    \textbf{(a)}
\end{minipage}
\hfill
\begin{minipage}[t]{0.48\textwidth}
    \centering
    \includegraphics[
        angle=-90,
        origin=c,
        height=0.9\linewidth
    ]{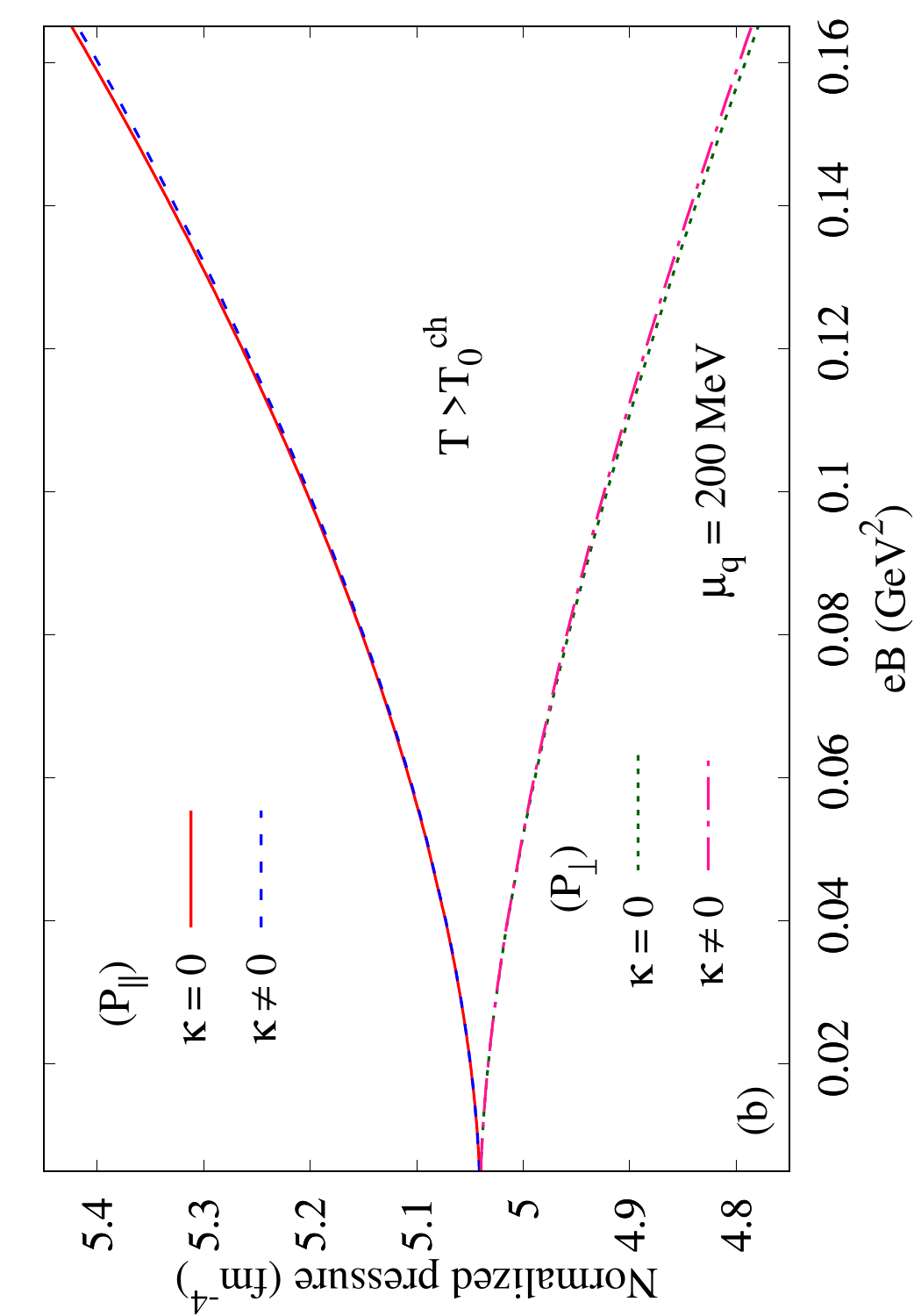}

    \vspace{-10mm}
    \textbf{(b)}
\end{minipage}
\vspace{-6mm}
\caption{Normalized longitudinal ($p_{||}$) and transverse ($p_{\perp}$) pressure as a function of $eB$ for temperature (a)$T < T^{\text{ch}}_{0}$ and (b) $T > T^{\text{ch}}_{0}$ for vanishing and finite values of AMM for quark chemical potential $\mu_q = 200~$MeV. Taken from Ref.\cite{Chaudhuri:2022oru}.}
\label{fig:pressure_par_perp_vs_eB}
\end{figure}
$\mathcal{Z}$ is the canonical partition function and $\phi_0$ denotes the field content of the lagrangian. The energy density is given by 
\begin{align}
    \epsilon = \frac{1}{\beta V}(\tilde{\tau}^{00}_{\psi} + \tilde{\tau}^{00}_{M}) = \Omega_f + TS + \mu N + \frac{B^2}{2}
\end{align}
The $p_{||}$ and $p_{\perp}$ is obtained as 
\begin{align}
    p_{||} = \frac{1}{\beta V}(\langle \tilde{\tau}^{33}_{\psi} \rangle + \langle \tilde{\tau}^{33}_{M} \rangle) = -\Omega_f - \frac{B^2}{2}
\end{align}

\begin{align}
    p_{\perp} = \frac{1}{\beta V}(\langle \tilde{\tau}^{11}_{\psi} \rangle + \langle \tilde{\tau}^{11}_{M} \rangle) = -\Omega_f - BM_f + \frac{B^2}{2}
\end{align}
where $M_f = -\partial \Omega_f/\partial B$ is the magnetization of the system. Ref. \cite{Yang:2021rdo} shows that the compressibility of the system also splits into parallel and perpendicular components as 
\begin{align}
\kappa^{\parallel}
&=
-\frac{1}{V}\left(\frac{\partial V}{\partial n_q}\right)
\left(\frac{\partial n_q}{\partial p^{\parallel}}\right)_T
=
\frac{1}{n_q^2}\left(\frac{\partial n_q}{\partial \mu}\right), \\
\kappa^{\perp}
&=
\frac{1}{n_q}
\left(\frac{\partial n_q}{\partial p^{\perp}}\right).
\end{align}
Moreover, in an anisotropic medium, it is essential to distinguish between longditudial and transverse speed of sound as 
\begin{align}
C^2_{s,\parallel,N} &= \frac{\partial P_{\parallel,N}}{\partial \epsilon}\Bigg|_{B}
\qquad\qquad &
C^2_{s,\perp,N} &= \frac{\partial P_{\perp,N}}{\partial \epsilon}\Bigg|_{B}
\end{align}
and for large magnetic fields, they are expected to differ, as shown in Ref.\cite{Ferraris:2025fva}.
\par A detailed analysis of the magnetic field dependence of $p_{||}$ and $p_{\perp}$ has been done in Ref.\cite{Chaudhuri:2022oru} in the context of PNJL model both in the presence and in the absence of the anomalous magnetic moment (AMM) of quarks. Fig.\ref{fig:pressure_par_perp_vs_eB} shows the variation of $p_{||}$ and $p_{\perp}$ as a function of magnetic field for the quark chemical potential $\mu_q = 200~$MeV for $T < T^{\text{ch}}_{0}$ and $T > T^{\text{ch}}_{0}$, $T^{\text{ch}}_{0}$ being the chiral restoration temperature. For small field strengths, $p_{\perp}$ almost coincides with $p_{||}$, irrespective of the AMM of quarks and only begin to be diverging from each other for $eB > 0.03~$GeV$^2$. At even higher values of $eB$, $p_{||}$ shows slight oscillatory behavior but its magnitude remains more or less unchanged when the quark AMM is switched off but increases for finite quark AMM. Conversely, the $p_{\perp}$, becomes highly oscillatory for large values of $eB$, similar to the results shown in Ref.\cite{Ferraris:2025fva,Ferraris:2025bzb} in the context of NNJL model, and the amplitude of oscillation increases for increasing quark AMM. For $T > T^{\text{ch}}_{0}$ these oscillations are absent and $p_{||}$ and $p_{\perp}$ diverge from each other even for small values of $eB$. The de Haas–van Alphen oscillations are reflected in oscillation of magnetization for large enough magnetic fields, similar to the results in Ref.\cite{Ferraris:2025fva}.

\begin{figure}[t]
\centering

\begin{minipage}[t]{0.48\textwidth}
    \centering
    \includegraphics[width=\linewidth]{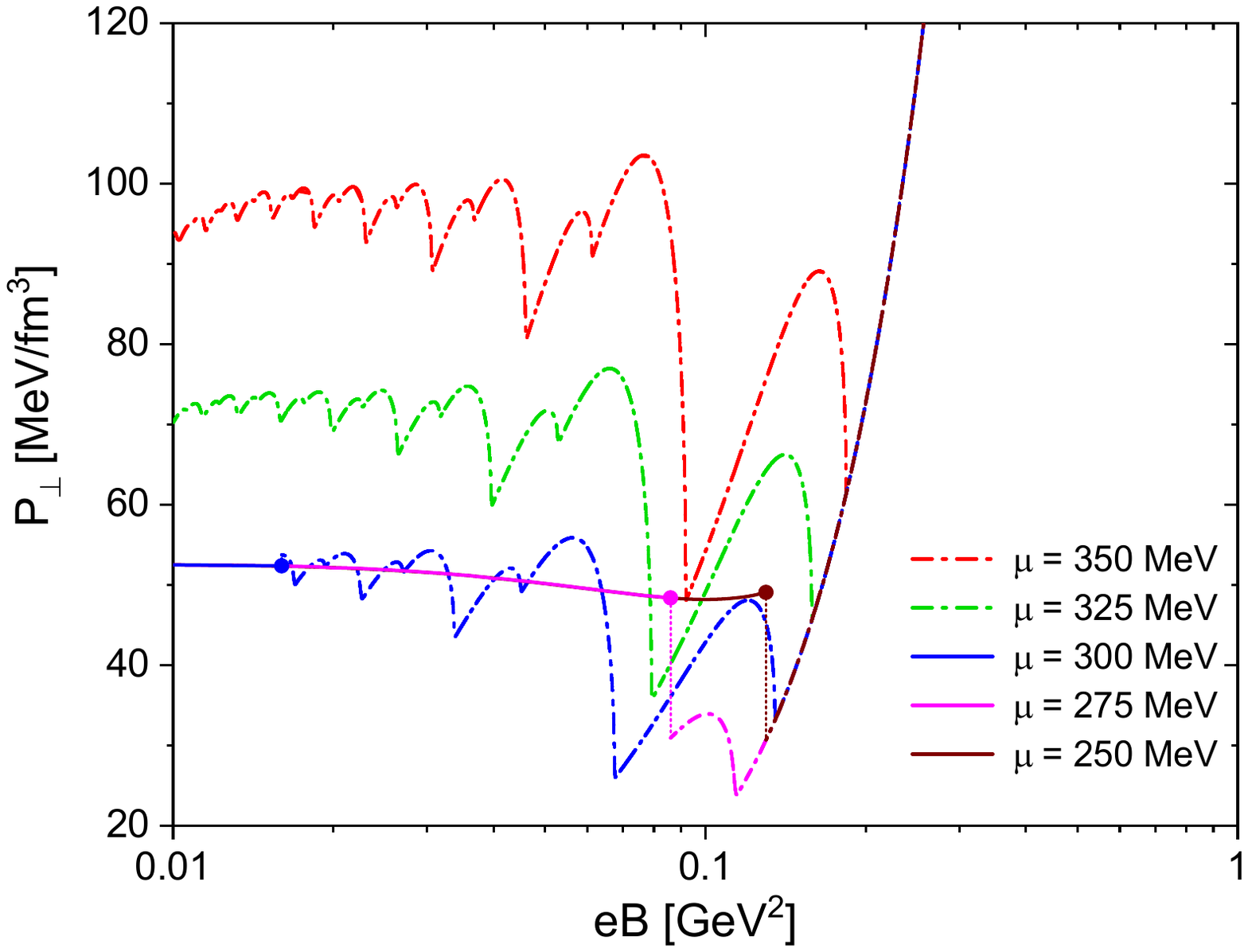}

    \vspace{2mm}
    \textbf{(a)}
\end{minipage}
\hfill
\begin{minipage}[t]{0.48\textwidth}
    \centering
    \includegraphics[width=\linewidth]{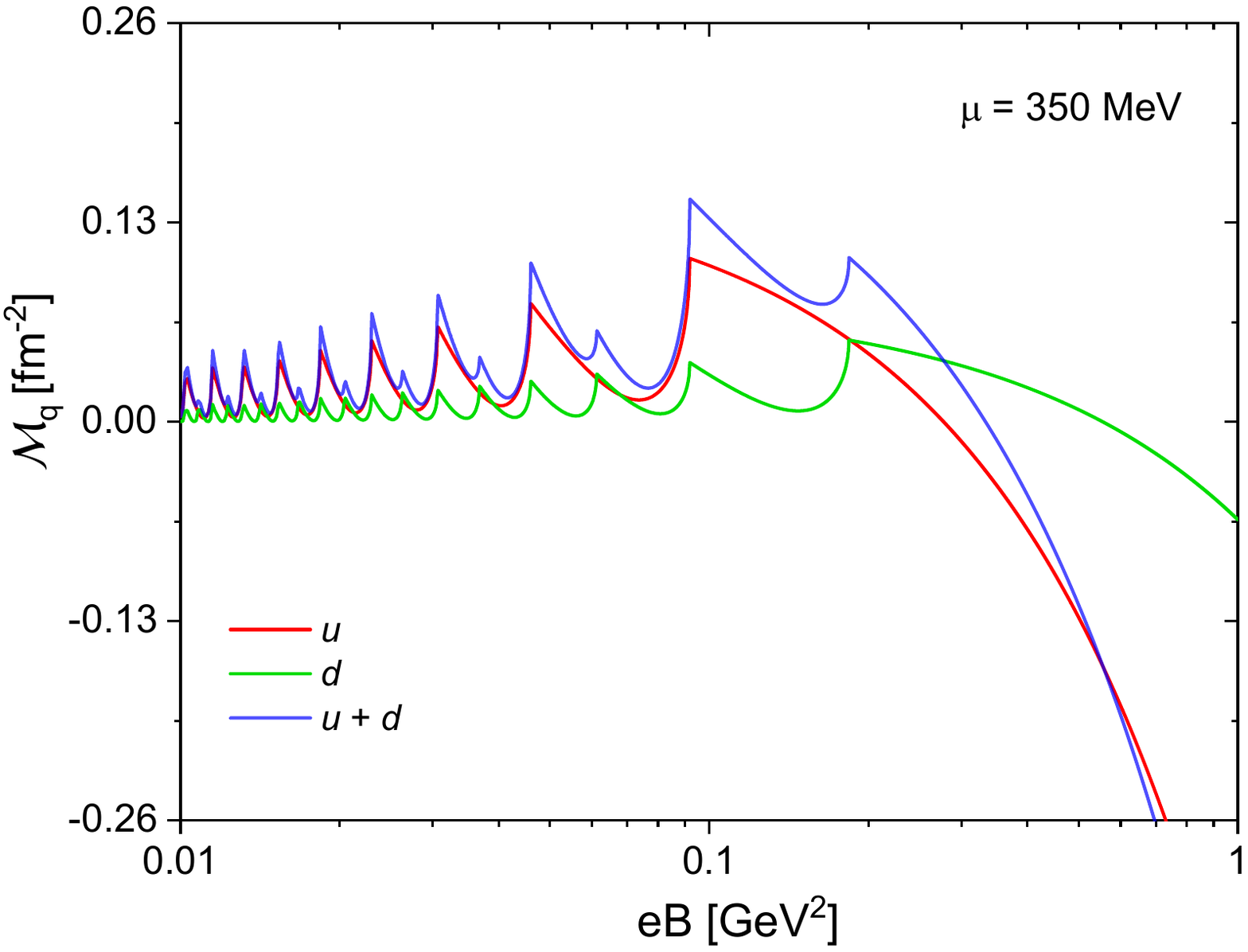}

    \vspace{2mm}
    \textbf{(b)}
\end{minipage}

\vspace{-2mm}
\caption{Caption describing both panels. (a) $p_\perp$
as functions of $eB$ for various benchmark values of quark chemical potential. Solid lines (broken lines) correspond to the chiral symmetry broken (restored phase) phase. (b) System magnetization $\mathcal{M}_q$
to the applied magnetic field $eB$ where De Haas–Van Alphen effect is manifested in oscillation of $\mathcal{M}_q$ as function of magnetic field. Taken from \cite{Ferraris:2025bzb}.}
\label{fig:combined}
\end{figure}

\section{Phases of QCD matter with magnetic fields}

\noindent Following the discussion of the NJL model and its variants, we now turn to the rich phase structure and emergent phenomena predicted within these frameworks. Of particular interest is the behavior of the chiral condensate in the presence of external magnetic fields. Effective theories generally predict MC (magnetic catalysis) - an enhancement of chiral symmetry breaking in the presence of a magnetic field. However, finite-temperature lattice QCD simulations reveal an opposite trend near the chiral crossover region, where the condensate is suppressed with increasing magnetic field - a phenomenon known as inverse magnetic catalysis or IMC~\citep{Li:2019nzj}. Reconciling these two competing effects is therefore essential for a consistent description of magnetized SIM. Moreover, strong magnetic fields can induce mesonic condensation channels, notably $\pi$, and perhaps somewhat controversially, $\rho$ condensates, which, depending on the density, could be favored or disfavored. 

\par The phase structure becomes even richer at high baryon density and low temperature, where various color superconducting phases are expected to emerge. At asymptotically high densities, the ground state is predicted to be the CFL phase, characterized by symmetric pairing among $u$, $d$, and $s$ quarks. In the presence of a magnetic field, the CFL phase is modified into the magnetic CFL (MCFL) phase, which retains the pairing structure but exhibits reduced symmetry. At intermediate densities, less symmetric phases such as the 2SC phase become relevant, dominated by pairing between up and down quarks. In electrically neutral 2SC matter, Fermi-surface mismatch between pairing quarks can generate imaginary Meissner masses for certain gluons, signaling chromomagnetic instability and potentially favoring inhomogeneous phases such as the CCS phase or the Larkin-Ovchinnikov-Fulde-Ferrell (LOFF) phase.
At even lower densities, where hadronic degrees of freedom become as important as quark degrees of freedom, quarkyonic phases dominate and understanding their magnetic properties could be relevant for better understanding of highly magnetized neutron star properties. The competition between these phases leads to a very complex phase structure where MC and IMC further modify the phase boundaries. The magnetic behavior of the chiral condensate is primarily governed by the phenomenon of MC, which predicts an enhancement of chiral symmetry breaking in the presence of a magnetic field. However, this expectation is challenged by lattice QCD results at finite temperature, which indicate a suppression of the condensate near the transition region, the IMC effect. Understanding the origin and interplay of these two competing mechanisms is therefore essential for a consistent description of phases of QCD.
\subsection{Magnetic Catalysis (MC) Effect}
\label{MC_section}
The MC effect is the enhancement of the expectation value $\langle \overline{\psi}\psi \rangle$ under conditions of a strong external magnetic field. Within the context of the NJL model, magnetic fields have a tendency to enhance spin-zero $\langle \overline{\psi}\psi\rangle$ condensates associated with the breaking of chiral symmetry, leading to dynamical generation of fermion masses. This behavior contrasts sharply with ordinary superconductivity. Unlike conventional superconductors, a state with nonzero $\langle \overline{\psi}\psi \rangle$ does not exhibit the Meissner effect. Furthermore, in a neutral spin-zero quark--antiquark pair, the magnetic moments of the fermion and antifermion align parallel to the external field, making the condensate energetically favored in strong magnetic backgrounds, unlike fermion--fermion Cooper pairs in ordinary superconductors. Before describing the varieties of works that have been done on MC, we want to introduce the concepts of dimensional reduction of quarks \cite{Shovkovy:2012zn} in the presence of strong magnetic fields and how MC can be obtained in simple models.
\par In the expression of $\epsilon_{f\text{k}}$, $\text{k} = n + s + 1/2$ where $n\in\mathbb{Z}_{\geq 0}$ is the orbital quantum number whereas $s = \pm~1/2$ is the spin projection quantum number in the direction of the applied magnetic field. The spin polarized lowest Landau level (LLL) corresponds to $n = 0$ and $s = -1/2$ only. For $q_f B >> m^2_f$, the minimum of the first excited LL has energy $E^2_1 \approx 2|q_f|B$, which is large for large magnetic field. On the other hand, the LLL has the energy $E^2_0 = p^2_z + M^2_f$. So for large magnetic fields, the energy gap between these two levels is approximately $\Delta E \approx \sqrt{2|q_f|B}$ for large magnetic field. Hence for large magnetic fields, the LLL decouples from the higher levels and the low energy physics is solely determined by the level $\text{k} = 0$. The spectrum of LLL is identical to that of a $(1+1)$-dimensional field theory with a single spatial coordinate identified with the $z$ direction. The dynamics of LLL changes to $1 + 1$-dimensional. This phenomenon is termed as dimensional reduction. It is essentially a consequence of restricted fermion motion in a plane perpendicular to the magnetic field.
\par Having explained the concept of dimensional reduction, we try to explain how dimensional reduction plays a role in MC. The easiest way to illustrate this effect is in $2 + 1$ dimensions. Consider the fermionic condensate in the presence of a magnetic field
\begin{align}
    |\langle \overline{\psi}\psi \rangle|\xrightarrow[B\neq 0]{}2N_CM_f\frac{|q_f|B}{(2\pi)}\mathlarger{\sum}_{_\text{k}}\frac{\alpha_\text{k}}{\sqrt{M^2_f + 2|q_f|B\text{k}}}
\end{align}

The replacement of the 2 dimensional momentum integral measure by just a summation on the LLs reflects dimensional reduction where the dynamics has shifted from $2 + 1$ to $0 + 1$ dimensions. Now in the limit of large magnetic field, i.e. $|q_f|B >> M^2_f$, the condensate is dominated by the lowest Landau level k $= 0$
\begin{align}
    |\langle \overline{\psi}\psi \rangle| \sim 2N_CM_f\frac{|q_f|B}{(2\pi)}\frac{1}{M_f} = \frac{N_C|q_f|B}{\pi} \sim~B
\end{align}
which is precisely the phenomenon of magnetic catalysis: the magnitude of the condensate increasing with the magnetic field, linearly in this specific case. Similar behavior can be shown in the most general $D + 1$ dimensions. 

\par The robustness of this phenomenon has been tested numerous times within the framework of NJL model \citep{Menezes:2009uc,Fayazbakhsh:2010bh,Chatterjee:2011yi,Avancini:2012ee,Ferrari:2012yw} and within LQCD frameworks as well \citep{Buividovich:2010tn, Cea:2011hu, DElia:2010abb, DElia:2011koc, Bali:2011qj, Bali:2011uf, Bali:2012zg, Frasca:2011zn, Bruckmann:2011zx, Fraga:2012fs}.
The phenomenon of MC appears in other formalisms of QCD as well. For example, within the Dyson-Schwinger framework, one finds in the mean-field, large-$N_c$, and weak-coupling ($G$) limit that the dynamical mass scales as, $m \approx GN_c(eB/2\pi)$, demonstrating that the magnetic field enhances chiral symmetry breaking through dimensional reduction and Landau-level dynamics \citep{Shovkovy:2012zn}. In the $3 + 1$ massless NJL model, the following non-trivial solution of the gap equation shows the MC effect

\begin{align}
m\approx \sqrt{\frac{|eB|}{\pi}}\text{exp}\bigg(\frac{\Lambda^2}{2|eB|}\bigg)\text{exp}\bigg(-\frac{2\pi^2}{G|eB|}\bigg).
\end{align} 
\par
\begin{figure*}[hbt]
\centering
\def\trimA{trim={0pt 0pt 0pt {.0375\height}},clip}
\def\trimB{trim={0pt 0pt 0pt 0pt}}
\def\trimC{trim={0pt 0pt 0pt {.0375\height}},clip}
\sbox0{\expandafter\adjustbox\expandafter{\trimA,height=10cm}{\includegraphics{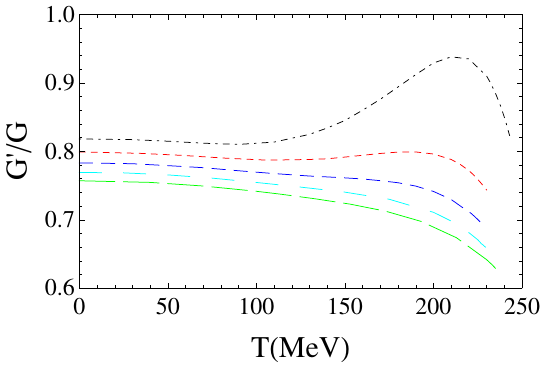}}}%
\sbox2{\expandafter\adjustbox\expandafter{\trimB,height=10cm}{\includegraphics{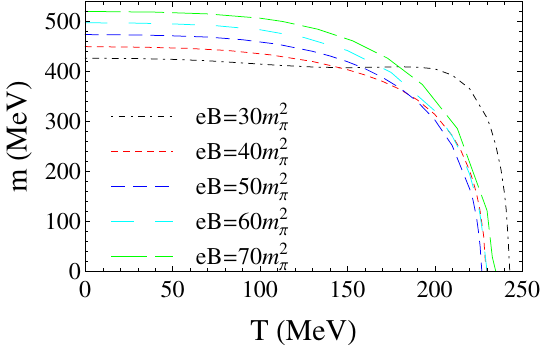}}}%
\sbox4{\expandafter\adjustbox\expandafter{\trimC,height=10cm}{\includegraphics{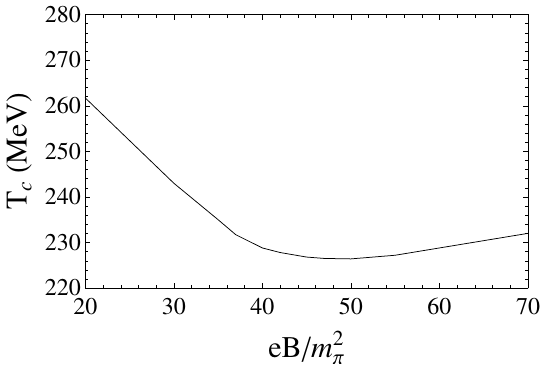}}}%
\dimen0=\dimexpr\textwidth-2em\relax
\dimen2=\dimexpr\wd0+\wd2+\wd4\relax
\edef\figH{\the\dimexpr 10cm*\dimen0/\dimen2\relax}%
\subfloat[\centering ]{\expandafter\adjustbox\expandafter{\trimA,height=\figH}{\includegraphics{fig11.pdf}}}%
\hfill
\subfloat[\centering ]{\expandafter\adjustbox\expandafter{\trimB,height=\figH}{\includegraphics{fig12.pdf}}}%
\hfill
\subfloat[\centering ]{\expandafter\adjustbox\expandafter{\trimC,height=\figH}{\includegraphics{fig2.pdf}}}
\caption{(a) Magnetic inhibition of the effective coupling $G^\prime$ due to the presence of mesons, (b) MC of the current quark mass and (c) Delayed MC of the transition temperature $T_c(B)$. These results have been taken from~\cite{Mao:2016lsr}.}
\label{DMC}
\end{figure*}

 Various nuances in the phenomena have been unraveled in further investigation of MC and IMC~\citep{Shi:2015ufa,Mueller:2015fka,Cao:2019res,Wang:2022xxp}. Using the NJL model, Ref. \citep{Fukushima:2012kc} shows that in the presence of a strong magnetic field, quark fluctuations would enhance the chiral symmetry breaking, while the effect of neutral mesons is to dampen it. This result is supported by Ref.~\citep{Mao:2016fha} where it was shown that in comparison with mean field approximation containing quarks only, including mesons in the model leads to a transition from inverse magnetic catalysis (IMC, to be discussed in the next subsection), to delayed MC at finite temperature~\citep{Mueller:2015fka}. Inclusion of mesons is reflected in the medium-dependent coupling $G^\prime(T,\mu_B, B)$, and it is shown that $G^\prime / G < 1$, indicative of magnetic inhibition that leads to IMC~\citep{Mao:2016lsr}, as shown in Figs.~\ref{DMC}(a) and \ref{DMC}(b). However, after a critical value of the magnetic field is reached, the MC effect dominates the magnetic inhibition; hence, MC is delayed.
\par Ref.~\citep{Ahmad:2016iez} demonstrates that a vector–vector contact interaction produces MC; however, when a lattice-inspired running coupling that decreases with magnetic field strength is introduced, IMC emerges instead. In a related development, Refs.~\citep{Ferrer:2013noa,Ferrer:2014rya} show that the explicit breaking of rotational symmetry by the magnetic field gives rise via Fierz transformations to new tensor interaction channels. This framework also predicts that MC induces not only the conventional chiral condensate but an additional spin-one condensate, both of which characterize the chirally broken phase.

{\renewcommand{\thesubfigure}{(\Roman{subfigure})}
\begin{figure*}[hbt]
\centering
\subfloat[\centering]{\includegraphics[width=0.485\textwidth]{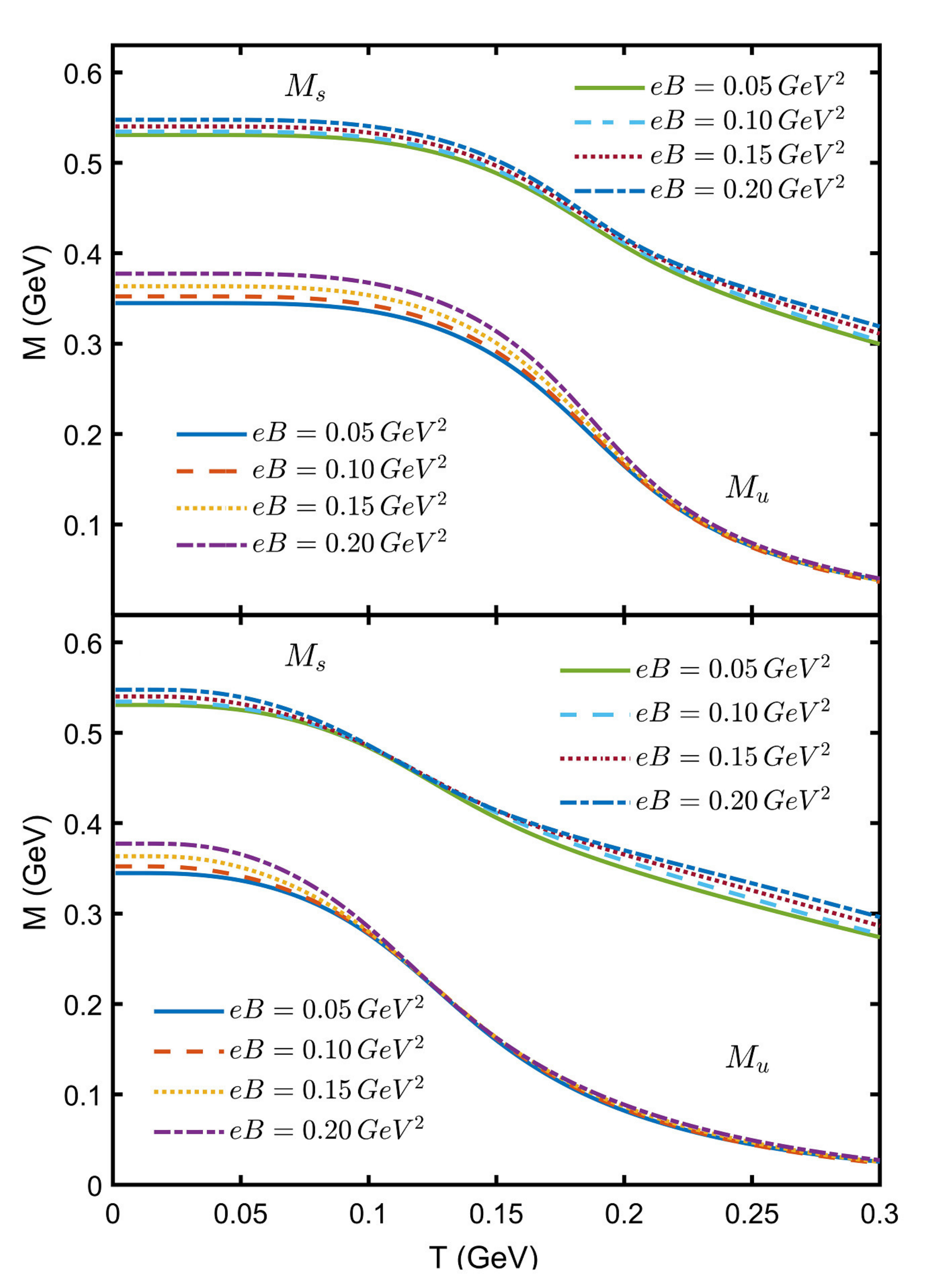}}
\hfill
\subfloat[\centering]{\includegraphics[width=0.485\textwidth]{fig2_TSP_lr.pdf}}
\caption{The dependence of dynamical quark mass M on temperature (T) for four different magnetic fields without considering TSP in (I) and with considering TSP in (II). MC upon inclusion of TSP is clearly observed. Figs.(a) is for the quark chemical potential $\mu = 0~\text{GeV}$ and (b) is for $\mu = 0.25~\text{GeV}$. These results are taken from \citep{Qiu:2023kwv}.}
\label{fig:TSP}
\end{figure*}
}

\begin{figure*}[t]
\centering
\def\trimB{trim={0pt 0pt 0pt {.037\height}},clip}
\sbox0{\includegraphics[height=10cm]{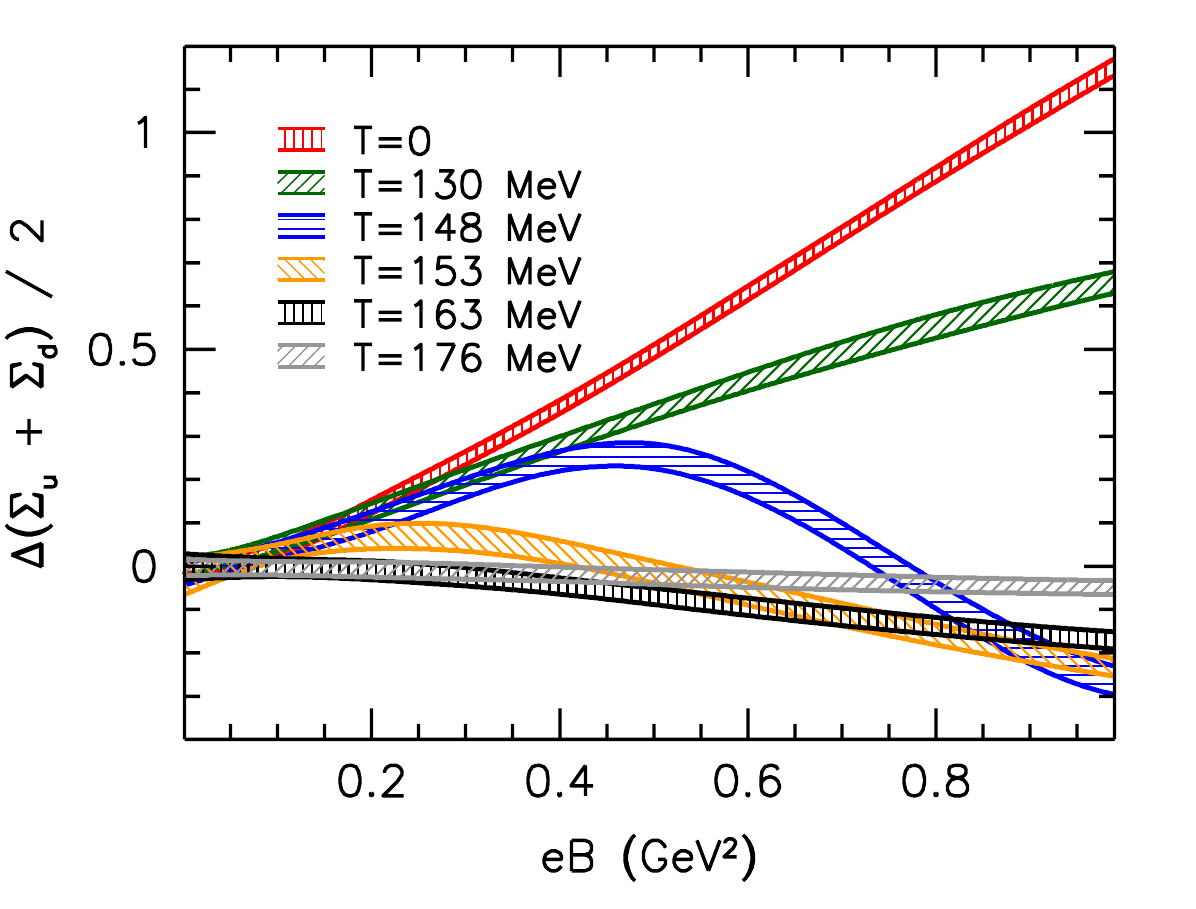}}%
\sbox2{\expandafter\adjustbox\expandafter{\trimB,height=10cm}{\includegraphics{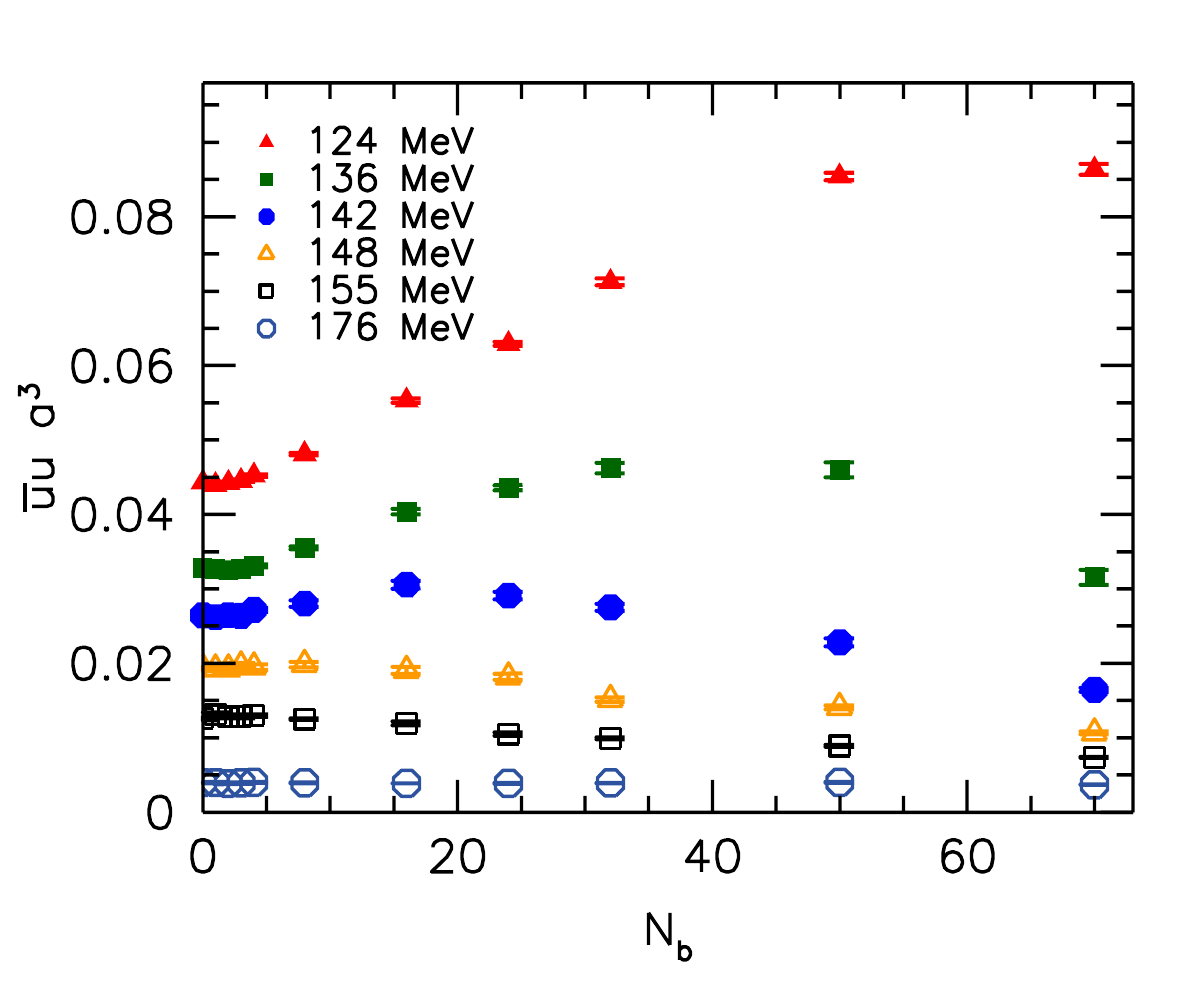}}}%
\dimen0=\dimexpr\textwidth-1em\relax
\dimen2=\dimexpr\wd0+\wd2\relax
\edef\figH{\the\dimexpr 10cm*\dimen0/\dimen2\relax}%
\subfloat[\centering ]{\includegraphics[height=\figH]{pbp_B_T.pdf}}%
\hfill
\subfloat[\centering ]{\expandafter\adjustbox\expandafter{\trimB,height=\figH}{\includegraphics{pbp_beta.pdf}}}
\caption{Results from (a) \citep{Bali:2012zg} and (b) \citep{Bali:2011qj}. Although they show MC at low temperatures, at sufficiently high temperatures the condensates develop a non-monotonic dependence on $eB$ and even decrease with it, showing IMC. Here $\Sigma_{u,d}(B,T) = (2m_{u,d}/M^2_\pi F^2)[\overline{\psi}\psi_{u,d}(B,T) - \overline{\psi}\psi_{u,d}(0,0)]$, and $N_b\in \mathbb{Z}$ is the number of elementary magnetic flux quanta through the lattice.}
\label{fig:IMC_Lattice}
\end{figure*}

\begin{figure*}[t]
\centering
\subfloat[\centering ]{\includegraphics[width=0.5\textwidth]{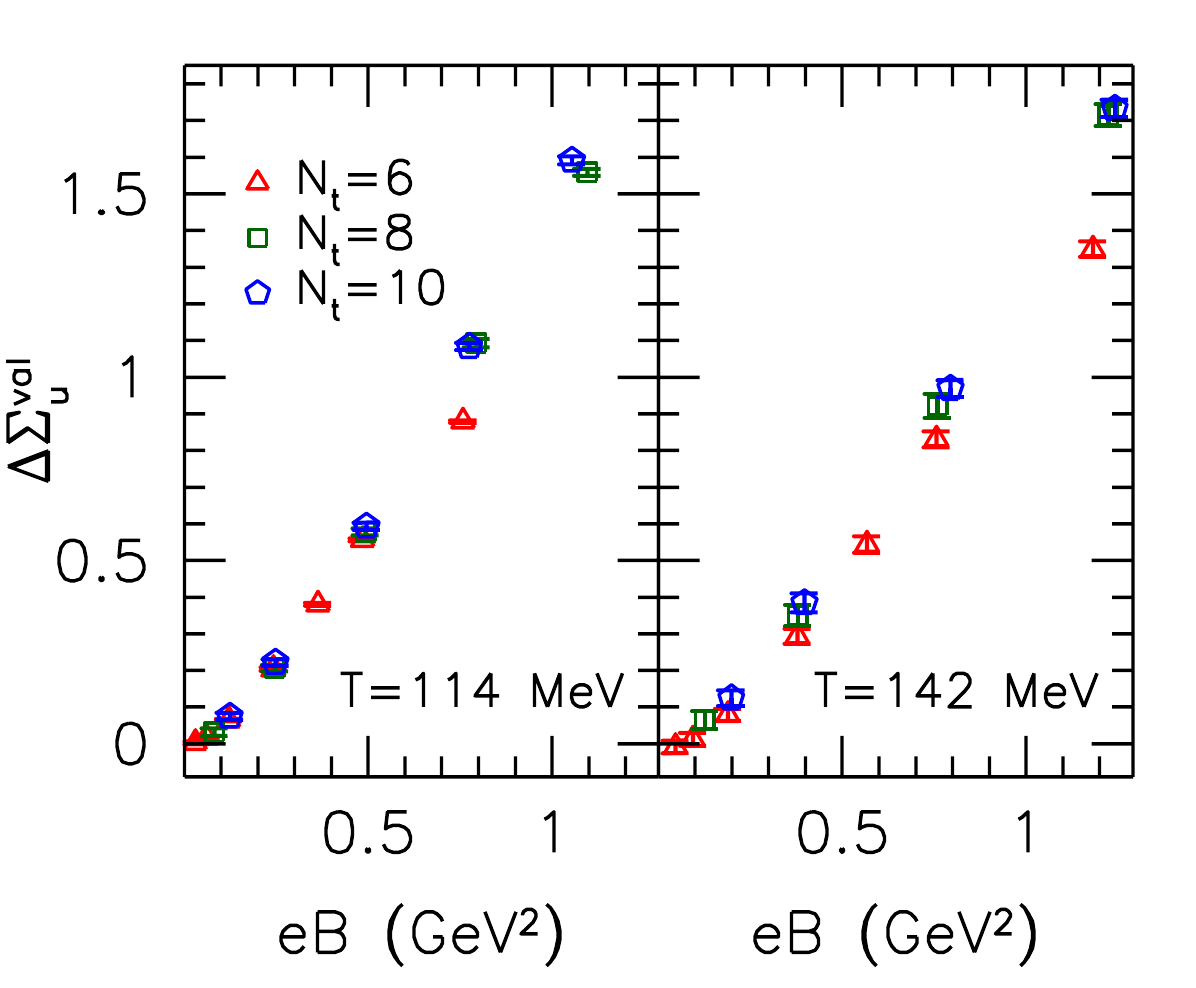}}%
\subfloat[\centering ]{\includegraphics[width=0.5\textwidth]{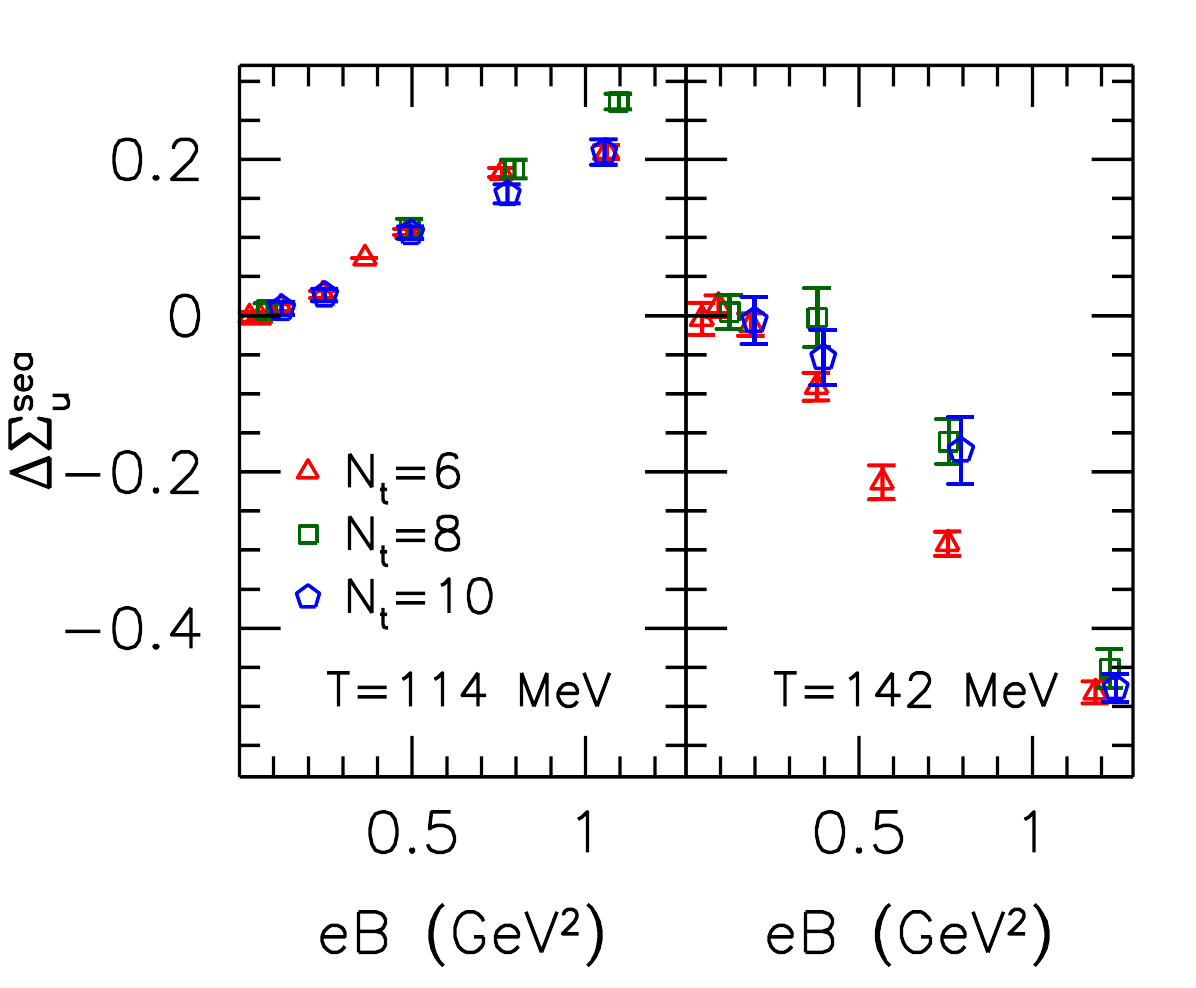}}
\caption{Valence (a) and sea (b) contributions to the up quark condensate as a function of $eB$, calculated at $T = 114~\text{MeV}$, which is well below $T_C$, and $T = 142~\text{MeV}$, very close to $T_C$. Near $T_C$, the decreasing negative sea contribution overwhelms the increasing positive valence contribution, which is interpreted as the mechanism behind IMC. Taken from \citep{Bruckmann:2013oba}.}
\label{fig:valence_sea_IMC}
\end{figure*}

The role of tensor structures is further emphasized in Ref.~\citep{Qiu:2023kwv}, where incorporating tensor spin polarization for heavy quarks enhances MC (see Fig.~\ref{fig:TSP}). Similarly, within a two-flavor NJL model, Ref.~\citep{Tsue:2016age} finds MC when tensor-type and axial-vector-type four-point interactions are included. The sensitivity of magnetic behavior to microscopic inputs is highlighted by Ref.~\citep{Kawaguchi:2022dbq}, which shows that the sign of the quark anomalous magnetic moment determines whether the interaction acts as a catalyzer or inhibitor.
\par MC has also been explored in non-extensive frameworks. For instance, Ref.~\citep{Islam:2023zpl} reports that, with a constant coupling, MC persists but with reduced strength compared to the extensive case. When the coupling depends on the magnetic field, a competition arises between the Tsallis non-extensive parameter $q$ and $eB$, leading to the suppression of IMC at sufficiently large $q$. Along similar lines, Ref.~\citep{Liu:2025uey} finds that, within Tsallis statistics, the chiral chemical potential can significantly modify the magnetic response, even driving a transition from MC to IMC.
\par Although MC emerges as a robust, model-independent feature, lattice QCD simulations at finite temperature have revealed a qualitatively different phenomena of IMC in the vicinity of the pseudo-critical region, where the chiral condensate is suppressed rather than enhanced as the magnetic field increases, indicating that additional dynamics beyond those captured in simple effective descriptions become relevant near the transition. We therefore turn now to a discussion of IMC. 
\subsection{Inverse Magnetic Catalysis}
\label{IMC_Section}
IMC~\citep{Andersen:2021lnk}, opposite to magnetic catalysis, is characterized by a decrease in the pseudo-critical temperature $T_{pc}$ with increasing magnetic-field strength, or equivalently, by the leftward shift of the chiral susceptibility peak $\chi_T$ as a function of temperature \citep{Farias:2014eca}. Although IMC is believed to arise from a subtle interplay of competing effects \citep{Endrodi:2019zrl}, active only within certain temperature ranges and predominantly affecting light quarks, its complete theoretical understanding remains unresolved. Most phenomenological models predict MC at all temperatures. However, high-precision lattice QCD simulations for $N_f = 2+1$ with physical pion masses, $m_\pi = 140~\text{MeV}$ \citep{Bali:2011qj,Bali:2012zg}, revealed a more intricate behavior (Fig.~\ref{fig:IMC_Lattice}): while MC persists at low temperatures, at sufficiently high temperatures the condensates exhibit a non-monotonic dependence on $eB$ and eventually decrease with increasing magnetic field.

\par Variations of the NJL model have motivated extensive efforts to explain the discrepancy between MC predicted by effective models and IMC observed in lattice QCD~\citep{Fukushima:2010fe,Ferreira:2013tba,Tawfik:2017cdx,Tawfik:2019rdd,Farias:2015eea,Ferreira:2014exa,Cao:2014uva,Pagura:2018eav,Yu:2014xoa,Li:2016dta,Ferreira:2014kpa,Ferrari:2012yw,Farias:2016ttj,Mao:2016fha,Mao:2017tcf,GomezDumm:2017iex,Pagura:2016pwr,Ayala:2016bbi,Feng:2015qpi,Chaudhuri:2019lbw}. One prominent explanation is the sea effect \citep{Bruckmann:2013oba}, illustrated in Fig.~\ref{fig:valence_sea_IMC}, where lattice results are decomposed into valence and sea contributions to the condensate \citep{DElia:2011koc,Sharpe:2006pu,Verbaarschot:2000dy}. While the valence contribution enhances the condensate at all temperatures, the sea contribution enhances it only well below $T_{pc}$ and suppresses it near $T_{pc}$. Close to the pseudo-critical temperature, this suppression dominates over the valence enhancement, leading to IMC \citep{Bali:2012zg,Allen:2013lda,Endrodi:2019whh,Tomiya:2019nym}. Several alternative mechanisms have also been proposed. Ref.\cite{Yu:2014xoa} adds vector-isovector channels 
\begin{widetext}
\begin{equation}
\mathcal{L}_{VA} =
\mathlarger{\sum}_{f = u, s}\bigg[G_V(\overline{\psi}_f \gamma^\mu \psi_f)^2 + G_A(\overline{\psi}_f \gamma^\mu \gamma^5 \psi_f)^2\bigg] + \bigg[G^\prime_{V}(\overline{s}\gamma^\mu s)^2 +
G^\prime_{A}(\overline{s}\gamma^\mu \gamma^5 s)^2\bigg]
\end{equation}
\end{widetext}
with $G^\prime_V << G_V$, $G^\prime_A << G_A$ ($\sim (m_u/m_s)^2$) on top of KMT corrected NJL model and shows a new a first-order phase transition around the $T_C$ toward a phase without chiral condensates, but with nontrivial dynamic chiral chemical potentials for the light quarks. The corresponding critical temperature of the transition decreases with the magnetic field and it gives a natural explanation to the inverse magnetic catalysis effect for light quarks when incorporating a reasonable value of the coupling constant in the isoscalar axial-vector channel. Ref. \cite{Yu:2014sla} adds repulsive iso-scalar axial-vector interactions channels

\begin{figure*}[hbt]
\centering
\subfloat[\centering ]{\includegraphics[width=0.5\textwidth]{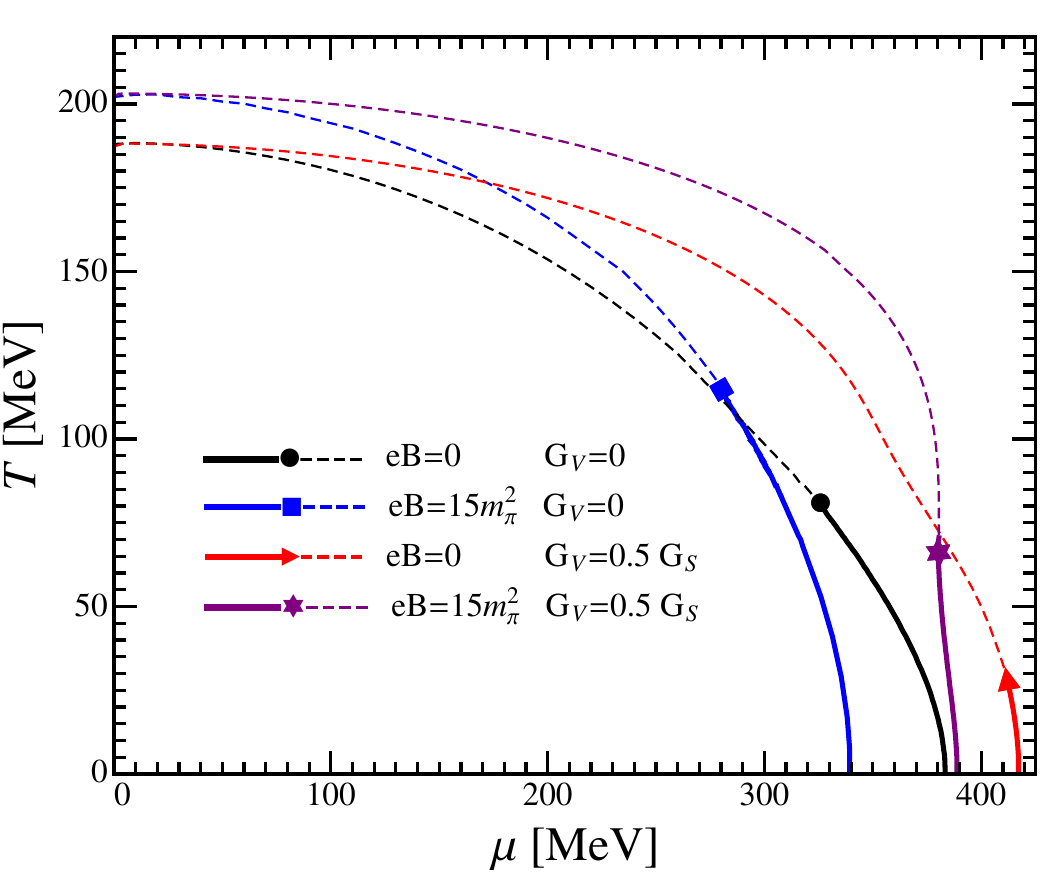}}%
\subfloat[\centering ]{\includegraphics[width=0.5\textwidth]{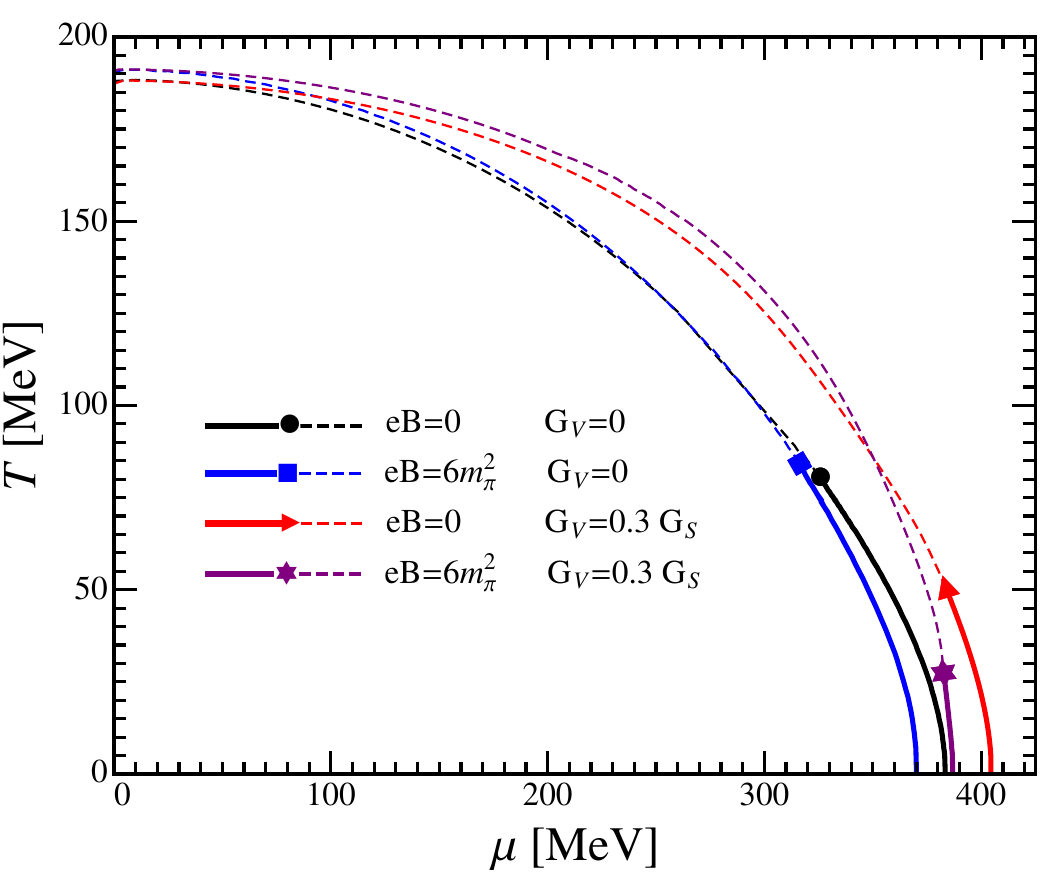}}
\caption{Phase diagram in the $T$--$\mu$ plane from \citep{Denke:2013gha}. The solid symbols represent the critical points. The action of the magnetic field and the repulsive vector interaction $G_V$ are in opposite directions.}
\label{fig:combined_Gv}
\end{figure*}

\begin{widetext}
\begin{equation}
\mathcal{L}_{\text{ISAV}} = \overline{\psi}i\slashed{D}\psi + G_S\bigg[(\overline{\psi}\psi)^2 + (\overline{\psi}i\gamma^5 \tau\psi)\bigg] - G_V(\overline{\psi}\gamma^\mu \psi)^2 - G_A(\overline{\psi}\gamma^\mu\gamma^5\psi)^2
\end{equation}
\end{widetext}
and shows that a reasonable strength of the repulsive interaction in the iso-scalar axial-vector channel can naturally explain the inverse magnetic catalysis around the critical temperature under external magnetic fields. Ref. \citep{Ferreira:2013tba} uses EPNJL model with the KMT interaction term and entangled scalar channel coupling as shown in Eq.\eqref{EPNJL_G} and shown that with a magnetic field dependent scale parameter $T_0$ of the form
\begin{align}
    T_0(eB) = T_0(eB = 0) + \zeta(eB)^2 + \xi(eB)^4
\end{align}
($\zeta$ and $\xi$ being model parameters) it is possible to obtain IMC due to entanglement between chiral and deconfinment transition. In Ref.~\citep{Farias:2014eca}, IMC was demonstrated in the NJL model framework, by arguing that the running of the model couplings with the magnetic field strength \citep{Ferreira:2017vzi,Li:2023rsy} was not considered. However, alternate prescriptions e.g. inclusion of screening effects \citep{Ayala:2021nhx} also remain valid. The authors of Ref.~\citep{Bali:2012zg} used the standard NJL model with scalar and pseudoscalar channels and used the following ansatz for the magnetic field-dependent coupling 
\begin{align}
G(eB) = \frac{G_0}{1 + \alpha \ln\bigg(1 + \beta \dfrac{eB}{\Lambda^2_{\text{QCD}}}\bigg)},
\end{align}
and 
\begin{align}
    G(eB, T) = G(eB)\bigg(1 - \gamma\frac{|eB|}{\Lambda^2_{\text{QCD}}}\frac{T}{\Lambda_{\text{QCD}}}\bigg)
\end{align}
with $G_0 = 5.033~\text{GeV}^{-2}$, and $\alpha$ and $\beta$ being free parameters to obtain a reasonable description of the lattice
average. Decreasing profile of $G(eB)$ produces IMC \citep{Moreira:2020wau,Ghosh:2019ubc}. In \citep{Ferraris:2021vun} within the nonlocal NJL approach, it is shown that whether IMC will be observed actually depends on the strength of the magnetic field, with IMC at low and moderately strong magnetic fields \citep{Abhishek:2018xml}, in contrast to LQCD results in \citep{Endrodi:2015oba} where it is observed that even at extremely strong magnetic fields, IMC prevails. Similar temperature and magnetic field dependent ansatz had been made for finite temperature scenarios 
\citep{Yang:2017xdd,Farias:2016gmy} which are in reasonable agreement with lattice results. Although various other fitting ansatz exist \citep{Ferreira:2014kpa,Endrodi:2019whh,Farias:2014eca,Farias:2016gmy} such expressions are essentially phenomenological and not unique. Any fitting model consistent with the lattice data would give the same qualitative results in suitable ranges of magnetic field and temperature e.g. done in Ref. \citep{Endrodi:2019whh} by using the running coupling obtained by fitting with LQCD data \citep{Bali:2012zg}.

\par Extensions of the conventional NJL model e.g. vector-vector contact interactions \citep{Ahmad:2016iez} significantly influence IMC. The work uses the standard NJL model with the standard scalar channel with the mass gap equations
\begin{align}
    \frac{M - m_0}{2G} = -\text{Tr}\int^{\Lambda}\frac{d^4k}{(2\pi)^4}\tilde{S}(k)
\end{align}
with $\tilde{S}(k)$ being full quark propagator in the presence of magnetic field
\begin{widetext}
\begin{align}
\tilde{S}(k)
={}& \int_{0}^{\infty} \frac{ds}{\cosh(|q_f B|s)}\,
\exp\!\bigg[-s\bigg(k_{\parallel}^{2} + k_{\perp}^{2}\,\frac{\tan(|q_f B|s)}{|q_f B|s} + M^{2}\bigg)\bigg] \nonumber\\
&\times\bigg[\big(\cosh(|q_f B|s) - i\gamma^{1}\gamma^{2}\sinh(|q_f B|s)\big)\big(M-\slashed{k}_{\parallel}\big)
- \frac{\slashed{k}_{\perp}}{\cosh(|q_f B|s)}\bigg].
\end{align}
\end{widetext}
Considering a lattice inspired running coupling which monotonically decreases with $eB$, IMC takes place in this model. Ref.~\citep{Denke:2013gha} shows that, in the presence of a repulsive vector coupling $G_V$, the IMC behavior observed at $G_V=0$ is strongly suppressed at low temperatures, since the vector interaction counteracts the magnetic-field effect (Fig.~\ref{fig:combined_Gv}). Several studies \citep{Chaudhuri:2019lbw,Chaudhuri:2020lga,Qiu:2023kwv,Farias:2021fci,Mao:2022dqn,Wen:2021mgm} have further demonstrated that including the anomalous magnetic moment (AMM) of quarks 
\begin{equation}
\mathcal{L} = \overline{\psi}\Big(i\slashed{D} - m + \frac{i}{2}\hat{a}\sigma^{\mu\nu}F_{\mu\nu}\Big)\psi
+ G\mathlarger{\sum}_{\Gamma}(\overline{\psi}\Gamma\psi)^2
\end{equation}
where $\Gamma \in \{1,\, i\gamma^5\vec{\tau}\}$ runs over the scalar and pseudoscalar channels, $\hat{a} = \hat{Q}\hat{\kappa}$, $\hat{Q}$ the charge matrix of quarks and $\kappa = \text{diag}(\kappa_u, \kappa_d)$, can generate IMC. However, Ref.~\citep{Farias:2021fci} found that for sufficiently small AMM, MC persists over the entire magnetic-field range. Using a Pauli-Villars regularized PNJL model, Refs.~\citep{Mao:2022dqn,Fayazbakhsh:2014mca} show that a sufficiently large quark AMM converts MC into IMC throughout the $\mu_B-T$ plane. In particular, Refs.~\citep{Mei:2020jzn,Mao:2022dqn} demonstrate that the AMM reduces the effective quark mass according to $m^{\star} = m_0 - \kappa_f|Q_fB|$, thereby mimicking the IMC effect. Other studies have explored IMC in diverse contexts, including the electric-charge (EC) phase transition and the BEC-BCS crossover \citep{Duarte:2017nzv}, the correlation between $U(1)_A$-IMC and IMC \citep{Wang:2021dcy}, and the influence of IMC on the Mott transition of pseudo-Goldstone modes \citep{Li:2023rsy,Mao:2019avr}, only to name a few.
\par MC and IMC underscore that the response of the chiral condensate to external magnetic fields is highly nontrivial and till date, no widely accepted consensus for these phenomena exists. Nevertheless, the chiral condensate captures only one facet of QCD matter. At sufficiently large baryon density, the dominant dynamics shifts to pairing dynamics at the Fermi surface, where quark degrees of freedom reorganize themselves into various pairing patterns. This leads to the emergence of color superconducting phases characterized by distinct gap structures. We now turn to a brief overview of these phases, their key properties, and their response in a magnetic field. 
\subsection{Magnetic color-flavor-locked phase}
\label{MCFL_Section}

We now discuss the $T << \mu$ regime of the QCD phase diagram. In this limit a very rich school of symmetry breaking patterns are observed. Moreover, cold neutron stars, whose environment is expected to reside in this regime, are an excellent candidate where occurrence of such exotic phases are expected to exist. In this regime, one expects to find quark Cooper pairing near the Fermi surface that spontaneously breaks the color $SU(3)_C$ symmetry, forming a color superconductor with various pairing patterns. 

In a color superconductor, the quark-quark cooper pair can be characterized by the condensate $\langle \psi^{\alpha}_{ia}\psi^{\beta}_{jb} \rangle = P^{\alpha\beta}_{ijab}\Delta$ where $\alpha$ and $\beta$ are color indices of quarks, $i$ and $j$ are flavor indices and $a$ and $b$ are spinor indices. The color-flavor-spin matrix $P$ characterizes a particular pairing channel, and $\Delta$ is the gap parameter which gives the strength of the pairing in this channel. Now at very high densities where $u$, $d$ and $s$ quarks are present, and the effect of strange quark mass can be neglected, the completely symmetric pairing in both color and the flavor space occur and produces the CFL phase~\citep{Alford:1997zt,Rapp:1997zu,Alford:1998mk,Gastineau:2001zke,BitaghsirFadafan:2018iqr}. In the CFL phase, quarks of all three flavors and all three colors form zero-momentum spinless Cooper pairs described by the two point correlation function 
\begin{align}
    \langle \psi_i^{\alpha} C \gamma_5 \psi_j^{\beta} \rangle = \Delta_{\text{CFL}}\,\epsilon^{\alpha\beta A}\epsilon_{ijA}
+ \Delta_{\text{CFL}}\kappa \left(\delta_i^{\alpha}\delta_j^{\beta}
+ \delta_j^{\alpha}\delta_i^{\beta}\right),
\end{align}
where $\Delta_{\text{CFL}}$ is called the CFL gap parameter. The condensate is invariant only under simultaneous rotations in both color and flavor spaces.

\begin{figure*}[t]
\centering
\subfloat[\centering ]{\includegraphics[width=0.5\textwidth]{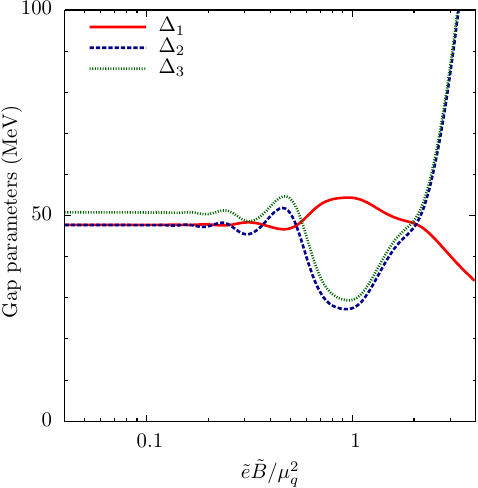}}%
\subfloat[\centering ]{\includegraphics[width=0.5\textwidth]{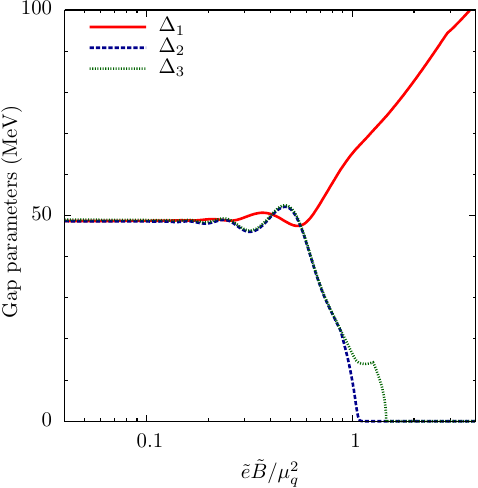}}
\caption{
    (a) Gap parameters $\Delta_1$ ($ds$), $\Delta_2$ ($su$) and $\Delta_3$ ($ud$) pairing as a function of $\widetilde{e}\widetilde{B}/\mu^2_q$ without neutrality.
    (b) Same with charge neutrality imposed. Taken from \citep{Fukushima:2007fc}.
    }
\label{fig:CFL_gap_splitting}
\end{figure*}

\par The structure of the CFL phase is significantly modified in the presence of strong magnetic fields. Since a rotated magnetic field can penetrate a spin-zero color superconductor \citep{Fukushima:2007fc,Ferrer:2006dx,Manuel:2006gu}, its effects on color-superconducting matter become particularly important. Ref.~\citep{Ferrer:2005vd} argues that neutral Cooper pairs may be formed either from $\widetilde{Q}$-neutral quarks or from quarks carrying opposite $\widetilde{Q}$ charges. When the magnetic length, $\ell_0 = 1/\sqrt{2eB}$ becomes smaller than the coherence length of the Cooper pair, the magnetic field significantly alters the pairing structure of the condensate.

In particular, Refs.~\citep{Ferrer:2005vd,Ferrer:2006vw,Ferrer:2005pu} show that the magnetic field splits the CFL gap into two distinct gaps: $\Delta_H$, which receives contributions from both oppositely charged and neutral quark pairs, and $\Delta$, which receives contributions only from neutral quark pairs. The resulting phase retains CFL-type pairing but possesses a reduced symmetry group, $SU(2)_{C+L+R}$, compared to the ordinary CFL phase. This less symmetric phase is known as the MCFL phase. While Cooper pairs composed of quarks with zero rotated charge $\widetilde{Q}$ remain largely unaffected by the magnetic field, condensates containing charged constituents are modified into the MCFL state \citep{Sen:2015cxa,Zhang:2015hca,Feng:2011fj,Fukushima:2007fc}, as illustrated in Fig.~\ref{fig:CFL_gap_splitting}. In the following, we present a brief digression on the thermodynamics of the MCFL phase \cite{Paulucci:2010uj}.
\par The rotated magnetic field $\tilde{H}$ couples to the charged quarks through the covariant derivative of the NJL Lagrangian. The magnetic field separates the original
$(u, d, s)$ quark representation to neutral, positively and negatively charged spinors according to the quark rotated
charges in units of $\tilde{e} = e\cos\theta$, with $\theta$ being the mixing angle of the rotated fields,

\begin{table}[h]
\centering
\begin{tabular}{|c|c|c|c|c|c|c|c|c|}
\hline
$u_r$ & $u_g$ & $u_b$ & $d_r$ & $d_g$ & $d_b$ & $s_r$ & $s_g$ & $s_b$ \\
\hline
0 & 1 & 1 & -1 & 0 & 0 & -1 & 0 & 0 \\
\hline
\end{tabular}
\end{table}
This separation makes it convenient to define separate Nambu-Gorkov spinors for positive, negative and neutral fields.
The detailed procedures on the Ritus method \cite{Elizalde:2000vz} used to transform the charged spinor fields to momentum space is presented in \cite{Ferrer:2006vw}. One can show that the MCFL thermodynamic potential can be written as the sum of contributions due to charged ($\Omega_{C}$) and neutral ($\Omega_N$) quarks: $\Omega_{\text{MCFL}} = \Omega_{C} + \Omega_N$ where $\Omega_{C}$ and $\Omega_{\text{N}}$ is given by the following expressions
\begin{align}
    \Omega_C = -\frac{\widetilde{e}\widetilde{H}}{8\pi^2} \mathlarger{\sum}_{\text{k} = 0}^{\infty} \alpha_{\text{k}} \int_0^{\infty} dp_3 \, e^{-(p_3^2 + 2\widetilde{e}\widetilde{H}\text{k})/\Lambda^2} \left[8|\varepsilon^{(c)}| + 8|\overline{\varepsilon}^{(c)}|\right],
\end{align}

\begin{align}
&\Omega_N = -\frac{1}{4\pi^2} \int_0^{\infty} dp\, p^2 \, \exp(-p^2/\Lambda^2) \Big[6\big(|\varepsilon^{(0)}| + |\overline{\varepsilon}^{(0)}|\big) \nonumber\\
&\quad + 2\mathlarger{\sum}_{j=1}^{2}\big(|\varepsilon_j^{(0)}| + |\overline{\varepsilon}_j^{(0)}|\big)\Big]
+ \frac{\Delta^2}{G} + \frac{2\Delta_H^2}{G}.
\end{align}
with 
\begin{align}
\genfrac{}{}{0pt}{}{\varepsilon^{(c)}}{\overline{\varepsilon}^{(c)}}
  &= \pm\sqrt{\left(\sqrt{p_3^{2}+2\widetilde{e}\widetilde{H}n}\mp\mu\right)^{2}+\Delta_H^{2}}, \\[4pt]
\genfrac{}{}{0pt}{}{\varepsilon_i^{(0)}}{\overline{\varepsilon}_i^{(0)}}
  &= \pm\sqrt{(p\mp\mu)^{2}+\Delta_i^{2}}, \qquad i = 0,a,b,
\end{align}

\begin{align}
\varepsilon^{(0)} &= \pm\sqrt{(p-\mu)^2+\Delta^2}, &
\overline{\varepsilon}^{(0)} &= \pm\sqrt{(p+\mu)^2+\Delta^2}, \\
\varepsilon_1^{(0)} &= \pm\sqrt{(p-\mu)^2+\Delta_a^2}, &
\overline{\varepsilon}_1^{(0)} &= \pm\sqrt{(p+\mu)^2+\Delta_a^2}, \\
\varepsilon_2^{(0)} &= \pm\sqrt{(p-\mu)^2+\Delta_b^2}, &
\overline{\varepsilon}_2^{(0)} &= \pm\sqrt{(p+\mu)^2+\Delta_b^2}.
\end{align}
being the dispersion relations of the charged (labled by $c$) and neutral (labled by $0$) quarks and
\begin{align}
    \Delta^2_{a/b} = \frac{1}{4}\bigg(\Delta \pm \sqrt{\Delta^2 + 8\Delta^2_H}\bigg)
\end{align} 
The MCFL gaps $\Delta$ and $\Delta_H$ correspond to the case where the $(d, s)$ pairing gap, which takes place only between neutral quarks, is $\Delta_1 = \Delta$, while the $(u, s)$ and $(u, d)$ pairing gaps, which receive contribution from pairs of charged and neutral quarks, become $\Delta_2 = \Delta_3 = \Delta_H$.
The separation of the gap in two different parameters in the MCFL case, as compared to the CFL, where $\Delta_1 =
\Delta_2 = \Delta_3$, reflects the symmetry difference between these
two phases \cite{Ferrer:2005vd}. The effect of confinement can be taken into account by a bag constant $\mathcal{B}$. On the other hand, in a magnetized system, the pure Maxwell contribution $\tilde{H}^2/2$ should also be added \cite{Ferrer:2010wz}, making the total thermodynamic potential as 
\begin{align}
    \Omega_H = \Omega_{\text{MCFL}} + \mathcal{B} + \frac{\tilde{H}^2}{2}
\end{align}
The gap equations for the system are given by 
\begin{align}
    \frac{\partial \Omega_H}{\partial \Delta_{CFL}} = \frac{\partial \Omega_H}{\partial \Delta} = \frac{\partial \Omega_H}{\partial \Delta_H} = 0
\end{align}
In the strong field limit, the MCFL gap has the following solution \citep{Ferrer:2005vd,Ferrer:2006vw,Ferrer:2005pu}
\begin{align}
\Delta_B \sim 2\sqrt{\delta \mu}~\exp\left[-\frac{3\Lambda^2 \pi^2}{g^2\left(\mu^2 + \widetilde{e}\widetilde{B}/2\right)}\right]
\end{align}
which reduces to the usual CFL gap in the $\widetilde{e}\widetilde{B}\to 0$ limit. Here, $\delta = \Lambda - \mu$, $\Lambda$ being the ultraviolet cutoff and $\mu$ is the baryon chemical potential. On the other hand, the gap $\Delta$, formed only by pairs of neutral quarks, in the strong field approximation, becomes \citep{Ferrer:2012wa}
\begin{align}
\Delta = \frac{1}{2^{7/34}} \exp\left(
-\frac{36}{17x} 
+ \frac{21}{17}\frac{1}{x(1 + y)} 
+ \frac{3}{2x}
\right) \Delta_{\text{CFL}}
\end{align}
where $x = g^2 \mu^2/\pi^2\Lambda^2$ and $y = \widetilde{e}\widetilde{B}/\mu^2$. The effect of the strong magnetic field $\widetilde{e}\widetilde{B}>> \mu^2$ is producing MCFL gap enhancement.

\par Numerous works on the stability of MCFL matter have been done \citep{Wen:2013yra,Cui:2015tvq,Felipe:2010vr,GonzalesFelipe:2011nc,Sahoo:2025cno,Celi:2022oxi} claiming that MCFL matter is more stable than other phases within a proper magnitude of the external magnetic field. Ref. \citep{Feng:2011fj} shows using the NJL model the emergence of a new condensate $\Delta_{M}$ associated with magnetic moment of the Cooper pairs with no de Haas-van Alphen oscillations \citep{Feng:2011fj}. The condensate $\Delta_M$ assists in enhancing the magnitude of $\Delta_B$ as well \citep{Noronha:2007wg} which, combined with the increase of $\Delta_M$ at strong fields, makes the MCFL phase more stable than the regular CFL. Other interesting works on the nature of MCFL matter within the context of NJL model include, but are not restricted to, the studies of the Debye and Meissner screening masses \citep{Feng:2012dqa}, quark dynamics and effect of regularization schemes \citep{Feng:2012dqa}, competition between MC and gap splitting \citep{Ren:2014xra}, effects on equations of state \citep{Ferrer:2017rlv} and new channels of interaction and chromomagnetic instability~\citep{Feng:2011fj}.
\par However, although appearance of MCFL in the core of NS is definitely a theoretical possibility and next generation gravitational wave detectors e.g. the Einstein Telescope and Cosmic Explorer may look for their signatures~\citep{Sahoo:2025cno}, there is no consensus yet in the community as to the MCFL phase in NS core. Results in Ref.~\citep{Paulucci:2013eia} are consistent with strange MCFL stars with low magnetic ﬁelds but do not seem consistent  with hybrid MCFL stars. On the other hand, Ref.~\citep{Ferrer:2021vuy} claims that the CFL/MCFL phases are the energetically favored phases only at densities which are just reachable in the inner core of NS. This motivates the search for other phases at intermediate densities where the system may favor less symmetric pairing patterns involving a reduced number of quark flavors, such as the 2SC phase. We conclude this section by touching upon the issue of absolute stability of MCFL phase.
\par One should check the stability and thermodynamic self-consistency of the MCFL phase by verifying that the baryon density at the point of minimum energy per baryon at zero temperature is equal to the baryon density at the zero-pressure point, formally established by the Hugenholtz–Van Hove (HVH) theorem \cite{Hugenholtz:1958zz,RCNayak:2011hjn}. However, barring a few exceptions, eg. \cite{Chu:2023von}, most studies of the NJL and PNJL models do not explicitly check this condition, and hence, there is always a chance of internal inconsistency in such results. However, one may make a fairly reasonable assumption that since the models described here derive all thermodynamic quantities from the same grand potential, the HVH theorem is satisfied. As long as all physical quantities are derived from the stationary approximation to the same grand potential, without ad-hoc phenomenological parameter values, this is a valid assumption. 

Nevertheless, a general statement about absolute stability of MCFL phase and that of SQM is difficult, mainly because such conclusions are sensitive to the underlying model used, so this warrants further investigation in the field. As an example, Ref.\cite{Paulucci:2010uj} takes into account pressure anisotropy due to strong magnetic field and shows that a stability window for MCFL matter exists only under very specific assumptions e.g. a magnetic field dependent bag constant. Also, \cite{Felipe:2010vr} shows in the MIT bag model setting, using variation of the strange quark mass, that MCFL matter is more stable compared to unpaired magnetized strange quark matter. Gluon effects complicate the situation even further. For further discussion of absolute stability of SQM or MCFL phases, interested readers are referred to\cite{Ferrer:2012wa,Isayev:2012sv,Wen:2013yra,Isayev:2014rch,Bozzo:2011vn}.

\subsection{Two-flavor and crystralline color superconductors, chromomagnetic instability}
\label{2SC_Section}
At intermediate densities, the CFL phase suffers from stresses induced by a non-zero strange quark mass; combined with $\beta$-equilibrium, the 2SC phase \citep{Reddy:2004my} is the most straightforward, less symmetrically paired form of quark matter in this regime~\citep{Bailin:1979nh,Bailin:1983bm,Alford:1997zt,Rapp:1997zu}. In the 2SC phase, quarks with two out of three colors, and only $u$ and $d$ quarks, pair in the standard BCS fashion, with the condensate $\langle \psi^{\alpha}_{i} C\gamma^5 \psi^{\beta}_{j} \rangle \propto \Delta_{2SC} \epsilon_{ij3}\epsilon^{\alpha\beta 3}$. The NJL models with strong couplings leave the possibility of a 2SC window in the non-CFL region of the phase diagram \citep{Ruester:2005jc}. It is a conductor with respect to the unbroken $U(1)_{\widetilde{Q}}$ gauge symmetry, as some of the ungapped blue and strange quarks are $\widetilde{Q}$-charged. Similar to the gapless CFL (gCFL) phase, the gapless 2SC (g2SC) phase~\citep{Shovkovy:2003uu} is unstable at zero temperature~\citep{Huang:2004am,Huang:2005pv,Iida:2006df}.

\par
\begin{figure*}[t]
\centering
\def\trimA{trim={0pt {-.040\height} 0pt {-.079\height}},clip}
\def\trimB{trim={0pt 0pt 0pt 0pt}}
\def\trimC{trim={0pt {-.074\height} 0pt {-.074\height}},clip}
\sbox0{\expandafter\adjustbox\expandafter{\trimA,height=10cm}{\includegraphics{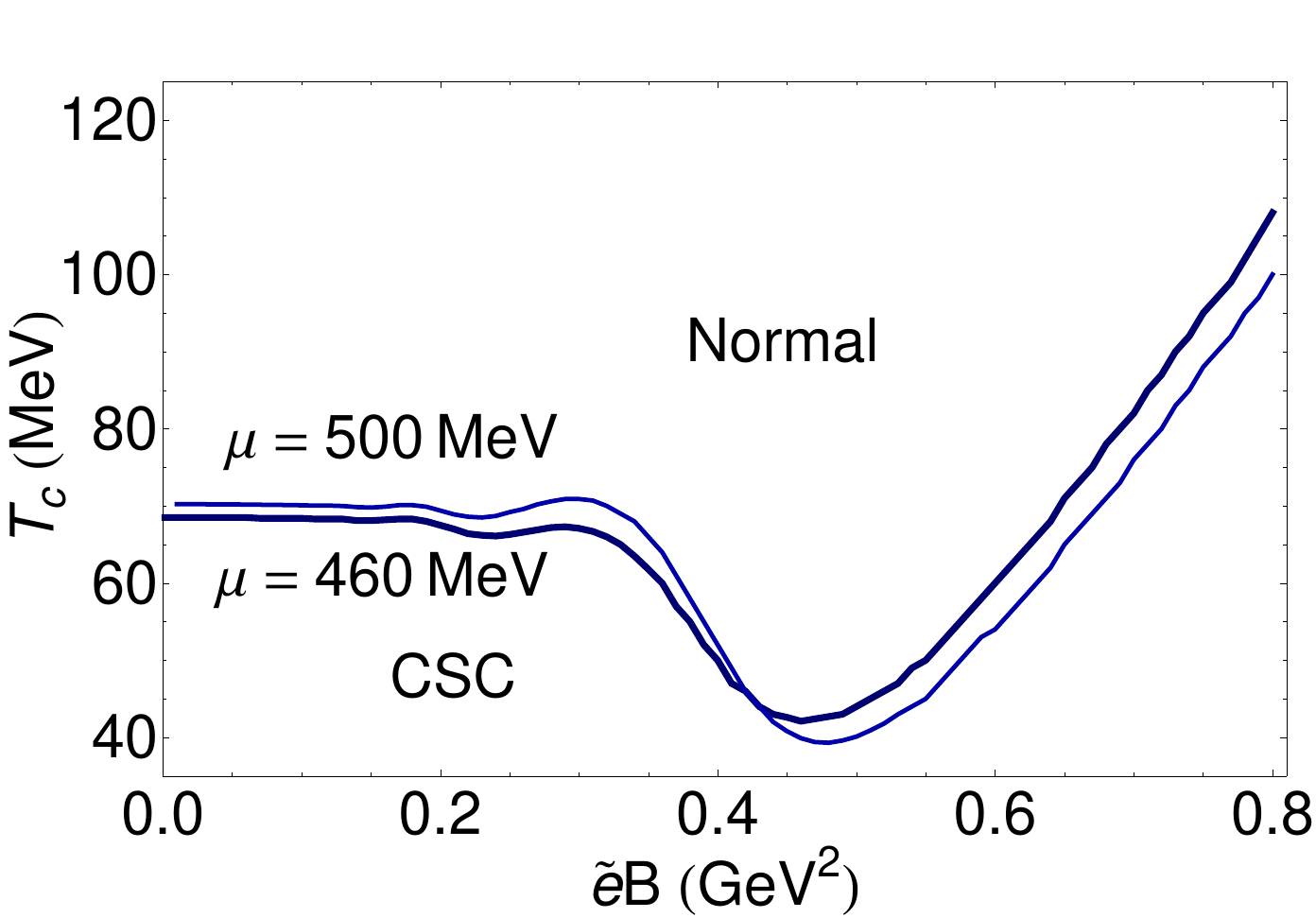}}}%
\sbox2{\expandafter\adjustbox\expandafter{\trimB,height=10cm}{\includegraphics{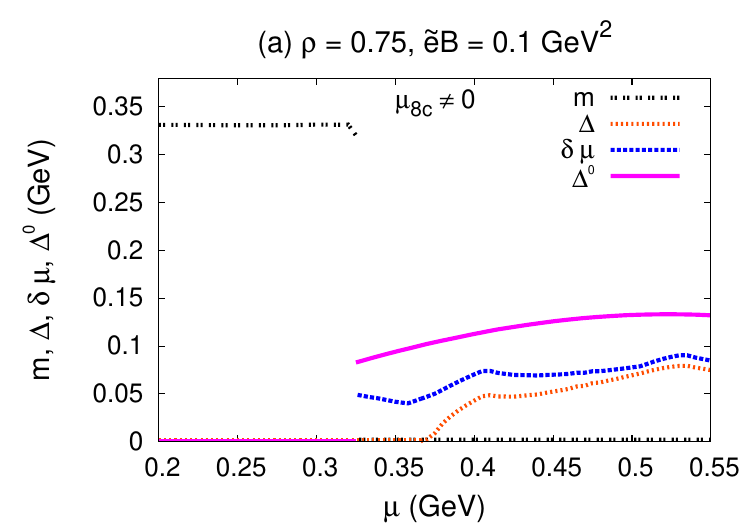}}}%
\sbox4{\expandafter\adjustbox\expandafter{\trimC,height=10cm}{\includegraphics{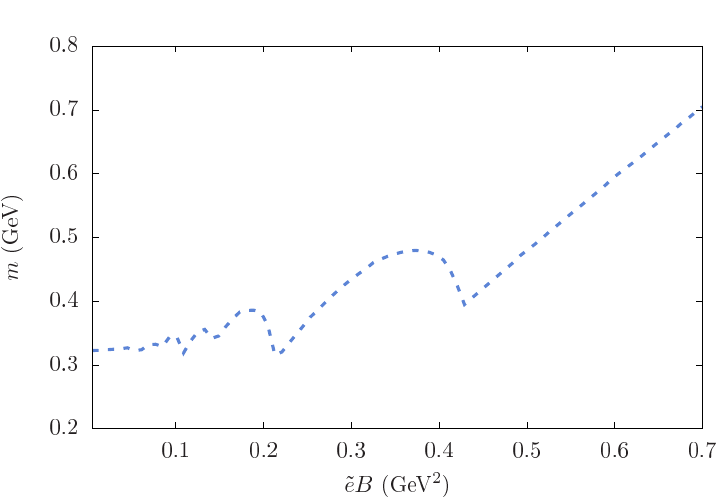}}}%
\dimen0=\dimexpr\textwidth-2em\relax
\dimen2=\dimexpr\wd0+\wd2+\wd4\relax
\edef\figH{\the\dimexpr 10cm*\dimen0/\dimen2\relax}%
\subfloat[\centering ]{\expandafter\adjustbox\expandafter{\trimA,height=\figH}{\includegraphics{T-eB-phase-mu-460-500-HSaa_lr.pdf}}}%
\hfill
\subfloat[\centering ]{\expandafter\adjustbox\expandafter{\trimB,height=\figH}{\includegraphics{mu_m_del_eB0_1_R0_75.pdf}}}%
\hfill
\subfloat[\centering ]{\expandafter\adjustbox\expandafter{\trimC,height=\figH}{\includegraphics{m_eB_magcat_lr.pdf}}}
\caption{
    (a) Illustration of delayed MC in magnetized 2SC matter \citep{Fayazbakhsh:2012bwd}.
    (b) Dependence of $m$, $\Delta$, $\delta\mu = \mu_e/2$ and $\Delta^0$ (diquark gap without charge neutrality) on $\mu$ for electrically and color neutral quark matter at $\widetilde{e}B = 0.1~\text{GeV}^2$ with $G_D/G_S = 0.75$ \citep{Mandal:2012fq}.
    (c) The constituent quark mass $m$ as a function of $eB$ at $\mu = 0.2~\text{GeV}$ in the chirally broken phase, showing MC \citep{Mandal:2016dzg}.
    }
\label{fig:combined_three_with_mandal}
\end{figure*}

 Studies of diverse nature have been performed on the 2SC phase under magnetic fields, including Ref.~\citep{Mandal:2012fq}, which shows delayed MC of the 2SC phase for large $G_D/G_S$ \citep{Fayazbakhsh:2012bwd} (Fig.~\ref{fig:combined_three_with_mandal}(a)); studies on electrical and thermal conductivities and color-magnetic flux tubes \citep{Alford:2014doa}; demonstrations of the roles of various regularization schemes and the removal of unphysical oscillations using the MFIR-MSS scheme of regularization, which are often wrongly claimed as de Haas–van Alphen oscillations \citep{Allen:2015paa,Azeredo:2026wgl}; compact star studies \citep{Coppola:2018urz}; mixed non-uniform phases \citep{Sadzikowski:2002iy}; coexistence of inhomogeneous chiral and the 2SC condensate \citep{Lakaschus:2020caq,Sadzikowski:2002iy}; and classification of vortices in a mixed phase of 2SC condensates and $P$-wave diquark condensates~\citep{Fujimoto:2020dsa,Fujimoto:2020cho}, which demonstrates both superconductivity and superfluidity, as well as a confinement mechanism in such a phase~\citep{Fujimoto:2021wsr}, only to name a few. Ref.~\citep{Mandal:2012fq} shows that in the charge-neutral g2SC, a large magnetic field drives the color-superconducting phase transition to a crossover, while the chiral phase transition is first order, as shown in Fig.~\ref{fig:combined_three_with_mandal}(b). Ref. \citep{Mandal:2016dzg} reports that for $|eB| \geq 0.02~\text{GeV}$, the 2SC gap in the strong-coupling regime is abruptly quenched at ultra-high magnetic fields due to the mismatched Fermi surfaces of up and down quarks imposed by charge neutrality, and oscillations of the gap due to Landau level quantization. The dynamical quark mass also displays de Haas–van Alphen oscillations and MC at high magnetic fields (supporting the result from \citep{Fayazbakhsh:2012bwd} and shown in Fig.~\ref{fig:combined_three_with_mandal}(c)), although the latter effect is suppressed by nonzero temperature.

\par Several works attempt to identify possible astrophysical implications of the 2SC phase as well~\citep{Shovkovy:2003ce,Yuan:2024ajk,Glampedakis:2017nqy,Alford:2010qf}. Refs. \citep{Iida:2002ev,Alford:2010qf} show that in the interior of compact magnetized stars, a stellar magnetic field stitching a CFL/2SC core is likely to do so by forming an array of color-magnetic quantized vortices. Ref. \citep{Glampedakis:2012qp} points out the possibility of creating a large deformation in a neutron star’s CFL/2SC core due to gluonic field degrees of freedom of the color-magnetic vortices. Ref. \citep{Glampedakis:2012qp} also investigates implications of the 2SC phase for gravitational waves from color-magnetic ``mountains'' in NSs. However, there are also counterclaims~\citep{Yuan:2023dxl} suggesting that the physically allowed parameter space for stable 2SC quark stars may not exist.

\par Having described the basic physics and relevance of the 2SC phase, we now comment on one of the primary threats to its stability: chromomagnetic instability \citep{Huang:2004bg,Huang:2004am} arising from the mismatch between the Fermi surfaces of pairing quarks and heralded by imaginary Meissner screening masses of certain gluons~\citep{Huang:2004am,Huang:2004bg,Rischke:2000cn}, thereby necessitating searches for possible remedies of the situation. Several attempts have been made to resolve this issue and reinterpret this phenomenon e.g. showing that the CCS phase is energetically favored over the 2SC phase \citep{Giannakis:2005sa}, resolution by minimal cylindrical gluonic phase II \citep{Kiriyama:2006ui,Hashimoto:2007tx}, inhomogeneous condensate of charged gluons \citep{Ferrer:2007uw} and charged gluon vortices \citep{Ferrer:2006ie}, only to summarize a few. However, a consensus explanation of this phenomenon is yet to be reached.

\begin{figure*}[t]
\centering
\subfloat[\centering ]{\includegraphics[width=0.333\textwidth]{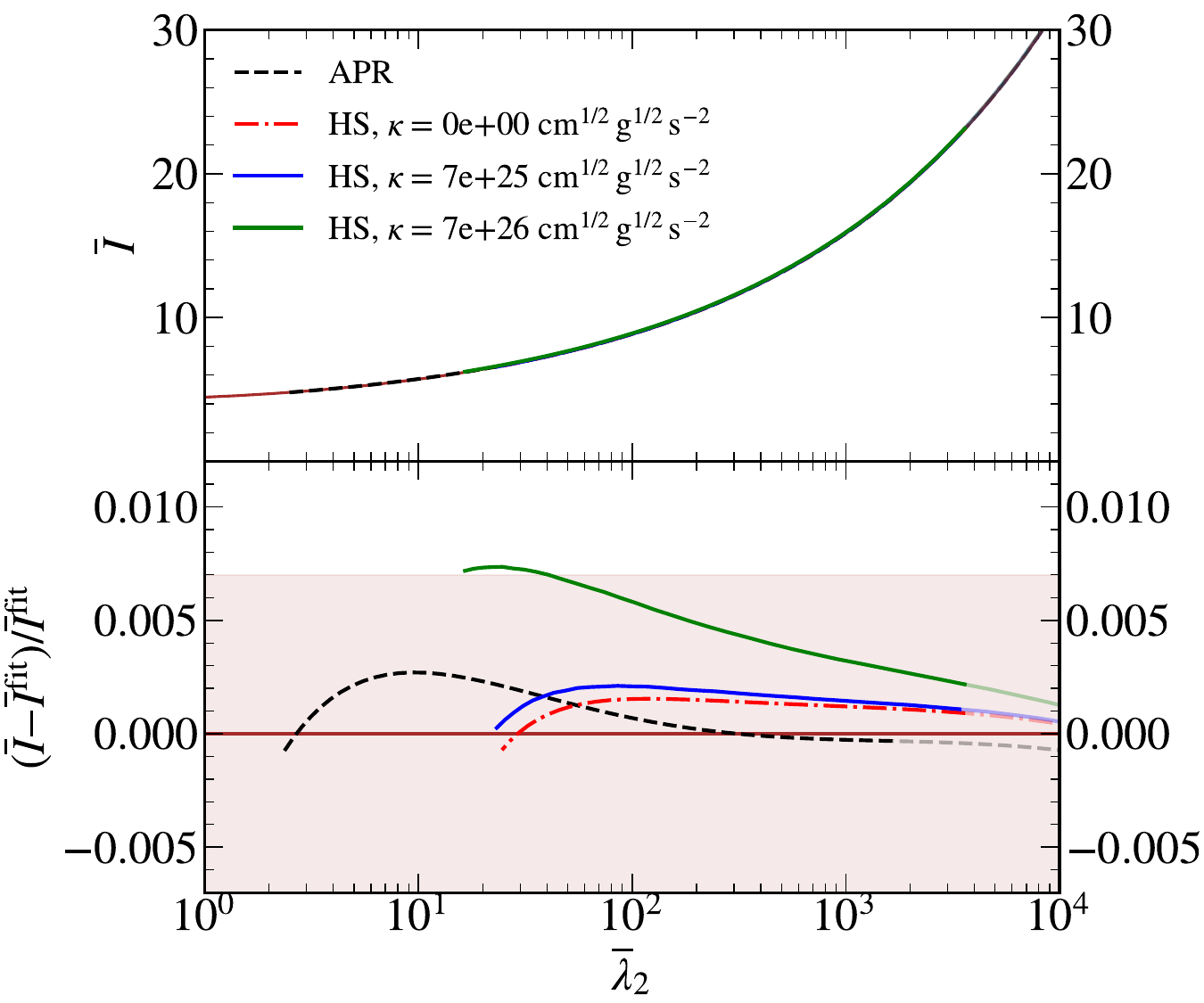}}%
\subfloat[\centering ]{\includegraphics[width=0.333\textwidth]{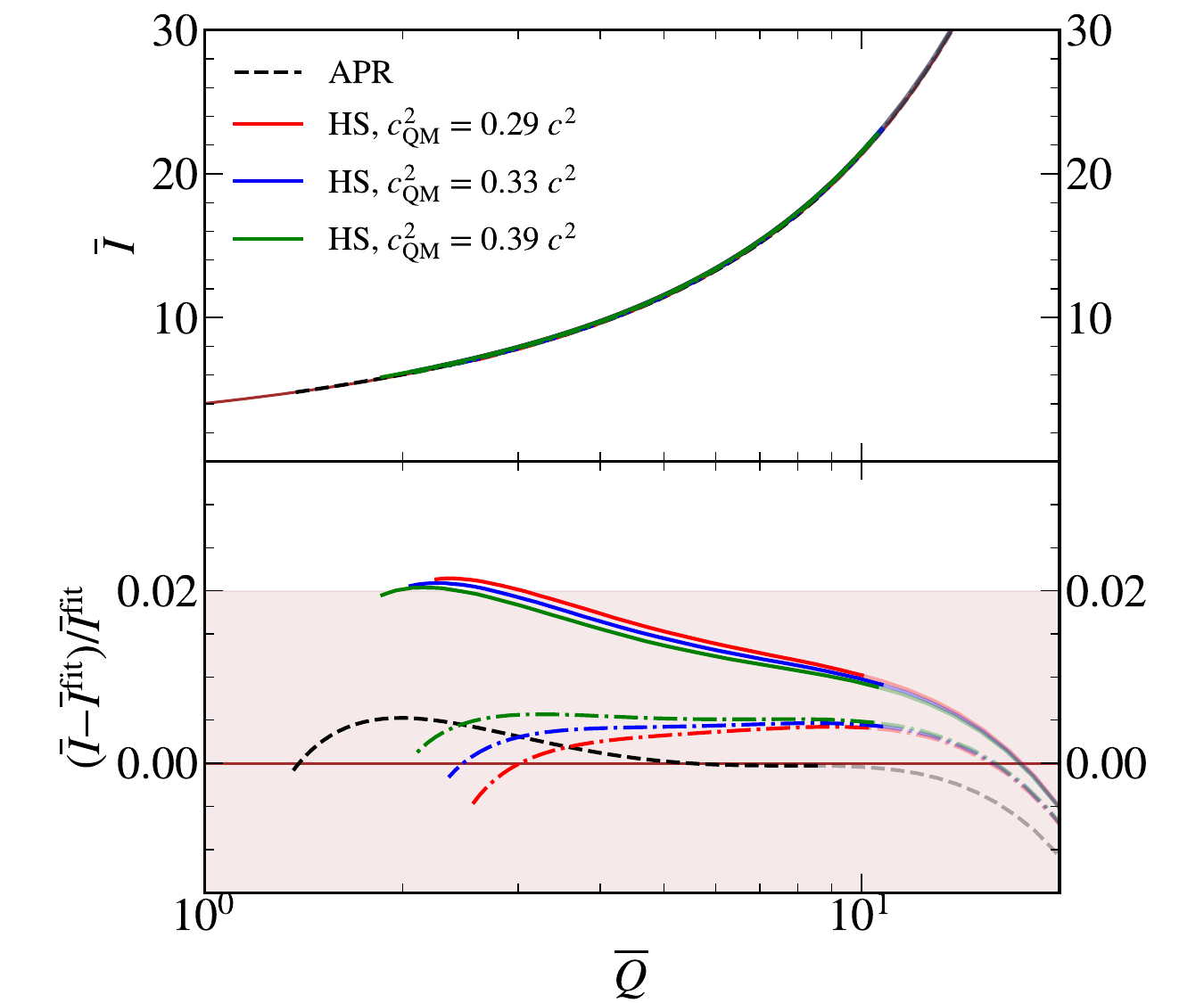}}%
\subfloat[\centering ]{\includegraphics[width=0.333\textwidth]{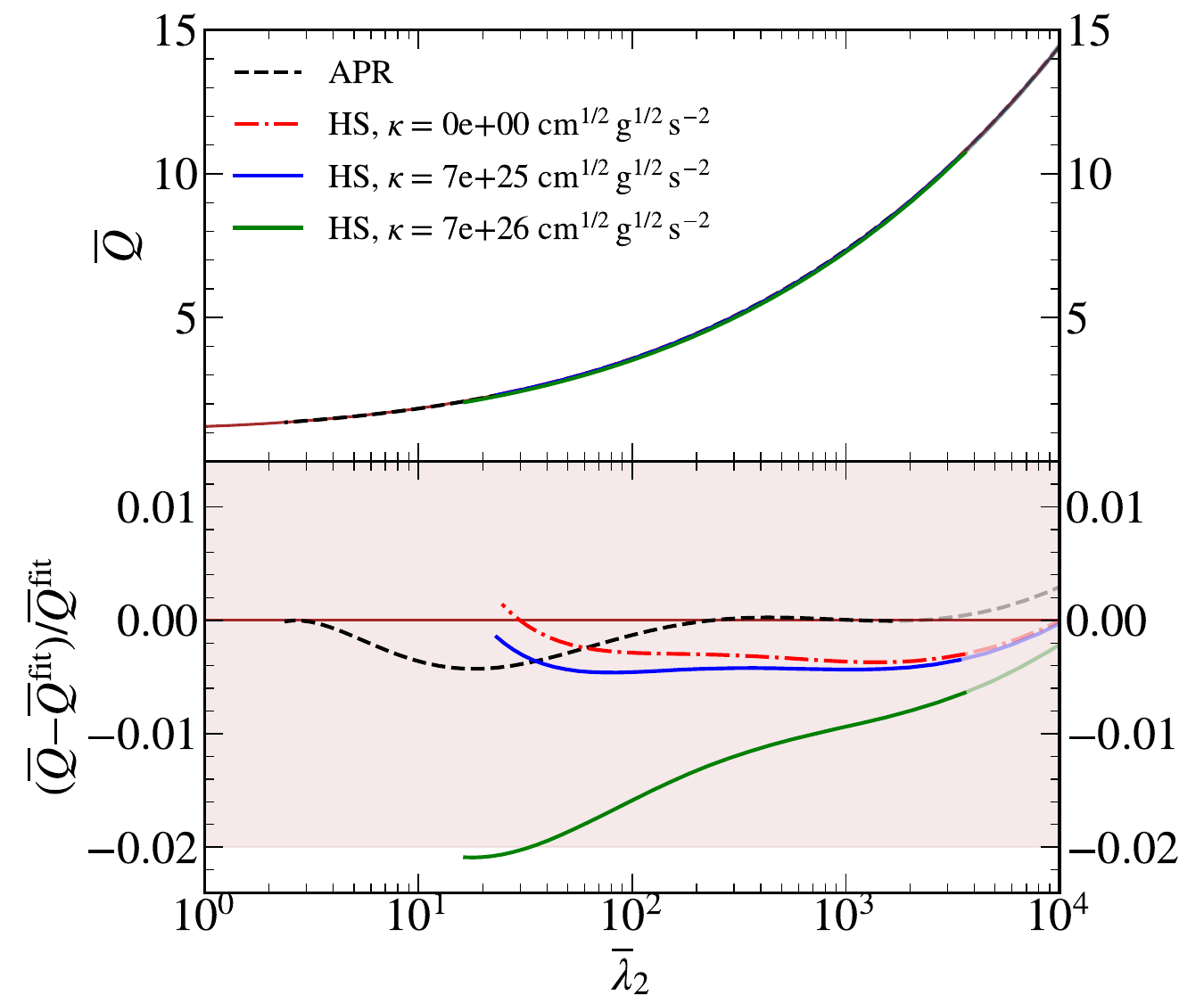}}
\caption{(a) $I$--$\lambda_2$, (b) $I$--$Q$ and (c) $Q$--$\lambda_2$ universal relations for APR and for models in which the shear modulus $\widetilde{\mu}$ is related to the energy density $\rho$ by $\widetilde{\mu} = \kappa \sqrt{\rho}$. Upper panels show the relations; lower panels show the fractional relative difference between the models and the fitting formulae \citep{Yagi:2016bkt}. The brown solid curves represent the fitting formulae, and the maximal variability of the fitting formulae for fluid NSs ($I$--$\lambda_2$: 0.7\%; $I$--$Q$: 2\%; $Q$--$\lambda_2$: 2\%) is shaded in the lower panels. These results have been taken from \citep{Yip:2025uzd}.}
\label{fig:IQL_relations}
\end{figure*}

\par
\begin{figure*}[hbt]
\centering
\subfloat[\centering ]{\includegraphics[width=0.5\textwidth]{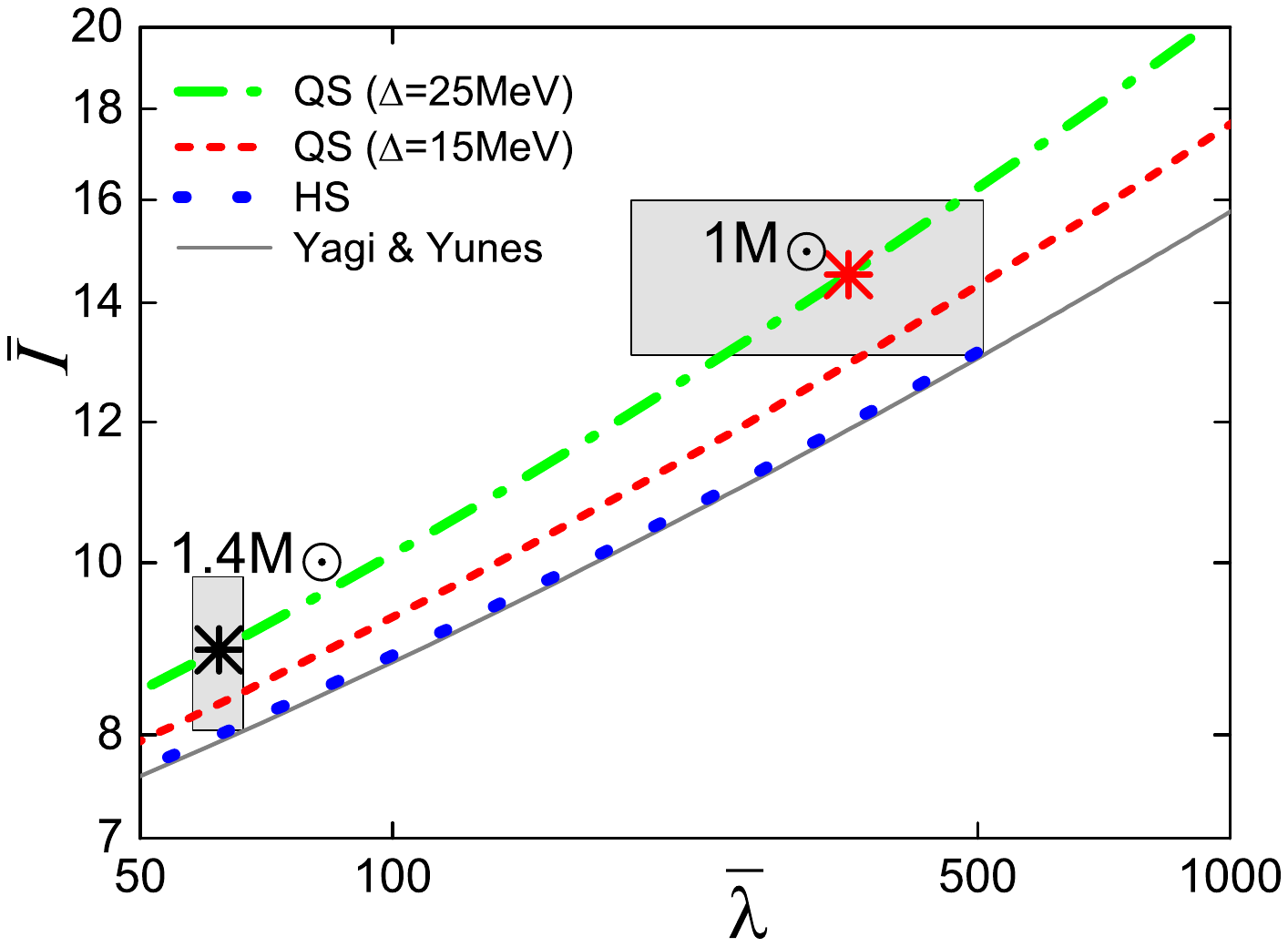}}%
\subfloat[\centering ]{\includegraphics[width=0.5\textwidth]{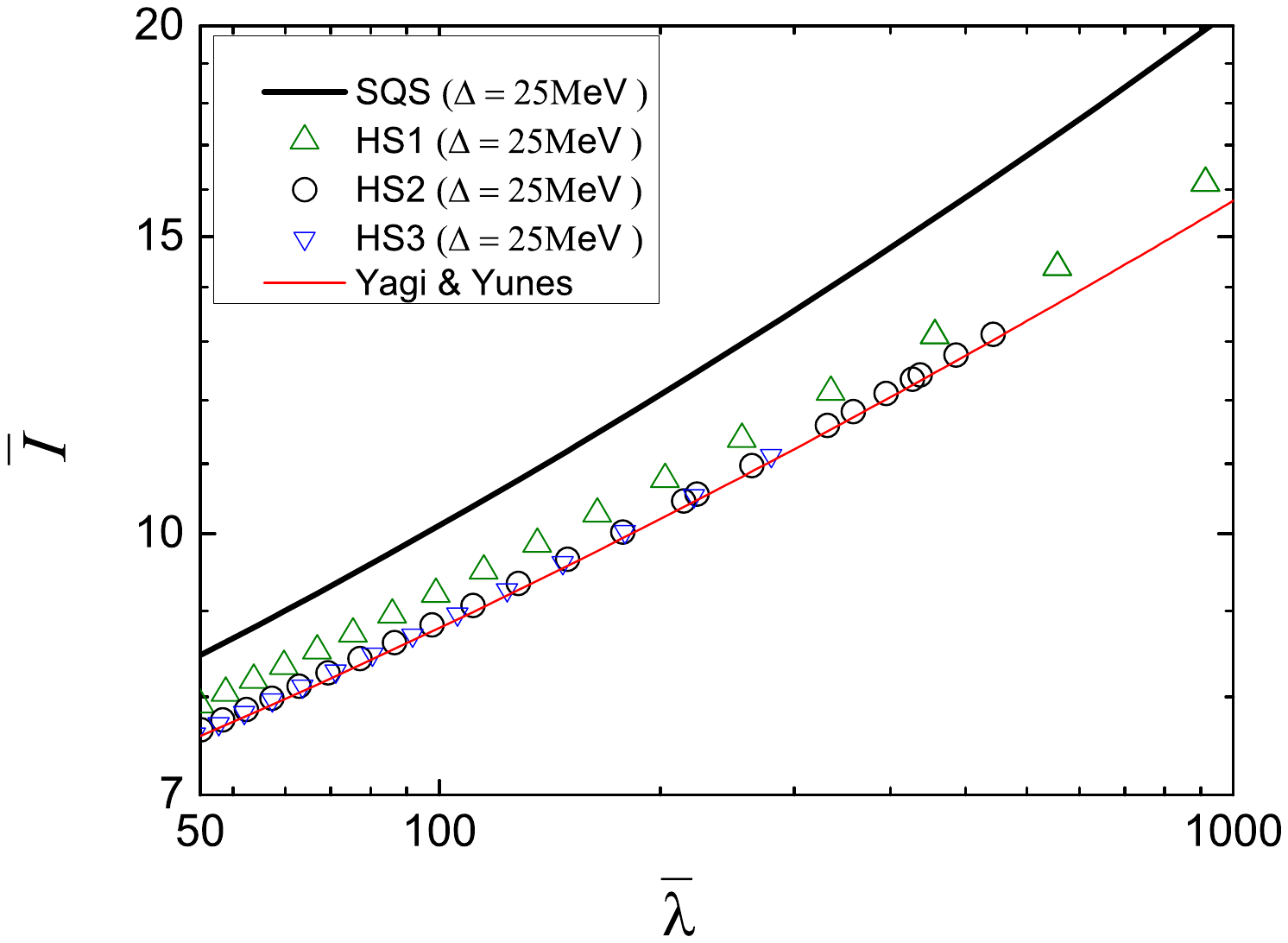}}
\caption{
    (a) $\overline{I}$ plotted against $\overline{\lambda}$ for bare solid quark stars and hybrid stars with a solid quark-matter core, compared with the universal relation proposed by Yagi and Yunes \citep{Yagi:2016bkt} (solid line). Two values of the gap parameter $\Delta$ are used for the quark star models, as opposed to a single value for the hybrid star models. These results are taken from \citep{Lau:2017qtz}.
    (b) Similar relations for the HS1--HS3 models (see \citep{Lau:2018mae} for the model parameters) and the solid quark star (SQS) model, shown together with the fitting formula of the universal relation \citep{Yagi:2016bkt}. These results are taken from \citep{Lau:2018mae}.}
\label{fig:IL_violation_combined}
\end{figure*}

 The CSS phase~\citep{Buzdin:2005zz} represents a possible resolution of the chromomagnetic instability, which allows different quark flavors to have different Fermi momenta so that stress of the strange quark mass could be tolerated, and to allow Cooper pairs with nonzero momentum. A significant amount of progress has been made in determining the properties of this phase. This includes the BCC structure of the LOFF phase \citep{Cao:2015rea,Cao:2015taa}, bosonic excitations in the cubic three-flavor LOFF phase and meson condensation phases \citep{Anglani:2007aa}, calculation of the shear viscosity for two-flavor plane-wave color superconducting quark matter \citep{Sarkar:2016gib}, and self-consistent evaluation of the constituent quark masses and stability window of the LOFF phase \citep{Ippolito:2007uz}. It also includes astrophysical studies such as \citep{Yip:2025uzd,Lau:2017qtz} on the consistency of inhomogeneous CCS phases and consistency with $I-\lambda_2-Q$ universal relations (with Fig.~\ref{fig:IQL_relations} and Fig.~\ref{fig:IL_violation_combined} showing contradictory results), glitch phenomena \citep{Zhou:2022cyp}, the possibility of the Einstein Telescope detecting torsional oscillation signals modified by the CCS/LOFF phase \citep{Lin:2013nza}, and cooling studies \citep{Sedrakian:2015qxa}, only to name a few.

\par The influence of strong magnetic fields on CCS remains a relatively unexplored compared to other phases. Given that magnetars can host magnetic fields as large as $10^{18}~\text{G}$, understanding how such strong fields affect the stability, pairing structure, and electromagnetic response of the LOFF phase is essential. Moreover, CSS remain rooted in the paradigm of only quark and gluonic degrees of freedom, which may be insufficient at intermediate densities where confinement effects persist and hadronic degrees are still present. This motivates the consideration of an alternative phases of QCD matter beyond conventional superconductivity: the quarkyonic phase.

\begin{figure*}[hbt]
\centering
\def\trimA{trim=0 -9 0 3,clip}
\def\trimB{trim={0pt 0pt 0pt {-.042\height}}}
\sbox0{\expandafter\includegraphics\expandafter[\trimA,height=10cm]{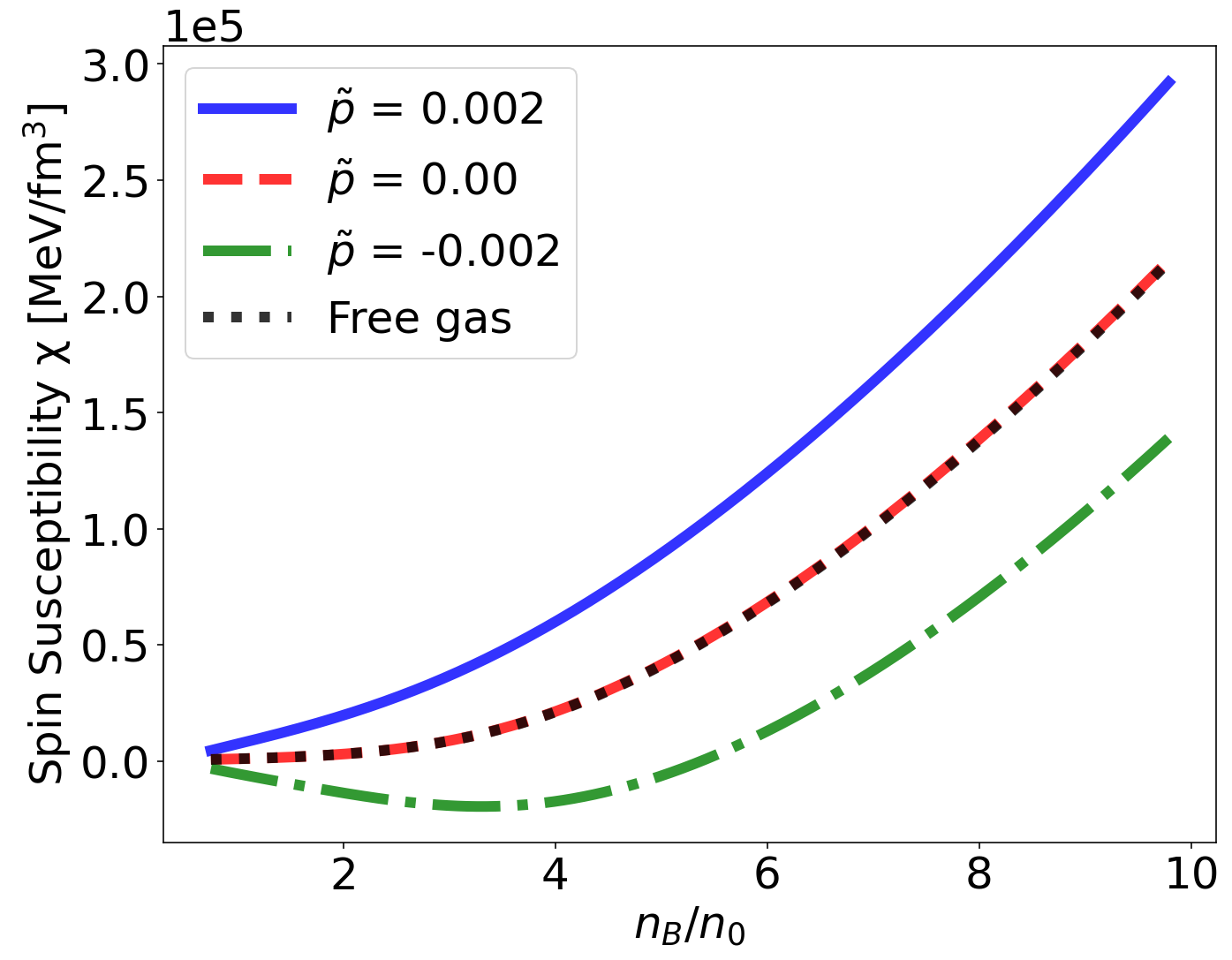}}%
\sbox2{\expandafter\adjustbox\expandafter{\trimB,height=10cm}{\includegraphics{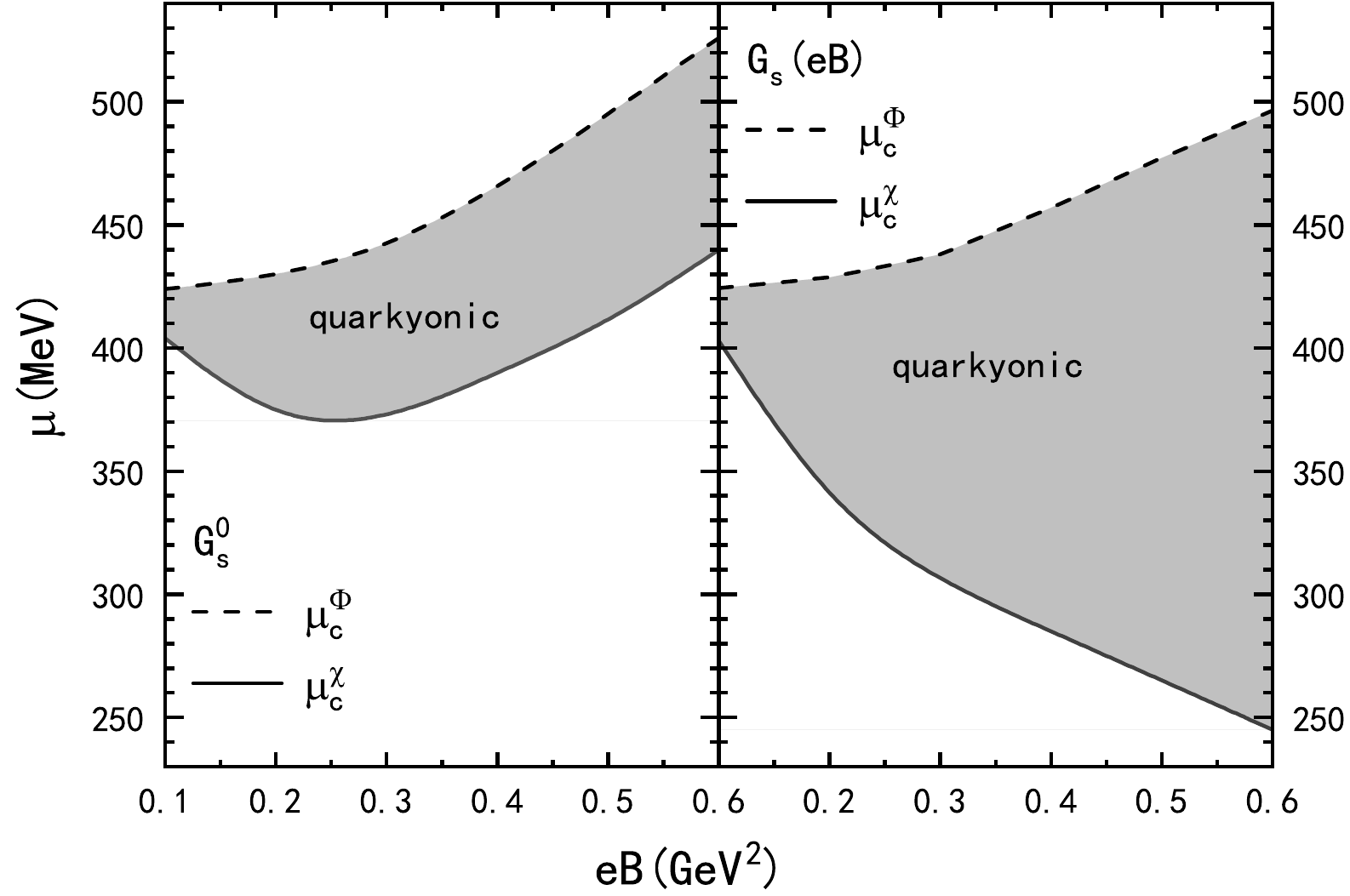}}}%
\dimen2=\dimexpr\wd0+\wd2\relax
\edef\figH{\the\dimexpr 10cm*\textwidth/\dimen2\relax}%
\subfloat[\centering ]{\expandafter\includegraphics\expandafter[\trimA,height=\figH]{p_term_crop.png}}%
\subfloat[\centering ]{\expandafter\adjustbox\expandafter{\trimB,height=\figH}{\includegraphics{Graph7_Enlarge_Quarkyonic.pdf}}}
\caption{
    (a) Spin susceptibility for spin-polarized quarkyonic matter. The parameter $\widetilde{p}$ represents the energy contribution from spin polarization. Ferromagnetic instability is observed from negative susceptibility for some choices of $\widetilde{p}$. Taken from \citep{Gao:2025rbq}.
    (b) Enlargement of the quarkyonic phase window under a running coupling interaction, with coupling varying with magnetic field. Taken from \citep{Wang:2022xxp}.
    }
\label{fig:quarky_ferro}
\end{figure*}

\subsection{Quarkyonic matter}
\label{Quarkyonic_Matter_Section}
\noindent An important measure for determining a change of phase from a non-relativistic fluid to an ultra-relativistic system is the speed of sound $c_S$. Around saturation density, matter is a non-relativistic liquid of nucleons with a sound velocity, $c_S^2 \sim 0.01 - 0.1$. However, there is sufficient amount of evidence in the literature \citep{Marczenko:2022jhl,Altiparmak:2022bke,Chatterjee:2023ecc} that $c_S^2$ should violate the conformal limit and attain a peak near $3-4\rho_0$. At such density regimes of transition, quarkyonic phase of matter~\citep{Mclerran:2015qgo,Fujimoto:2023mzy,Folias:2024rjr,Folias:2024skj} becomes relevant where quarks reside deep within in the Fermi sea while hardons live near the Fermi surface. Since the foundational work~\citep{McLerran:2008ux,McLerran:2009ve,McLerran:2007qj,McLerran:2007qj}, different theoretical frameworks have been used to explain various aspects of quarkyonic matter related to, for eg., its Bremsstrahlung photon spectrum during dynamical evolution~\citep{Torrieri:2013mk,Vogel:2014dra}, thermal properties~\citep{Sen:2020peq} and various astrophysical consequences~\citep{Margueron:2021dtx,Cao:2022inx,Fujimoto:2024doc,Jeong:2019lhv,Duarte:2020kvi,Gartlein:2025zhd,Sun:2026vxk,Pang:2023dqj,Sun:2026vxk}.

Literature on magnetic properties of quarkyonic matter is limited, providing only a fragmented picture. For instance, Ref. \citep{Chaudhuri:2022oru} analyzes the magnetic response of quarkyonic matter within the context of two-flavor PNJL model with AMM interaction and it is found to be paramagnetic in nature, with positive normalized susceptibility. From a more microscopic perspective, Ref.~\citep{Gao:2025rbq} demonstrates that incorporating spin-one polarization can impart ferromagnetic instabilities at low densities, as shown in Fig.~\ref{fig:quarky_ferro}(a) before reverting to paramagnetic behavior at higher densities. In Ref.~\citep{Wang:2022xxp}, in the context of the PNJL model, it is shown that strong magnetic fields enlarge the possible quarkyonic phase window where the coupling runs with the field, as shown in Fig.~\ref{fig:quarky_ferro}(b).

\subsection{Pion Condensate}
\label{Pion_condensate_Section}
\begin{figure*}[hbt]
\centering
\def\trimA{trim={0pt {-.008\height} 0pt 0pt},clip,valign=B}
\def\trimB{trim={0pt 0pt 0pt 0pt},valign=B}
\def\trimC{trim={0pt {-.033\height} 0pt {.119\height}},clip,valign=B}
\sbox0{\expandafter\adjustbox\expandafter{\trimA,totalheight=10cm}{\includegraphics{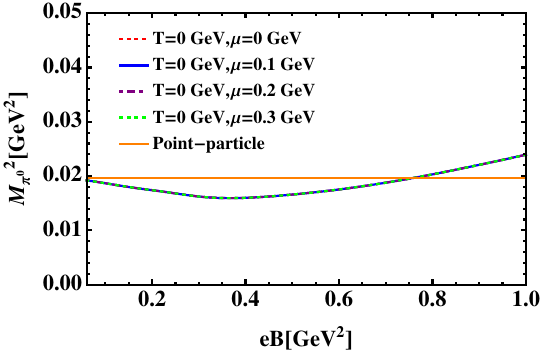}}}%
\sbox2{\expandafter\adjustbox\expandafter{\trimB,totalheight=10cm}{\includegraphics{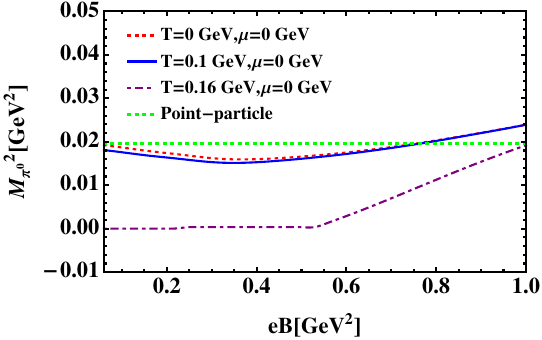}}}%
\sbox4{\expandafter\adjustbox\expandafter{\trimC,totalheight=10cm}{\includegraphics{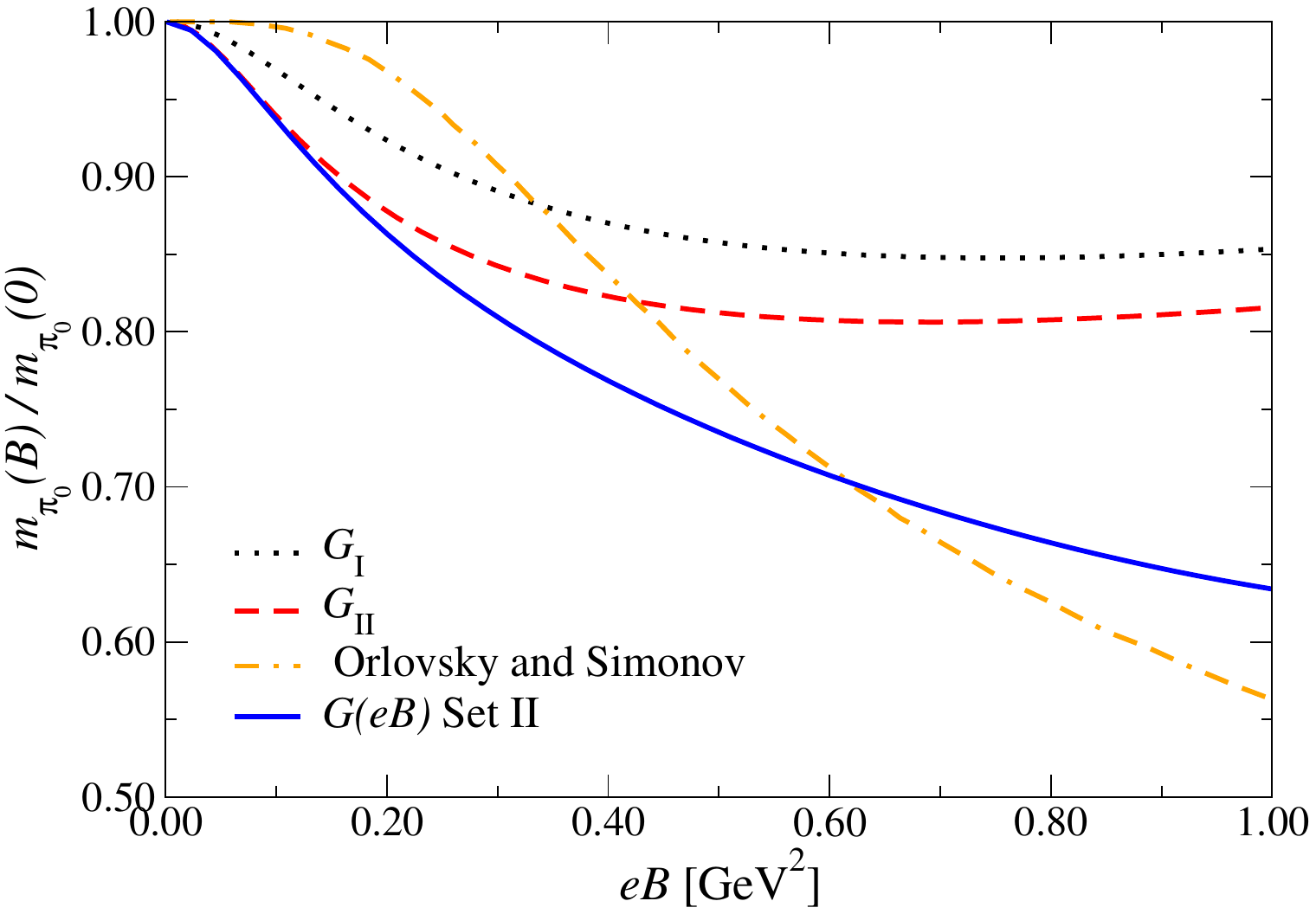}}}%
\dimen0=\dimexpr\textwidth-2em\relax
\dimen2=\dimexpr\wd0+\wd2+\wd4\relax
\edef\figH{\the\dimexpr 10cm*\dimen0/\dimen2\relax}%
\subfloat[\centering ]{\expandafter\adjustbox\expandafter{\trimA,totalheight=\figH}{\includegraphics{neutrpimuchange.pdf}}}%
\hfill
\subfloat[\centering ]{\expandafter\adjustbox\expandafter{\trimB,totalheight=\figH}{\includegraphics{nuetrapiTchange.pdf}}}%
\hfill
\subfloat[\centering ]{\expandafter\adjustbox\expandafter{\trimC,totalheight=\figH}{\includegraphics{mpi0run.pdf}}}
\caption{
    (a) Squared mass of the $\pi^0$ meson at $T = 0$ for different quark chemical potentials $\mu$. Taken from \citep{Liu:2018zag}.
    (b) Squared mass of the $\pi^0$ meson at $\mu = 0$ for different temperatures $T$. Taken from \citep{Liu:2018zag}.
    (c) The blue curve shows the normalized $\pi^0$ mass from \citep{Avancini:2016fgq} as a function of $eB$ in the NJL model with $G(eB)$. Similar results are reported in \citep{GomezDumm:2017jij}.}
\label{fig:neutral_pion_combined}
\end{figure*}

\noindent Pions play a central role in low-energy QCD \citep{Cao:2015xja} as the pseudo-Goldstone bosons associated with the spontaneous breaking of chiral symmetry, with small masses compared to other hadrons. In the NJL model, pions emerge naturally, providing a consistent framework to describe both their Goldstone nature, their in-medium modifications and how pion properties evolve under strong magnetic fields. Among various studies related to pions in a hot, dense and magnetized medium, the in-medium behavior of pion masses has been extensively studied in the literature showing that charged pion masses usually increase with magnetic field strength, while the neutral pion mass exhibits a much weaker dependence, as shown in Fig.~\ref{fig:neutral_pion_combined}. Ref.\cite{Mao:2020xvc} calculates the pion masses using NJL polarization function and uses $\langle \overline{\psi}\gamma^5\tau^1\psi\rangle$ as the order parameter for pion superfluid phase transition. It starts from the normal phase with neutral chiral condensate and determine the critical point of $\pi$ superfluid phase transition by the appearance of Goldstone boson, massless $\pi^+$ meson. Mesons are constructed above the mean mean field model by through quark bubble summations in the frame of random phase approximation \cite{Buballa:2003qv,Hatsuda:1994pi}. Following Ref.\cite{Mao:2018dqe},  meson propagator $D_M$ can be expressed in
terms of the meson polarization function $\Pi_M$ as
\begin{align}
    D_M(\bar{k}) = \frac{G}{1 - G\Pi_{M}(\bar{k})}
\end{align}
$\bar{k}$ being the conserved Ritus momentum. The meson pole mass $m_M$ is defined as
\begin{align}
    1 -G\Pi(k_0 = m_M) = 0 
\end{align}
Based on the Goldstone’s theorem, massless Goldstone mode $\pi^+$ exists in the pion superfluid phase. Therefore, the critical isospin chemical potential $\mu^{c\pi}_I$ for pion superfluid can be identified by the condition $m_{\pi^+}(B, T, \mu_B, \mu^{c\pi}_{I}) = 0$. For $\pi^+$ meson, one can show that $\Pi_{+}(k^0) = J_1 + J_2(k^0)$ where
\begin{equation}
J_1 = 3\mathlarger{\sum}_{f,\text{k}}\alpha_{\text{k}}\frac{|q_f|B}{2\pi}\int\frac{dp_z}{2\pi}\,\frac{1 - f(\epsilon_{f\text{k}} + \mu_f) - f(\epsilon_{f\text{k}} - \mu_f)}{\epsilon_{f\text{k}}}
\end{equation}
where $f = 1/(1 + e^{x/T})$ represents the Fermi Dirac distribution and 
\begin{widetext}
\begin{align}
    J_2(k^0) = \mathlarger{\sum}_{\text{k}, \text{k}^\prime}\int\frac{dp_z}{(2\pi)}\frac{j_{\text{k},\text{k}^\prime}(k^0)}{4\epsilon_{d\text{k}}\epsilon_{u\text{k}^\prime}}\Bigg[\frac{f(-\epsilon_{u\text{k}^\prime} - \mu_u) - f(\epsilon_{d\text{k}} - \mu_d)}{k^0 + \mu_I  + \epsilon_{d\text{k}} + \epsilon_{u\text{k}^\prime}} + \frac{f(\epsilon_{u\text{k}^\prime} - \mu_u) - f(-\epsilon_{d\text{k}} - \mu_d)}{k^0 + \mu_I  - \epsilon_{d\text{k}} - \epsilon_{u\text{k}^\prime}}\Bigg]
\end{align}
\end{widetext}
where
\begin{align}
&j_{\text{k}, \text{k}^\prime}(k^0) = \bigg[\frac{(k^0 + \mu_I)^2}{2} - \text{k}^\prime |Q_uB| - \text{k}|Q_uB|\bigg]j^{+}_{\text{k}, \text{k}^\prime} \nonumber\\
&\quad - 2\sqrt{\text{k}^\prime |q_uB|\,\text{k}|q_dB|}\; j^{-}_{\text{k}, \text{k}^\prime}
\end{align}
The expressions for $j^{\pm}_{\text{k},\text{k}^\prime}$ are given in detail in Ref.\cite{Mao:2018dqe,Mei:2026xlj}.

\begin{figure*}[hbt]
\centering
\subfloat[\centering ]{\includegraphics[width=0.325\textwidth]{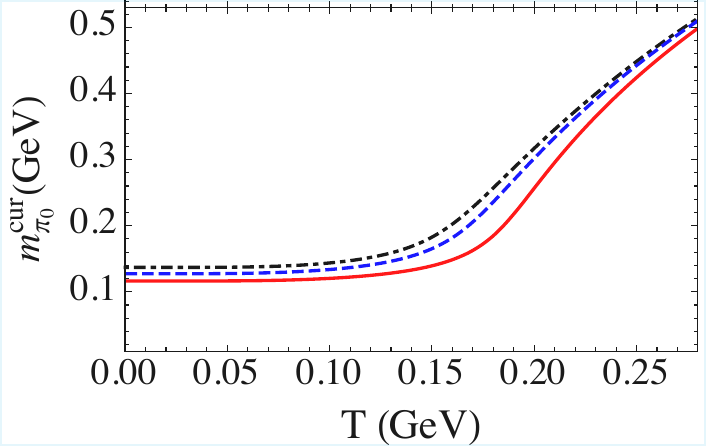}}%
\subfloat[\centering ]{\includegraphics[width=0.356\textwidth]{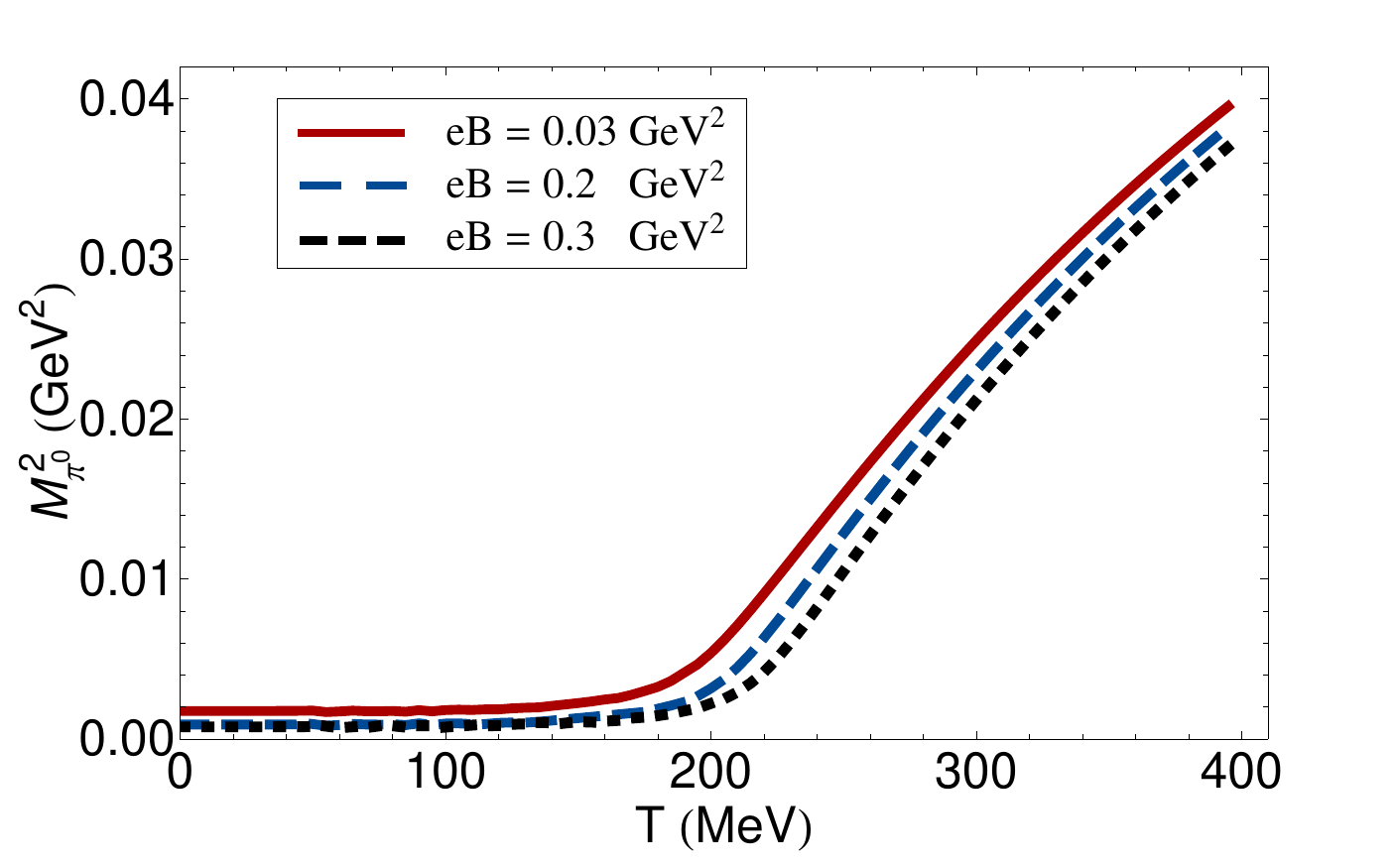}}%
\subfloat[\centering ]{\includegraphics[width=0.319\textwidth]{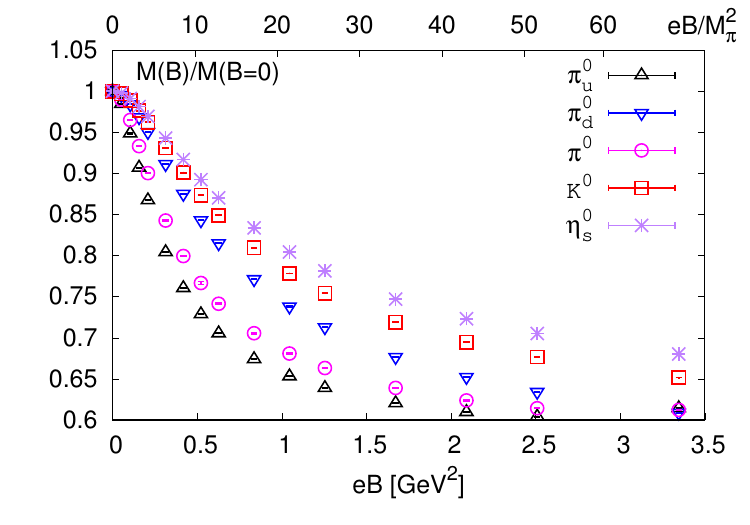}}
\caption{
    (a) Temperature variation of curvature mass of $\pi^0$. Taken from \citep{Wang:2017vtn}.
    (b) Same quantity from \citep{Fayazbakhsh:2012vr}.
    (c) Masses of $\pi^0_u$, $\pi^0_d$, $K^0$, $\eta^0$ as a function of $eB$ normalized by their corresponding masses at $eB = 0$. Taken from \citep{Ding:2020hxw}.
    }
\label{fig:Pion_Mass_vs_T}
\end{figure*}

\begin{figure*}[hbt]
\centering
\subfloat[\centering ]{\includegraphics[width=0.333\textwidth]{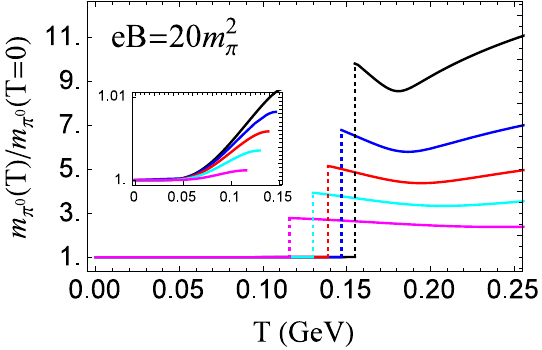}}%
\subfloat[\centering ]{\includegraphics[width=0.333\textwidth]{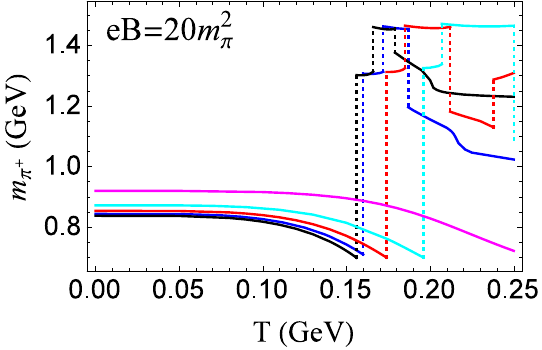}}%
\subfloat[\centering ]{\includegraphics[width=0.333\textwidth]{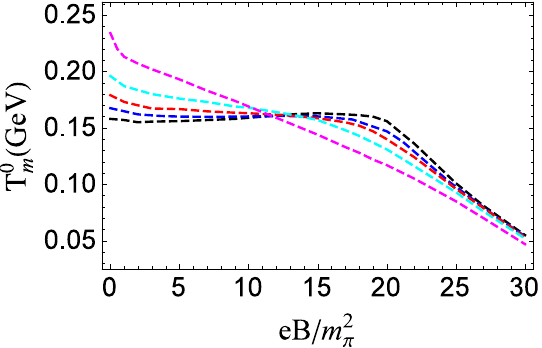}}
\caption{
    (a) Normalized $\pi^0$ mass $m_{\pi^0}$ (upper panel) as a function of temperature at fixed magnetic field $eB = 20\,m^2_\pi$ for different current quark masses in the NJL model. Black, blue, red, cyan and magenta lines correspond to $m_0 = 2, 5, 10, 20, 50$ MeV, respectively. Taken from \citep{Li:2023rsy}.
    (b) Similar results for the $\pi^+$ mass $m_{\pi^+}$. Taken from \citep{Li:2023rsy}.
    (c) Mott transition temperature $T^0_m$ of $\pi^0$ as a function of $eB$ for different current quark masses $m_0$, with the same color coding as in (a). Taken from \citep{Li:2023rsy}.
    }
\label{fig:pion_Mott_combined}
\end{figure*}

These trends are shown to be consistent for curvature, pole, and screening masses using the two-flavor NJL model \citep{Wang:2017vtn} at $T = 0$, where it is shown that the curvature and pole masses of $\pi^{+}$ increase with $B$, the same for $\pi^{0}$ remain almost the same, and the screening masses for $\pi^{0}$ decrease, supported by LQCD results \citep{Ding:2020hxw}, as shown in Fig.~\ref{fig:Pion_Mass_vs_T}(c). The approximate constancy of the neutral pion mass with magnetic field is also demonstrated in Ref.~\citep{Liu:2018zag} using the random phase approximation at leading order in $1/N_c$ within the two-flavor NJL model, as shown in Figs. \ref{fig:neutral_pion_combined}(a) and (b). Similar results have been confirmed in \citep{Li:2020hlp} and \citep{Coppola:2018vkw}. Moreover, Ref.~\citep{Fayazbakhsh:2012vr} shows that, because of the explicit breaking of Lorentz invariance by the magnetic field, the screening mass of neutral mesons exhibits a certain anisotropy in the transverse and longitudinal directions with respect to the direction of the external magnetic field. Ref. \citep{Liu:2018zag} also shows that the $\pi^\pm$ mass increases with magnetic field due to quark-antiquark loop contributions, the same as in \citep{Wang:2017vtn}. 

The inclusion of IMC modifies this picture substantially. By introducing a magnetic-field-dependent coupling $G(eB)$ that decreases with $eB$, Ref.~\citep{Avancini:2016fgq} found that the neutral-pion pole mass decreases with magnetic field (Fig.~\ref{fig:neutral_pion_combined}(c)), a trend also observed in \citep{GomezDumm:2017jij} and related models. At extremely strong magnetic fields, however, the behavior changes again. Ref.~\citep{Das:2019ehv} showed that for $eB\sim14m_\pi^2$, the neutral-pion mass increases with $eB$ and exhibits delayed MC, contrasting with earlier results where $m_{\pi^0}$ remained nearly constant or displayed IMC, while $m_{\pi^+}$ consistently showed MC.

\par Moreover, at finite temperature, depending on $B$, the curvature masses of $m_{\pi^{\pm}}$, as a function of temperature, develop minima as a function of temperature, whereas the curvature mass of $\pi^0$ increases, as shown in 
Fig.~\ref{fig:Pion_Mass_vs_T}. A similar qualitative trend of the temperature profile of $\pi^0$ mass has been found in Ref.~\citep{Sheng:2021evj} for studies of screening masses of neutral pions in hot, magnetized quark matter in the presence of IMC in the context of the lattice-improved PNJL model, although the nature of the temperature variation has been found to be different compared to the standard NJL model.

\par The NJL model is capable of describing chiral restoration, where the dynamical quark masses drop as a function of temperature and chemical potential, thus restoring the mass degeneracy of the chiral partner states. At the same time, the continuum thresholds for quark-antiquark scattering channels drop, which results in a lowering of the binding energy for the pseudoscalar meson bound states and, finally, in their dissociation when they enter the continuum and change their character to resonances with a finite lifetime. This is called the Mott transition effect. This is usually identified by mass poles for mesons becoming complex, with the real part being the ``mass'' of the resonance and the imaginary part being related to a finite width due to its decay into the quark constituents. Following early works in pion dissociation in quark matter \citep{Hufner:1994ma,Zhuang:1994dw,Blaschke:2013zaa,Dubinin:2013yga,Wergieluk:2012gd,Yamazaki:2012ux,Yamazaki:2013yua,Blaschke:2014zsa,Torres-Rincon:2016ahl}, numerous works have been done on the Mott transition of pions. Some relatively recent works include, but are not restricted to, Refs.~\citep{Mao:2017wmq,Sheng:2020hge,GomezDumm:2020bxj,Mei:2022dkd,Li:2023rsy}, where in the presence of a strong magnetic field a sudden jump in the mass of the Goldstone mode at the Mott transition temperature is observed, as shown in Fig. \ref{fig:pion_Mott_combined}.

\par
\begin{figure*}[t]
\centering
\def\trimA{trim={0pt 0pt 0pt 0pt},valign=B}%
\def\trimB{trim={0pt 0pt 0pt 0pt},valign=B}%
\def\trimC{trim={0pt {-.010\height} 0pt 0pt},clip,valign=B}
\sbox0{\expandafter\adjustbox\expandafter{\trimA,totalheight=10cm}{\includegraphics{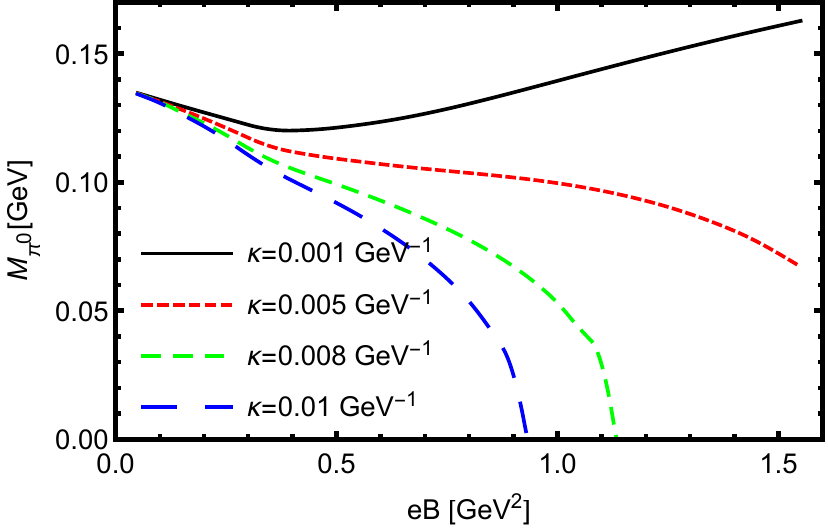}}}%
\sbox2{\expandafter\adjustbox\expandafter{\trimB,totalheight=10cm}{\includegraphics{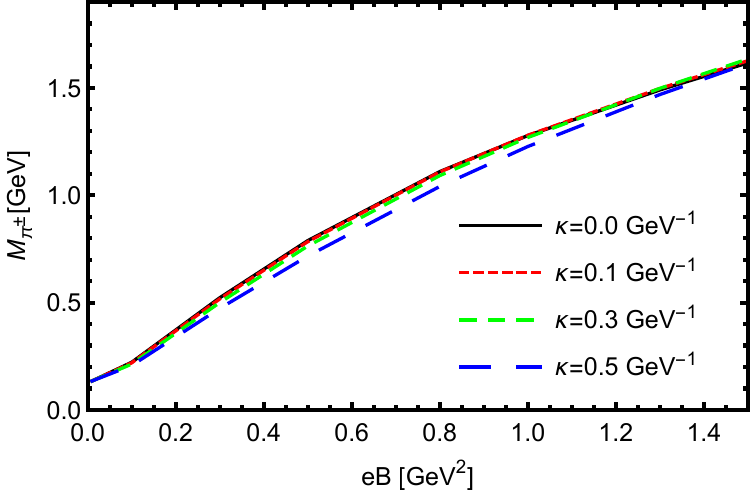}}}%
\sbox4{\expandafter\adjustbox\expandafter{\trimC,totalheight=10cm}{\includegraphics{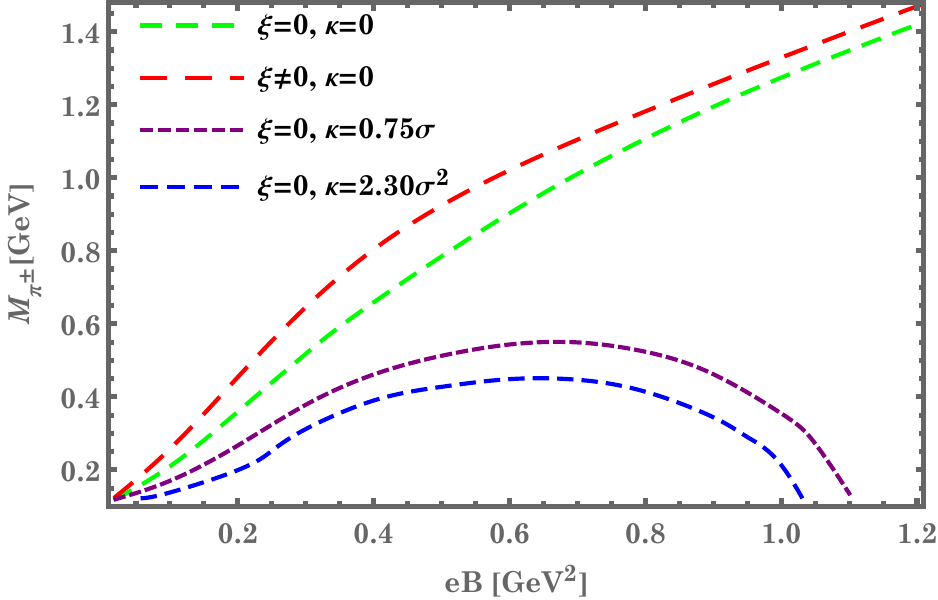}}}%
\dimen0=\dimexpr\textwidth-2em\relax
\dimen2=\dimexpr\wd0+\wd2+\wd4\relax
\edef\figH{\the\dimexpr 10cm*\dimen0/\dimen2\relax}%
\subfloat[\centering ]{\expandafter\adjustbox\expandafter{\trimA,totalheight=\figH}{\includegraphics{neutralpionmass.pdf}}}%
\hfill
\subfloat[\centering ]{\expandafter\adjustbox\expandafter{\trimB,totalheight=\figH}{\includegraphics{chargedpionmass.pdf}}}%
\hfill
\subfloat[\centering ]{\expandafter\adjustbox\expandafter{\trimC,totalheight=\figH}{\includegraphics{Cpion.pdf}}}
\caption{
(a) Neutral and (b) charged pion mass as functions of the magnetic field for different AMM couplings $\kappa$. Taken from \citep{Xu:2020yag}.
(c) Charged pion mass $m_{\pi^+}$ as a function of $eB$ with TSP and AMM. Taken from \citep{Lin:2022ied}.
}
\label{fig:pion_AMM_TSP}
\end{figure*}

 It has been shown many times in the literature that inclusion of AMM can significantly affect the pion mass profiles (although there are counterclaims as well \citep{Ferrer:2015wca}), as shown in Fig.~\ref{fig:pion_AMM_TSP}. Some of the recent works include Ref. \citep{Xu:2020yag}, which discusses the effects of AMM and IMC and shows that the neutral pion is very sensitive to the AMM term, with its mass decreasing rapidly with magnetic field and even leading to collapse of the neutral pion mass under sufficient conditions \citep{Aguirre:2021ljk,Yang:2025zcs} (Fig.~\ref{fig:pion_AMM_TSP}(a)), whereas the charged pion mass is much less sensitive to AMM (Fig.~\ref{fig:pion_AMM_TSP}(b)). Ref. \citep{Lin:2022ied}, which also includes contributions from the tensor-type spin polarization (TSP), shows a similar decreasing behavior of the neutral pion mass, while the charged pion mass is shown to behave non-monotonically (Fig.~\ref{fig:pion_AMM_TSP}(c)), similar to the former results, whereas the TSP induces MC of both neutral and charged pion masses. Ref. \citep{Chaudhuri:2019lbw} shows that, for non-zero values of the magnetic field, the strength of the Mott transition decreases substantially with increasing magnetic field when the AMM is included.

\par
\begin{figure*}[hbt]
\centering
\def\trimA{trim={0pt {-.008\height} 0pt {-.004\height}},clip,valign=B}%
\def\trimB{trim={0pt 0pt 0pt {-.004\height}},clip,valign=B}%
\def\trimC{trim={0pt {-.010\height} 0pt {-.015\height}},clip,valign=B}%
\def\trimD{trim={0pt {-.010\height} 0pt {-.015\height}},clip,valign=B}%
\def\trimE{trim={0pt {-.083\height} 0pt 0pt},clip,valign=B}%
\def\trimF{trim={0pt {-.070\height} 0pt 0pt},clip,valign=B}%
\def\pnl#1#2#3{\expandafter\adjustbox\expandafter{#1,totalheight=#2}{\includegraphics{#3}}}%
\sbox0{\pnl\trimA{10cm}{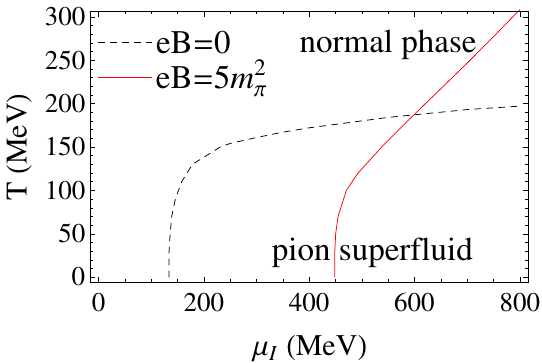}}\sbox2{\pnl\trimB{10cm}{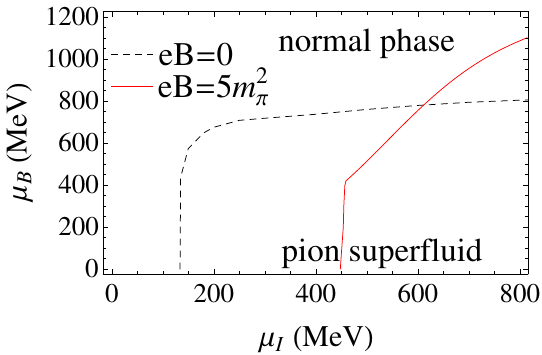}}\sbox4{\pnl\trimC{10cm}{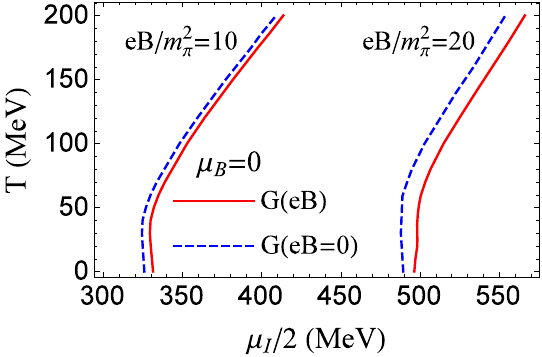}}%
\dimen2=\dimexpr\wd0+\wd2+\wd4\relax
\sbox0{\pnl\trimD{10cm}{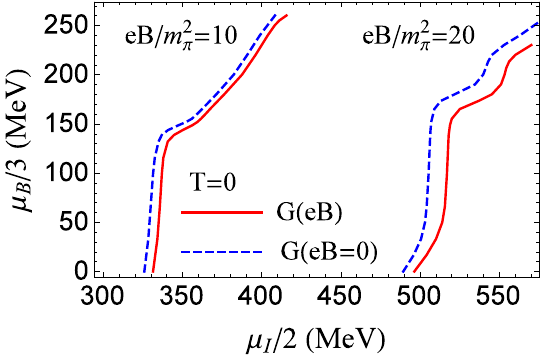}}\sbox2{\pnl\trimE{10cm}{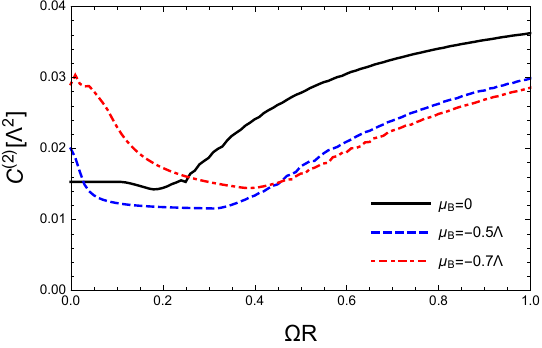}}\sbox4{\pnl\trimF{10cm}{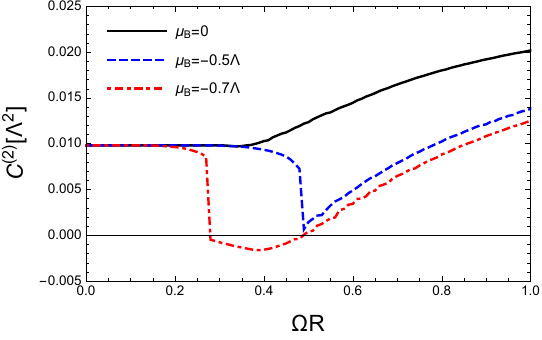}}%
\dimen4=\dimexpr\wd0+\wd2+\wd4\relax
\ifdim\dimen4>\dimen2 \dimen2=\dimen4 \fi
\dimen0=\dimexpr\textwidth-2em\relax
\edef\figH{\the\dimexpr 10cm*\dimen0/\dimen2\relax}%
\subfloat[\centering ]{\pnl\trimA\figH{fig2new.pdf}}\hfill
\subfloat[\centering ]{\pnl\trimB\figH{fig32020923.pdf}}\hfill
\subfloat[\centering ]{\pnl\trimC\figH{tmui.pdf}}\\
\subfloat[\centering ]{\pnl\trimD\figH{muimub.pdf}}\hfill
\subfloat[\centering ]{\pnl\trimE\figH{G11C.pdf}}\hfill
\subfloat[\centering ]{\pnl\trimF\figH{G17C.pdf}}
\caption{
(a) Pion superfluid phase diagram in the $\mu_I$--$T$ plane at $\mu_B = 0$ and fixed magnetic field \citep{Mao:2020xvc}. (b) The same diagram in the $\mu_I$--$\mu_B$ plane at $T = 0$ and fixed magnetic field \citep{Mao:2020xvc}. (c) Pion superfluid phase diagram in the $\mu_I$--$T$ plane at $\mu_B = 0$ and fixed magnetic field \citep{Mao:2022nfs}. The normal phase lies to the left of the transition line and the pion superfluid phase to the right. (d) The same diagram in the $\mu_I$--$\mu_B$ plane \citep{Mao:2022nfs}. As $eB$ increases, the critical isospin chemical potential for the pion superfluid transition increases, implying that the magnetic field disfavors the pion superfluid phase. (e) The Ginzburg--Landau coefficient $C^{(2)}$ as a function of the angular velocity for $G = 11\Lambda^{-2}$ and $|qB| = 0.1\Lambda^2$ at zero temperature, where $\Lambda$ is the UV cutoff \citep{Chen:2019tcp}. (f) $C^{(2)}$ for $G = 17\Lambda^{-2}$ \citep{Chen:2019tcp}. A negative $C^{(2)}$ signals the onset of pion superfluidity.
}
\label{fig:pion_superfluid}
\end{figure*}

 Also, pion superfluid phase \citep{Mu:2010zz,Xia:2013caa,Ueda:2013sia,Mao:2014hga,Adhikari:2018cea,Dantas:2019rgp} has attracted much attention due to its impact on compact stars, isospin asymmetric nuclear matter, and heavy-ion collisions at intermediate energies. 
A few recent works include Ref.~\citep{Mao:2020xvc} on pion superfluid phase diagram within the Pauli--Villars regularized NJL model (Fig.~\ref{fig:pion_superfluid}(a) and (b)) which applies Goldstone’s theorem to determine and obtain the phase diagram in $B$, $T$, $\mu_I$ and $\mu_B$
space, Ref.~\citep{Kang:2013bea} which studies the de Hass van Alphen oscillation of pion superfluid phase, and Ref.~\citep{Chen:2019tcp} which includes, on top of external magnetic field, the effect of rotation (Figs.~\ref{fig:pion_superfluid}(e) and (f)) finding that charged-pion condensation takes place only when both a strong coupling constant and negatively large baryon chemical potential are applied. Ref. \cite{Liu:2017spl} shows that the combined effects of a rotation and a magnetic field can cause charged pion condensation. Ref.\cite{Guo:2024ocw} studies vortex lattice structures in the $\pi$ condensates under conditions of small rotation and strong coupling constant in the context of a relativistic complex scalar field model with a repulsive self-interaction. These vortex lattices can be thought of as electrical superconductors which are created to better adapt to changes in rotation and interaction. However, there are counterclaims against the salability of the magnetic pion superfluid as well, e.g. Ref.~\citep{Mao:2022nfs}  (Figs.~\ref{fig:pion_superfluid}(c) and (d)) which shows that magnetic field disfavors the pion superfluid phase. 

\begin{figure*}[hbt]
\centering
\subfloat[\centering ]{\includegraphics[width=0.333\textwidth]{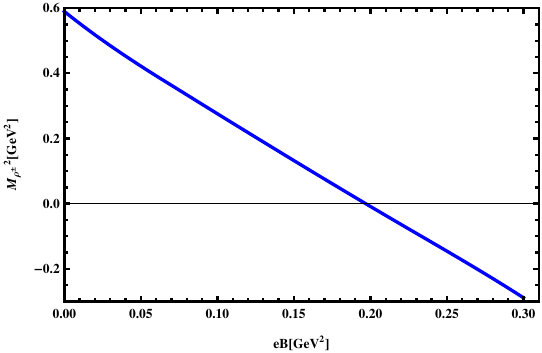}}%
\subfloat[\centering ]{\includegraphics[width=0.333\textwidth]{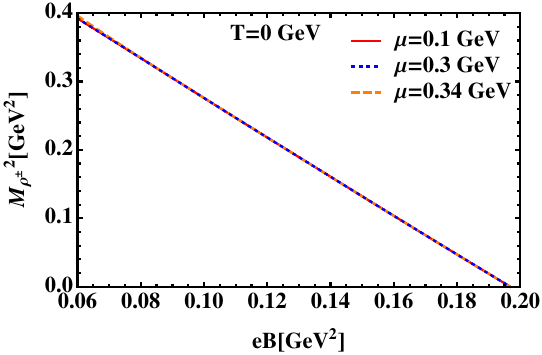}}%
\subfloat[\centering ]{\includegraphics[width=0.333\textwidth]{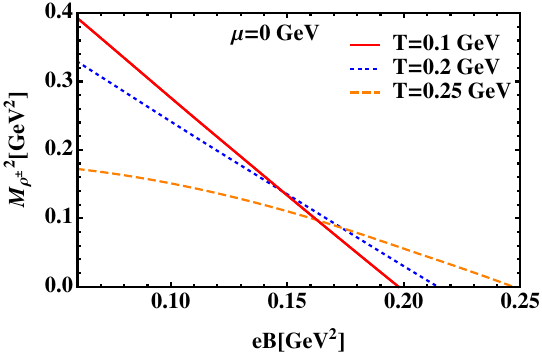}}
\caption{
(a) Squared mass of the charged $\rho^\pm$ with spin component $s_z = \pm 1$ as a function of $eB$, obtained in the NJL model, where mesons are constructed as infinite sums of quark-loop chains in the random phase approximation. Taken from \citep{Liu:2014uwa}.
(b) Mass of the $\rho^\pm$ states with $s_z = \pm 1$ as a function of $eB$ at $T = 0$ for different quark chemical potentials $\mu$. Taken from \citep{Liu:2015pna}.
(c) Same quantity at $\mu = 0$ for different temperatures. Taken from \citep{Liu:2015pna}.
}
\label{fig:rho_condensation}
\end{figure*}

Different works \cite{Sawyer:1972cq,Scalapino:1972fu} have tried to come up with the possible mechanisms of charged pion condensation by a softening of the pion spectrum by the p-wave pion-nucleus interaction. A transition to a superfluid phase was expected just above the nuclear saturation density. However, Ref. \cite{Migdal:1973zm} that the $\pi-\pi$ repulsive interaction would tilt the preference towards $\pi^0$ condensation. The problem of $\pi$-condensation is still an open problem. However, outside the so called $\beta$-equilibrium sheet $\mu_I = f(\mu_B, T)$, finite $\mu_I$ QCD has been studies very extensively \cite{Brandt:2017zck,Brandt:2017oyy} using chiral perturbation theory ($\chi PT$) starting from pioneering work by Son and Stephanov \cite{Son:2000xc}, a low energy EFT of QCD which describes the dynamics of the pseudo-Goldstone bosons resulting from spontaneous symmetry breaking of global symmetries in the QCD vacuum. At large and positive isospin chemical potential $\mu_I$, on expects to populate the $u$ and $\bar{d}$ states with the color interaction inducing the formation of $\pi^+$ states. On the flip side, a large negative $\mu_I$ will instead favor the $\pi^-$ states. It is established from both LQCD and $\chi PT$ that at $\mu_I = m_{\pi}$ at $T = 0$ \cite{Son:2000xc,Kogut:2001id}, the vacuum makes a second order phase transition to the pion condensate phase. On the other hand, for $T \approx 170~$MeV, thermal fluctuations destroy the condensate phase.
\par In principle, any information on the QCD phase diagram can be obtained by introducing in the QCD action a chemical potential for a certain charge by changing the covariant derivative as $D_\mu = \partial_{\mu} + ig_S G^{a}_{\mu}T^{a} - i\mu\delta_{\mu 0}$, $\mu$ being the chemical potential matrix $\mu$ is defined as 
\begin{align}
    \mu = \text{diag}(\mu_u, \mu_d, \mu_s) = \frac{\mu_B - \mu_S}{3}\mathbb{I}_3 + \mu_I T^3 + \frac{2\mu_S}{\sqrt{3}}T^8
\end{align}
and $\mu_{u,d} = \mu_B/3 \pm \mu_I/2$ and $\mu_s = \mu_B/3 - \mu_S$ where $\mu_B$, $\mu_I$ and $\mu_S$ are baryon, isospin and strangeness chemical potential. The pion condensates are defined as $\pi^a = \langle\overline{\psi}\sigma^a\gamma^5 \psi\rangle$ as a functions of $T$, $\mu_B$ and $\mu_I$ and $\sigma_a$ being Pauli matrices. Now, because studies of pions correspond to non-perturbative regime of QCD, the QCD lagrangian is practically is of no use here. Instead of effective model e.g. linear sigma models\cite{Dominguez:1993kr,Tawfik:2019tkp} and $\chi PT$ \cite{Kogut:2001id,Birse:2001sn,Loewe:2002tw,Loewe:2004mu,Loewe:2011tm,Mammarella:2015pxa,Loewe:2016wsk,Carignano:2016lxe,Lepori:2019vec}, NJL models \cite{Barducci:2004nc,He:2005nk,Ebert:2005cs,Ebert:2005wr,He:2006tn,Andersen:2007qv,Abuki:2008tx,Abuki:2008wm,Xia:2013caa,Chao:2018ejd,Khunjua:2019nnv}, quark-meson models \cite{Adhikari:2018cea,Andersen:2018osr}, random matrix models \cite{Klein:2003fy,Klein:2004hv}, ADS/QCD \cite{Lv:2018wfq} and pQCD \cite{Graf:2015pyl} are used. Apart from strong guidance from LQCD for $\mu_S = \mu_B = 0, \mu_I \neq 0$, the most widely used $\chi PT$ and NJL model give fairly good qualitative description of the meson condensed phase. However, the controversy around meson condensation revolve around three basic requirements: the particles must be bosons, the system has to be sufficiently cold and the particles must be stable. Mesons satisfy the first two requirements in sufficiently cold system. However, all mesons are unstable in vacuum. Then the next question is whether medium effects and/or presence of magnetic field can change the situation.
\par Before discussing the condensation mechanism we point out that group theoretic considerations provide a very simple explanation for condensation. As pions are part of an isospin triplet with $I = 1$, they are the three eigenstates with different $I_3$ values. The degeneracy of vacuum pion masses is
lifted by $\mu_I$. In fact, we expect the dispersion relations to be $\epsilon_{\pi^0} = \sqrt{p^2 + m^2_\pi}$ and $\epsilon_{\pi^\pm} = \mp \mu_I + \sqrt{p^2 + m^2_\pi}$. These expression are valid only for $|\mu_I| < m_\pi$, otherwise, one of the charged pions becomes massless and further increase of $\mu_I$ makes the energy of the particle negative. This suggest a spontaneous symmetry breaking with the massless pion associated to the Nambu-Goldstone boson associated to a broken generator. This means as $\mu_I$ hits the vacuum pion mass, it becomes energetically favorable for the vacuum to rearrange itself and the charged pion field acquired a non-zero expectation value, indicating condensation. Now we explain two different approaches to meson condensation.
\par The global symmetry of QCD for constructing the $\chi PT$ model is the chiral symmetry $SU(N_f)_L\times SU(N_f)_R$, $N_f$ being the relevant number of quark flavors and the meson fields are collected in a unimodular field $\Sigma \in SU(2)$ transforming as $\Sigma \xrightarrow[]{} L\Sigma R^\dagger$ where $L(R)\in SU(2)_{L(R)}$. The leading order chiral perturbation theory Lagrangian is given by
\begin{align}
    \mathcal{L}_{\chi PT} = \frac{f^2_\pi}{4}\text{Tr}(D_\mu\Sigma D^\mu\Sigma^\dagger) + \frac{f^2_\pi}{4}\text{Tr}(X\Sigma^\dagger + \Sigma X^\dagger)
\end{align}
where the trace is taken in the flavor space, the auxiliary field X transforms as $\Sigma$ and the covariant derivative $D_\mu$ is defined as 
\begin{align}
    D_\mu\Sigma = \partial_\mu \Sigma + \frac{i}{2}[v_\mu,\Sigma]
\end{align}
where $v_\mu = 2\mu\delta_{\mu 0}$ is the external current. This leading order $\chi PT$ model is enough to describe the QCD phase for $\mu_I \leq m_\pi$. The RMF lagrangian is obtained by replacing $\Sigma \rightarrow \overline{\Sigma}$, the RMF expectation value of $\Sigma$ and $X \rightarrow \langle X \rangle = 2\mathcal{M}\mathcal{B}$, where $\mathcal{M}$ is the quark mass matrix and $\mathcal{B}$ is a free parameter
\begin{align}
    \mathcal{L}_{\chi PT, RMF} = -\frac{f^2_\pi}{4}\text{Tr}([\mu,\overline{\Sigma}][\mu,\overline{\Sigma}^\dagger]) + \frac{Bf^2_\pi}{2}\text{Tr}(\mathcal{M}(\overline{\Sigma} + \overline{\Sigma}^\dagger))
\end{align}
In terms of the parameterization $\overline{\Sigma} = e^{i\alpha\vec{n}\cdot\vec{\lambda}} = \cos\alpha + i\hat{n}\cdot\sigma \sin\alpha$, in principle, one should try to extremize $\mathcal{L}_{\chi PT, RMF}$ with the 8-dimensional vector $\vec{n}$, called the tilting vector. However, as it can be shown that each meson could be rotated by different $\vec{n}$, attempting this problem could be computationally challenging. For pions however ($N_f = 2$ and $\vec{n} = (n_1, n_2, n_3)$), one can show that
\begin{align}
    \mathcal{L}_{\chi PT, RMF} = f^2_\pi m^2_\pi \cos\alpha + \frac{f^2_\pi}{2}\sin^2\alpha~ 
    \mu^2_I (1 - n^2_3)
\end{align}
which is independent of $n_1$ and $n_2$. Then to maximize $\mathcal{L}_{\chi PT, RMF}$, one can set $n_3 = 0$ and one has the freedom of setting $n_1 = 0$ and $n_2 = 1$. Extremizing with these values leads to two vacuum solutions: $\sin\alpha = 0$ which leads to $\bar{\Sigma} = \text{diag}(1, 1, 1)$ the trivial QCD vacuum, and the non-trivial vacuum $\text{cos}\alpha = m^2_\pi/\mu^2_I$ which forces the condition $|\mu_I| > m_\pi$, which is precisely the pion condensation condition. More details on the condensation mechanism can be found in Ref.\cite{Mannarelli:2019hgn}. In the presence of sufficiently large magnetic fields, and at low-to-moderate isospin chemical potential, the ground state of QCD is expected to be a spatially modulated condensate of neutral pions termed as the chiral soliton lattice (CSL). In this setting, the model is modified by inclusion of the Wes-Zumino-Witten term \cite{Gronli:2022cri}
\begin{align}
    \mathcal{L}_{WZW} = \frac{e\mu_I}{8\pi^2}(\vec{B}\cdot\vec{\nabla})\phi 
\end{align}
where $\phi = \phi^0/f_\pi$ being the dimensionless neutral pion field. The chiral soliton lattice (CSL) is a spatially inhomogeneous $\pi^0$ condensate. One can parametrize this solution as $\Sigma = e^{i\tau_3\phi}$ and the solution for $\Sigma$ ultimately show elliptic modulations with period $\ell = 2kK(k)/m_\pi$ \cite{Gronli:2022cri,Brauner:2016pko} where $K(k)$ is the complete elliptic integral of the first kind and $0 < k < 1$. Pion condensation can also be studied using the NJL model \cite{Mannarelli:2019hgn}
\begin{align}
\mathcal{L}_{\text{NJL}} = \overline{\psi}(i\gamma^\mu D_\mu - M)\psi + G\Big[(\overline{\psi}\psi)^2 + (\overline{\psi}\gamma^5 \sigma \psi)^2\Big]
\end{align}
To obtain the properties of the vacuum, one bosonizes the model and introduces collective auxiliary variables $\sigma_f = -(4G/\Lambda)(\overline{\psi}_f\psi_f)$ and $\pi_a = -(2G/\Lambda)(\overline{\psi}\gamma^5 i\lambda_a\psi + \text{h.c.})$ corresponding to scalar and pseudoscalar channels respectively. At vanishing temperature, the grand potential has a fairly simple expression
\begin{align}
\Omega = G(\sigma^2 + \langle \pi \rangle^2) - \frac{3}{2\pi^2}\int_{0}^{\Lambda}dk~k^2 (\epsilon_+ + \epsilon_-)
\end{align}
where $\epsilon_{f,\pm} = \sqrt{(\epsilon_f \pm \mu_I/2)^2 + 4G^2\langle \pi\rangle^2}$. So, the effect of the pion condensate is to create a gap between the quasiparticle energies. This is the typical signature of formation of correlated pairs of fermions as now it costs additional energy $\propto \langle \pi \rangle^2$ to break the quark-antiquark pairs to produce quasi-particle fermionic excitations.

\subsection{Vector meson condensate}
\label{rho_condensate_section}
Despite counterclaims from QCD~\citep{Hidaka:2012mz} as well as sign-problem-free LQCD simulations~\citep{Hidaka:2012mz,Luschevskaya:2016epp,Bali:2017ian}, the QCD vacuum is shown to be unstable against very strong magnetic fields and to make a transition to neutral and charged rho mesons condensates \citep{Liu:2014uwa,Liu:2015pna}, as shown in Fig.~\ref{fig:rho_condensation}, in contradiction to Vafa-Witten Theorem~\citep{Hidaka:2012mz}. In fact, there are works which suggest that electromagnetic superconductivity of vacuum in strong magnetic background is actually not in conflict \cite{Chernodub:2012zx} with the Vafa-Witten and Elitzur theorem and results of \cite{Hidaka:2012mz} are consistent with inhomogeneous $\rho$-meson condensation \cite{Chernodub:2013uja}. This can be understood in a simple way: the dispersion relation of a free particle of the mass $m$ in a magnetic
field involves three parameters: the Landau level index $\text{k}\in\mathbb{Z}_{\geq 0}$, the spin projection in the direction of the magnetic field $s_z = -s \cdot\cdot\cdot\cdot s$, and the particle momentum along the
field’s axis, $p_z$: $\epsilon^2_{\text{k}} = m^2 + p^2_z + |qB|(2k + 1 -2s_z)$. Therefore, for pions and $\rho$ mesons, we have the magnetic dependent masses as
\begin{equation}
\begin{aligned}
m^2_{\pi^{\pm}}(B) &= m^2_{\pi^{\pm}} + eB
\end{aligned}
\qquad\qquad
\begin{aligned}
m^2_{\rho^{\pm}}(B) &= m^2_{\rho^{\pm}} - eB
\end{aligned}
\end{equation}
where the zero-field vacuum masses of $\pi^\pm$ and $\rho^\pm$ are given by $m_\pi = 139.6~\text{MeV}$ and $m_\rho =  775.5~\text{MeV}$. For the $\rho^\pm$ meson at the lowest Landau level with zero longitudinal momentum ($p_z = 0$), the meson mass becomes purely imaginary when $B$ exceeds $m^2_\rho/e \approx 10^{16}~\text{Tesla}$, suggesting that QCD vacuum is unstable against charged $\rho$ condensation~\citep{Callebaut:2011ab,Chernodub:2010qx,Chernodub:2011mc}. 

\begin{figure*}[hbt]
\centering
\def\trimA{trim={0pt {-.034\height} 0pt {-.004\height}},clip,valign=B}%
\def\trimB{trim={0pt {-.036\height} 0pt {-.004\height}},clip,valign=B}%
\def\trimC{trim={0pt {-.047\height} 0pt {-.009\height}},clip,valign=B}%
\def\trimD{trim={0pt {-.008\height} 0pt 0pt},clip,valign=B}%
\def\trimE{trim={0pt {-.028\height} 0pt {-.004\height}},clip,valign=B}%
\def\trimF{trim={0pt 0pt 0pt {-.023\height}},clip,valign=B}%
\def\pnl#1#2#3{\expandafter\adjustbox\expandafter{#1,totalheight=#2}{\includegraphics{#3}}}%
\sbox0{\pnl\trimA{10cm}{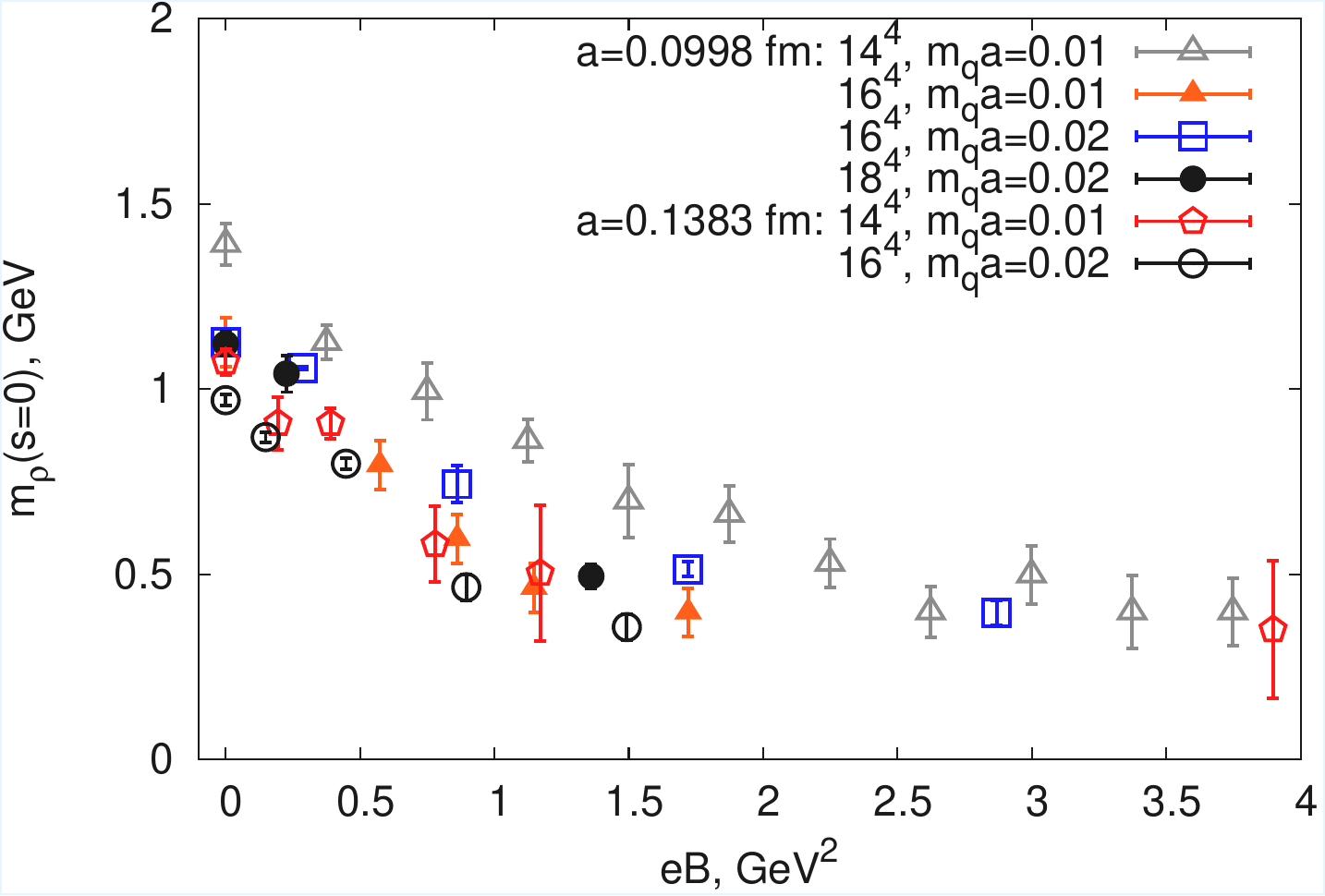}}\sbox2{\pnl\trimB{10cm}{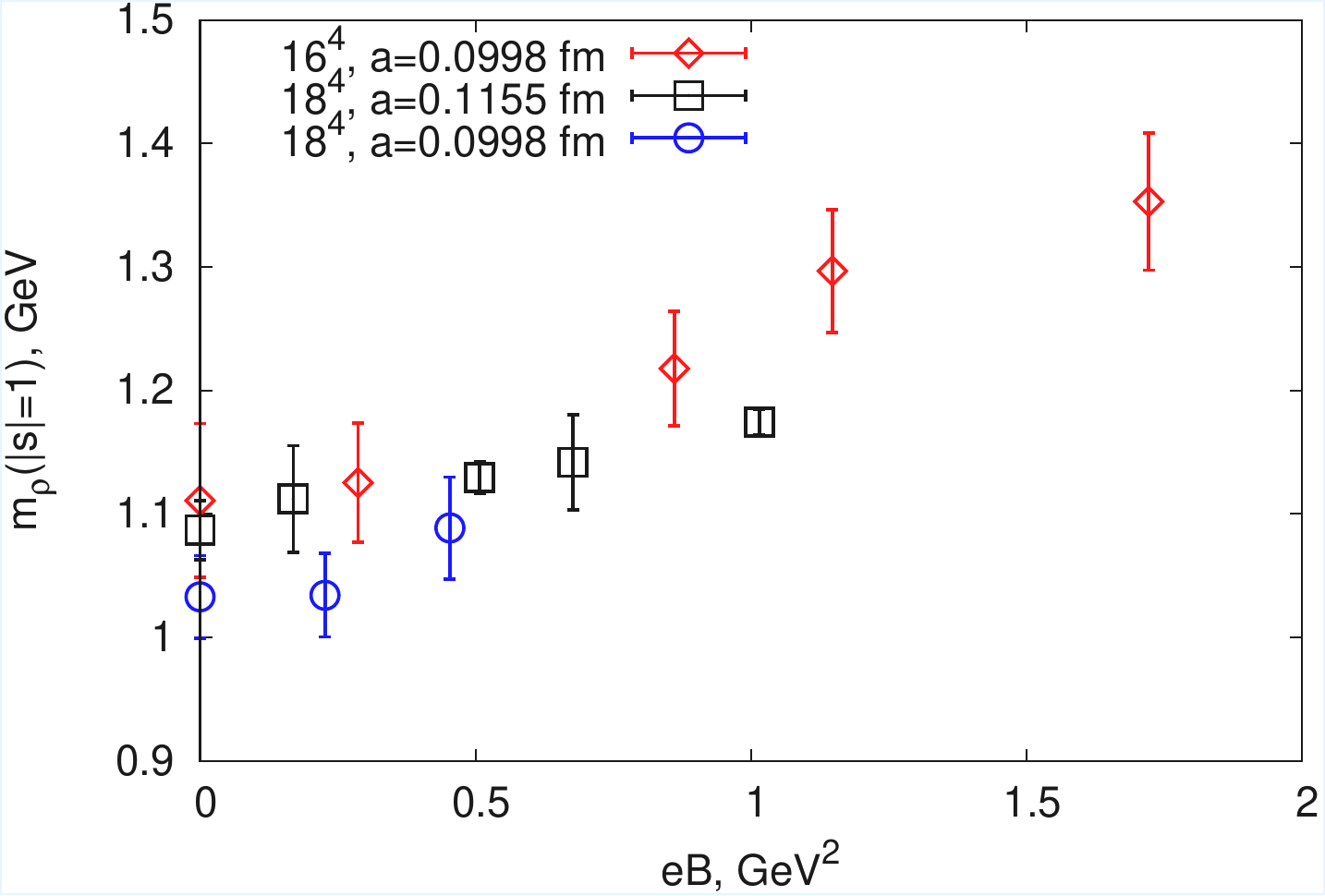}}\sbox4{\pnl\trimC{10cm}{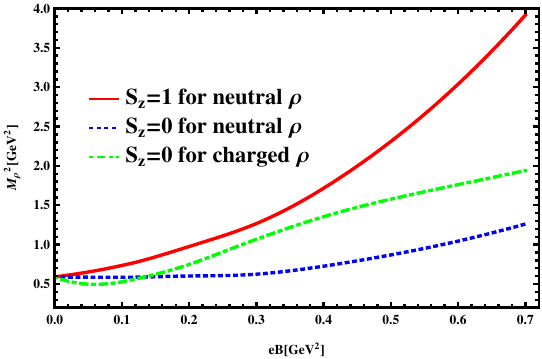}}%
\dimen2=\dimexpr\wd0+\wd2+\wd4\relax
\sbox0{\pnl\trimD{10cm}{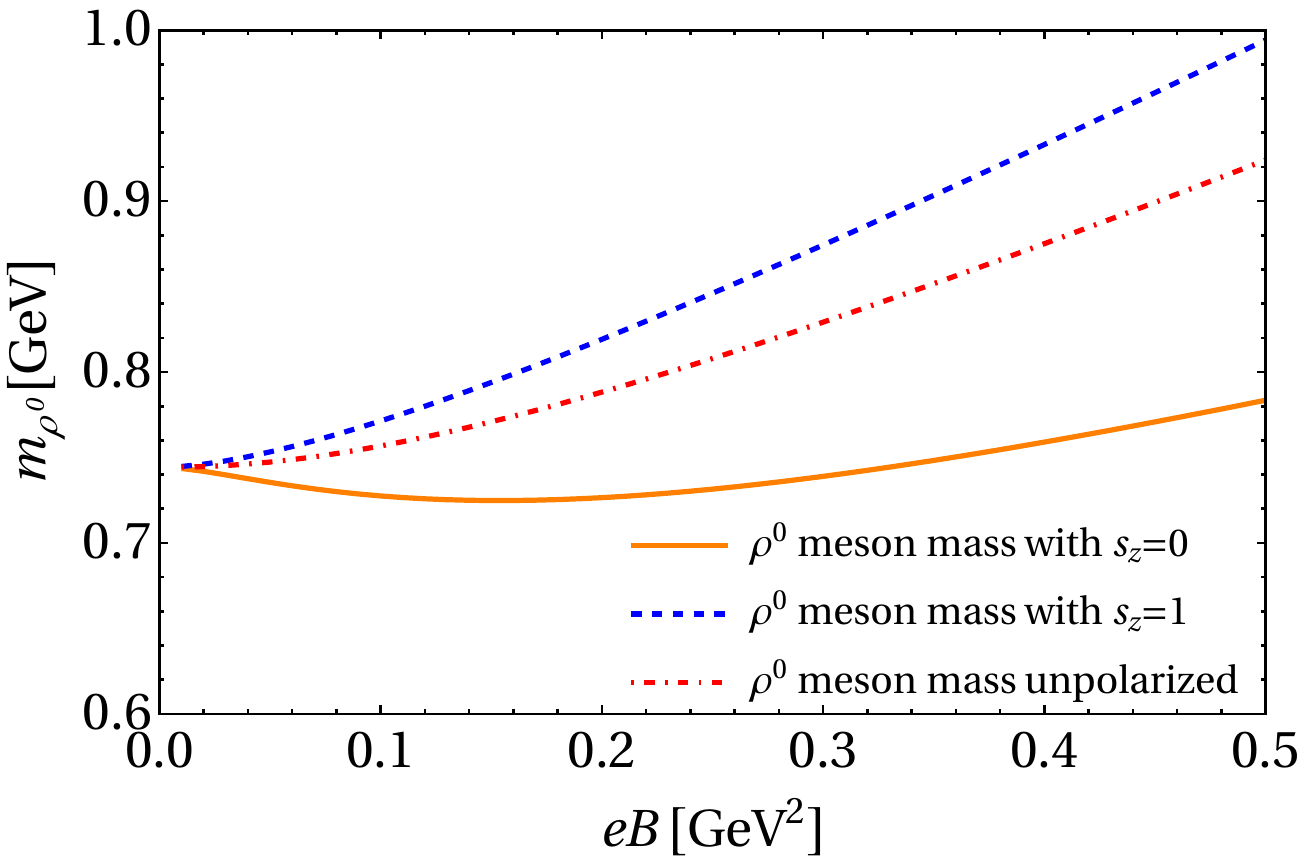}}\sbox2{\pnl\trimE{10cm}{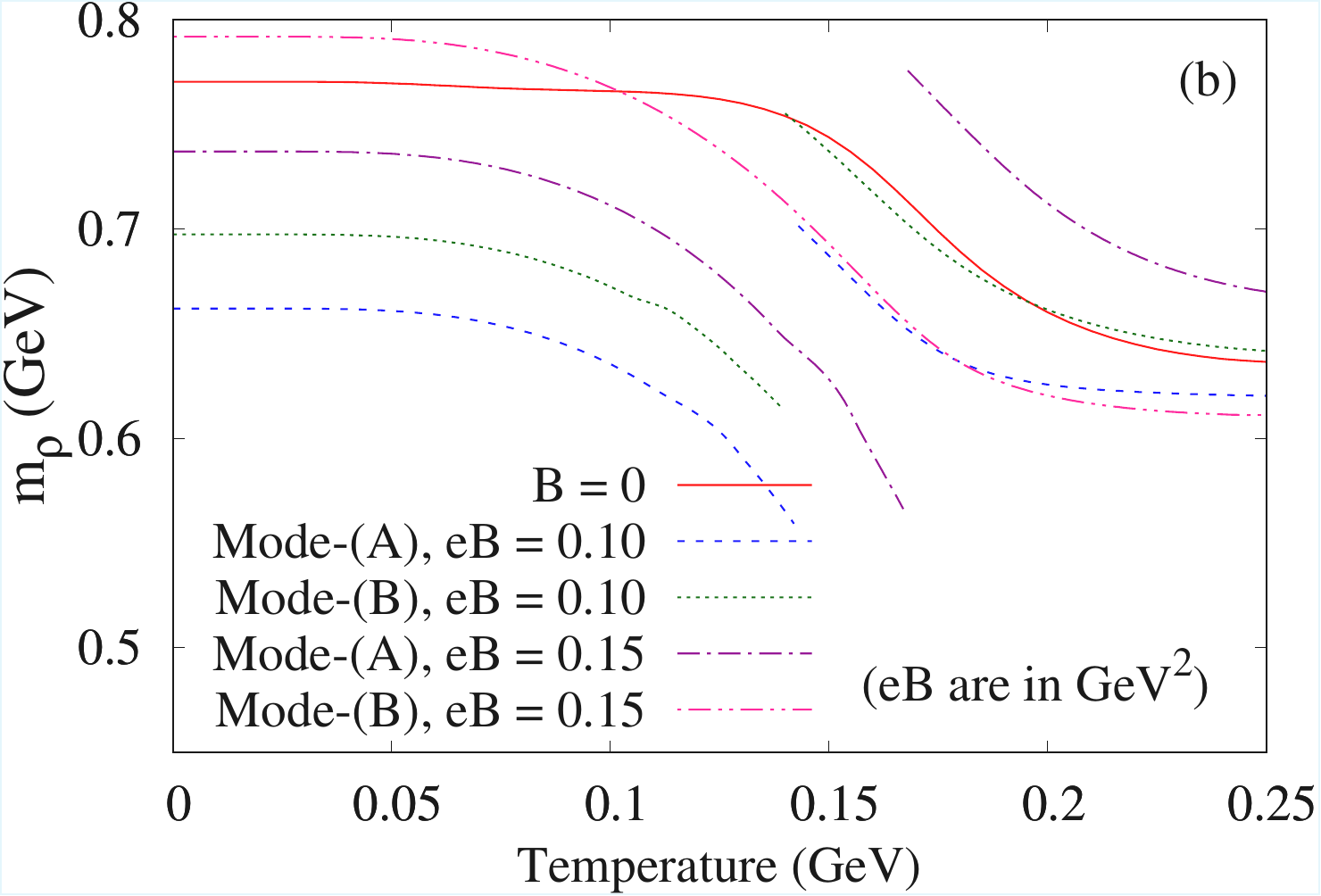}}\sbox4{\pnl\trimF{10cm}{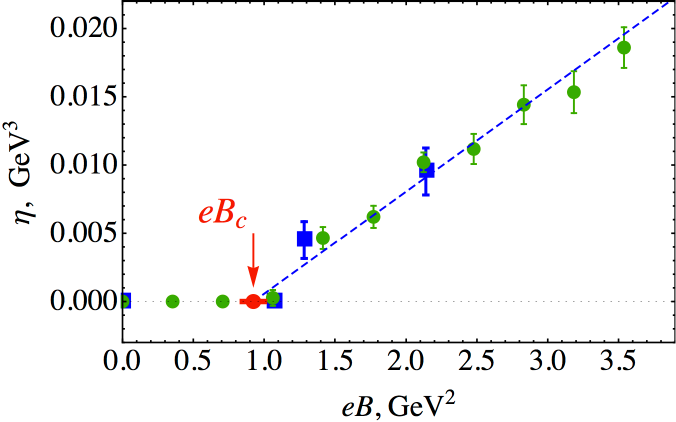}}%
\dimen4=\dimexpr\wd0+\wd2+\wd4\relax
\ifdim\dimen4>\dimen2 \dimen2=\dimen4 \fi
\dimen0=\dimexpr\textwidth-2em\relax
\edef\figH{\the\dimexpr 10cm*\dimen0/\dimen2\relax}%
\subfloat[\centering ]{\pnl\trimA\figH{ro_meson_B_combined_MEM_cropped.pdf}}\hfill
\subfloat[\centering ]{\pnl\trimB\figH{rho_H_B2_all_s1_cropped.pdf}}\hfill
\subfloat[\centering ]{\pnl\trimC\figH{neutral.pdf}}\\
\subfloat[\centering ]{\pnl\trimD\figH{rho_sz0_contradiction.pdf}}\hfill
\subfloat[\centering ]{\pnl\trimE\figH{meson5_rho_Mott_cropped.pdf}}\hfill
\subfloat[\centering ]{\pnl\trimF\figH{condensate2_vs_B.pdf}}
\caption{(a) Mass of the $\rho^0$ meson with spin projection $s_z = 0$ for various lattice volumes and lattice spacings, obtained using the Maximum Entropy Method \citep{Luschevskaya:2012xd}. (b) Same masses with $s_z = \pm 1$ \citep{Luschevskaya:2012xd}. (c) Masses of the charged vector meson $\rho^\pm$ with $s_z = 0$ and the neutral vector meson $\rho^0$ with $s_z = 0, \pm 1$, calculated in the NJL model \citep{Liu:2014uwa}. (d) $\rho^0$ mass for different spin projections at finite magnetic field in the two-flavor NJL model \citep{Avancini:2022qcp}. (e) $\rho^0$ mass as a function of temperature for different magnetic fields, including vector and axial-vector channels in the NJL model, showing the Mott transition \citep{Ghosh:2020qvg}. (f) The superconducting condensate $\eta = |\langle \rho \rangle|$ of the charged $\rho$ mesons. The red arrow marks the insulator--superconductor phase transition \citep{Braguta:2011hq}.}
\label{fig:Rho_Mass_2}
\end{figure*}

\begin{figure*}[hbt]
\centering
\sbox0{\includegraphics[width=0.5\textwidth]{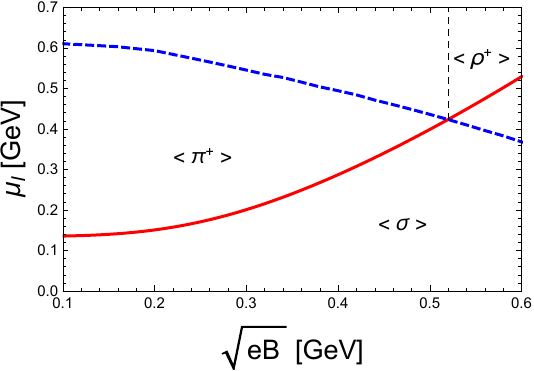}}%
\edef\figHa{\the\ht0}%
\subfloat[\centering ]{\includegraphics[width=0.5\textwidth]{PT.pdf}}%
\subfloat[\centering ]{\includegraphics[width=0.5\textwidth,height=\figHa]{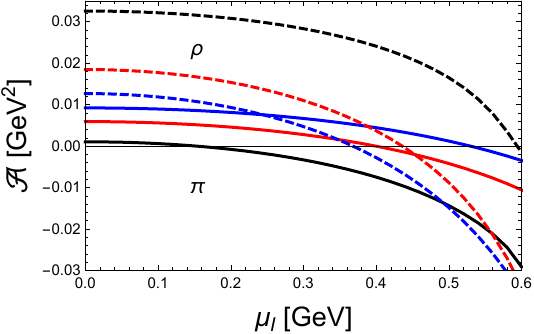}}\\
\subfloat[\centering ]{\includegraphics[width=0.5\textwidth]{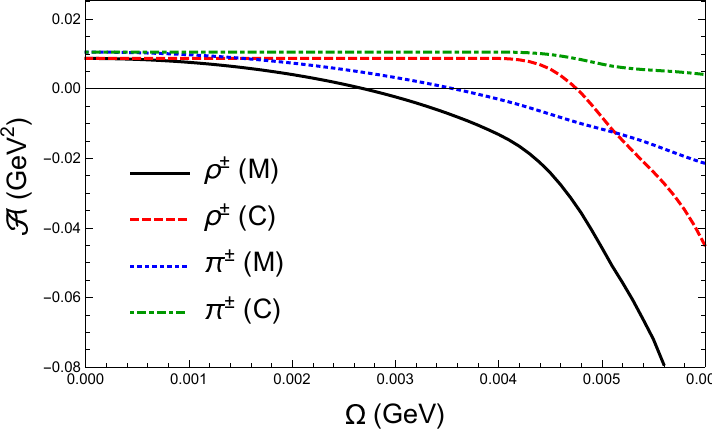}}%
\subfloat[\centering ]{\includegraphics[width=0.5\textwidth]{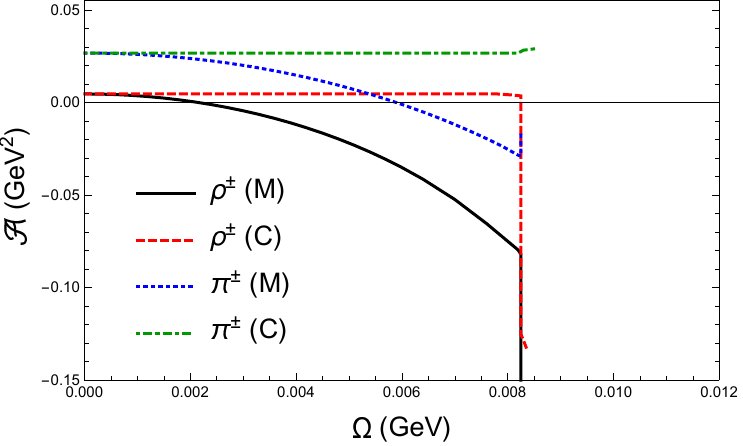}}
\caption{(a) Phase transition lines for pion superfluidity (red solid) and $\rho$ superconductivity (blue dashed) in the $\mu_I$--$B$ plane. The labels $\langle \sigma \rangle$, $\langle \pi^+ \rangle$ and $\langle \rho^+ \rangle$ denote the normal chiral-symmetry-breaking phase, pion superfluidity and $\rho$ superconductivity, respectively. The thin dashed line only indicates the boundary between pion superfluidity and $\rho$ superconductivity. For large $\mu_I$, $\pi$ superfluidity is favored at small $eB$, while $\rho$ superconductivity is favored at large $eB$ \citep{Ke:2026npb}. (b) Ginzburg--Landau (GL) coefficients $A_{\pi^+\pi^+}$ (solid lines) and $A_{\rho_1^+\rho_1^+}$ (dashed lines) as functions of the isospin chemical potential $\mu_I$ for $\sqrt{eB} = 0.2$ GeV (black), $0.5$ GeV (red) and $0.6$ GeV (blue), calculated in the NJL model with vector and axial-vector interactions \citep{Ke:2026npb}. For large $eB$, the GL coefficient of $\rho^+$ vanishes at smaller $\mu_I$ than that of $\pi^+$, confirming the results of Fig.~\ref{fig:Rho_sup}(a). (c) GL coefficients of charged $\rho$ condensation compared with those of charged $\pi$, for the Schwinger phase in Minkowski space and in curved space, denoted ``M'' and ``C'', respectively, at $eB = 0.5~\text{GeV}^2$. The $\rho^\pm$ condensate with the Minkowski Schwinger phase is favored over all other condensates \citep{Cao:2020pmm}. (d) Same as (c) for $eB = 1.5~\text{GeV}^2$. As $eB$ increases, the $\rho^\pm$ condensate sets in at smaller angular velocity $\Omega$ \citep{Cao:2020pmm}.}
\label{fig:Rho_sup}
\end{figure*}

\begin{figure*}[hbt]
\centering
\subfloat[\centering ]{\includegraphics[width=0.5\textwidth]{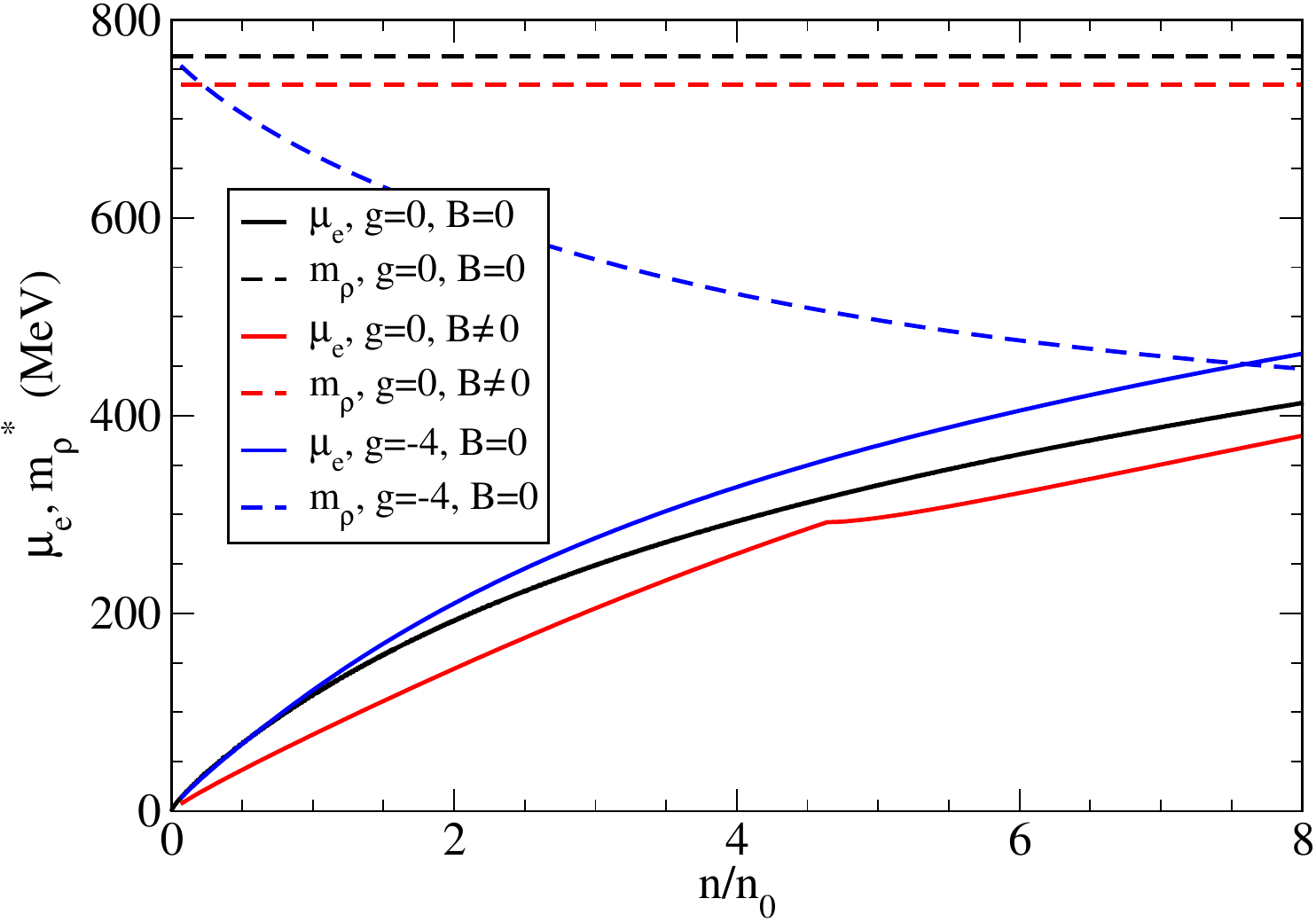}}%
\subfloat[\centering ]{\includegraphics[width=0.5\textwidth]{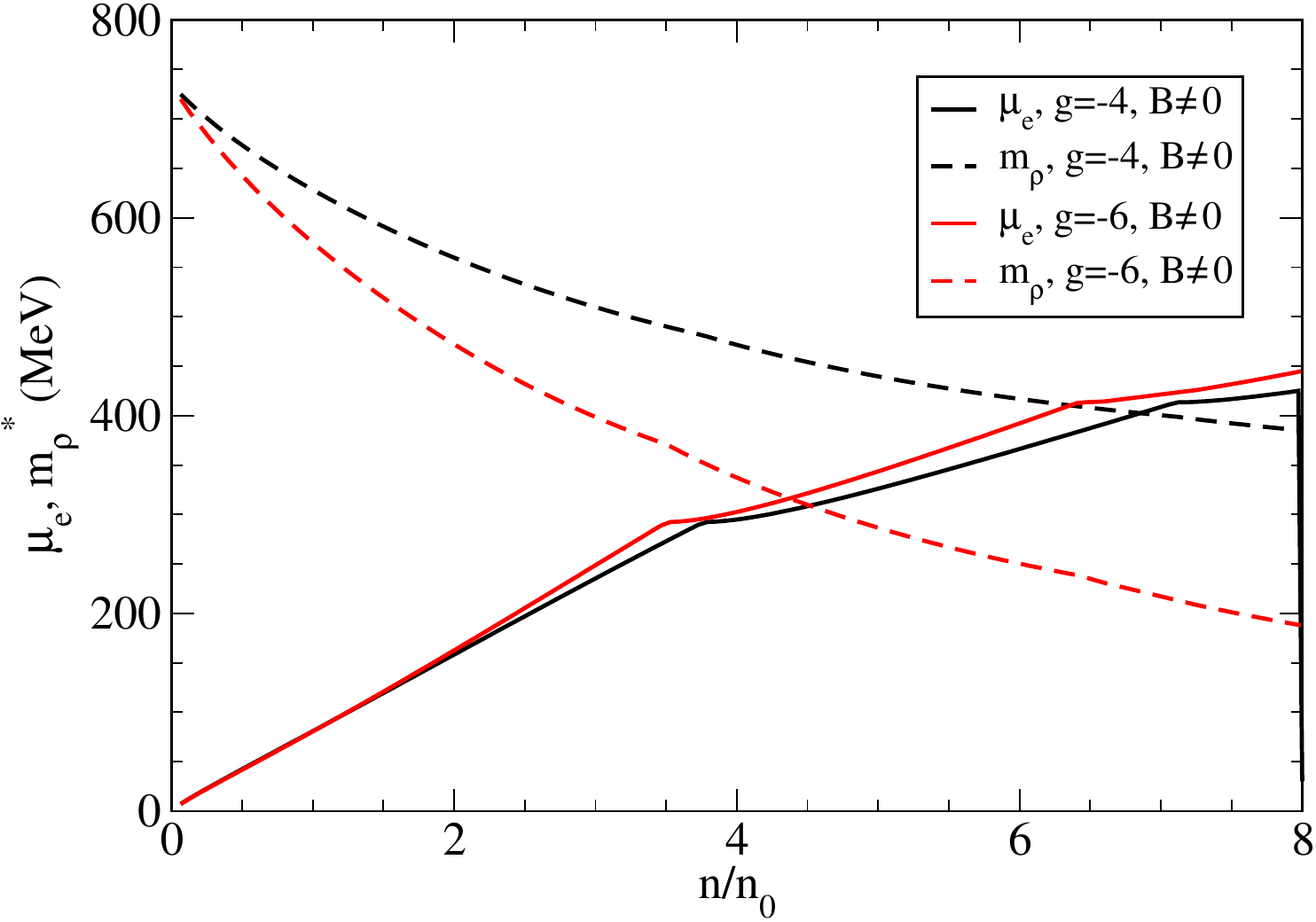}}
\caption{The electron chemical potential $\mu_e$ and the effective $\rho^-$ mass (dashed lines) as functions of the normalized density $n/n_0$ \citep{Mallick:2014faa}. In the relativistic mean-field approach, the effective $\rho$ meson mass is assumed to vary as $m^\star_\rho = m_\rho - g\sigma$, where $\sigma$ is the scalar meson field, $g$ is a dimensionless constant and $m_\rho = 776$ MeV is the vacuum $\rho$ mass. In a constant magnetic field, the effective mass becomes $m^{\star 2}_{\rho^-}(B) = m^{\star 2}_{\rho^-} - eB$. The crossing points of $m^\star_\rho$ with the $\mu_e$ lines mark the density at which $\rho$ meson condensation sets in.
}
\label{fig:rhomeson_NS_ritam}
\end{figure*}

\par It is shown that even if the $\rho$-meson mass increases with the magnetic field because of MC, for sufficiently large value of the magnetic field, $m_{\rho}(B)$ can actually vanishes and a transition to a superconducting state \citep{Chernodub:2011gs} that is built out of charged as well as neutral $\rho$-meson condensates can occur. But before discussing $\rho$-meson condensation, we first present a brief survey of how the existing literature describes the behavior of the $\rho$-meson mass in a magnetic field.

\par Using $SU(2)$ lattice gauge theory, Ref.~\citep{Luschevskaya:2012xd} calculates the masses of $\rho^0$ mesons with different spin projections along the magnetic-field direction. It finds that the mass of the $\rho^0$ state with spin projection $s_z=0$ decreases with increasing magnetic field (Fig.~\ref{fig:Rho_Mass_2}(a)). However, this behavior is contradicted by NJL-model calculations in Ref.~\citep{Liu:2014uwa}, where the $s_z=0$ $\rho^0$ mass increases with $B$ (Fig.~\ref{fig:Rho_Mass_2}(c)), and by Ref.~\citep{Avancini:2022qcp}, which reports a non-monotonic dependence (Fig.~\ref{fig:Rho_Mass_2}(d)).

In contrast, there is broader agreement regarding the $\rho^0$ states with $s_z=\pm1$. Ref.~\citep{Luschevskaya:2012xd} shows that their masses increase with $eB$ (Fig.~\ref{fig:Rho_Mass_2}(b)), consistent with the results of Refs.~\citep{Liu:2014uwa,Avancini:2022qcp} (Figs.~\ref{fig:Rho_Mass_2}(c) and (d)). Ref.~\citep{Liu:2014uwa} further shows that the masses of charged vector mesons $\rho^\pm$ with $s_z=0$ increase with magnetic field, whereas the masses of $\rho^+$ ($\rho^-$) with $s_z=1$ ($s_z=-1$) decrease approximately linearly with $eB$ and vanish at $eB\approx0.2~\text{GeV}^2$ (Fig.~\ref{fig:rho_condensation}(a)). This behavior suggests the possibility of charged vector-meson condensation, a controversial conjecture discussed later. Similar conclusions are obtained in Ref.~\citep{Liu:2015pna} (Fig.~\ref{fig:rho_condensation}(b)).

On the other hand, Ref.~\citep{Ghosh:2016evc}, including an additional $\rho\pi\pi$ interaction in weak magnetic fields, finds that both $\rho^0$ and $\rho^\pm$ masses decrease with magnetic field in vacuum. Furthermore, Ref.~\citep{Ghosh:2020qvg} reports discontinuities in the masses of mesonic excitations in a thermo-magnetic background for nonzero magnetic field, signaling a Mott transition (Fig.~\ref{fig:Rho_Mass_2}(e)), analogous to the pion mass behavior shown in Fig.~\ref{fig:pion_Mott_combined}.

\par The existing literature on magnetized $\rho$ mesons shows no consensus regarding their behavior in strong magnetic fields. While some studies find that the $\rho$-meson mass decreases and may vanish at sufficiently large $B$, others report an increasing mass. Nevertheless, results supporting a vanishing $\rho$-meson mass motivate investigations of vacuum instability and possible $\rho$-meson condensation. Since its proposal, $\rho$ condensation remains controversial, with some works \citep{Hidaka:2012mz} arguing that $\rho$ mesons cannot condense even in the presence of a magnetic field. QCD inequalities imply that massless $\rho$ mesons are possible only if the neutral pion mass also vanishes. Consequently, the existence of a nonzero minimum in the $m_{\pi^0}(B)$ profile, discussed earlier, disfavors $\rho$-meson condensation.

However, studies of quenched overlap quarks in $SU(2)$ QCD, based on spatial vector correlators, report a critical magnetic field of approximately $0.9~\text{GeV}^2$, above which $\rho$ condensation appears \citep{Braguta:2011hq} (Fig.~\ref{fig:Rho_Mass_2}(f)). Despite persistent doubts about its existence, the possibility of magnetized $\rho$-meson condensation continues to attract considerable attention. Ref.~\citep{Chernodub:2011mc} demonstrates within the NJL model that the QCD vacuum can transit into an anisotropic and inhomogeneous electromagnetically superconducting phase carrying the quantum numbers of charged $\rho$ mesons. Ref.~\citep{Liu:2015pna} further shows, after including axial-vector channels, that the charged $\rho$-meson mass decreases with magnetic field and vanishes at a critical field, signaling vector-meson condensation that survives even at high temperature and density (Figs.~\ref{fig:rho_condensation}(b) and (c)). Ref.~\citep{Ke:2026npb} finds that increasing isospin chemical potential favors $\pi$ superfluidity at weak magnetic fields but $\rho$ superconductivity at strong magnetic fields (Figs.~\ref{fig:Rho_sup}(a) and (b)). In contrast, Ref.~\citep{Cao:2020pmm} identifies the $\rho^\pm$ superconducting phase as the robust ground state of SIM only in the simultaneous presence of strong magnetic fields and rapid rotation (Figs.~\ref{fig:Rho_sup}(c) and (d)).
\par Despite controversies, given a considerable body of evidence in the support of $\rho$ meson condensation under sufficiently strong magnetic fields, whether alone or combined with rotation, this raises a natural question of whether the onset of this condensation can be captured by an effective field theoretic description. The conventional superconductivity could be modeled in the framework of both microscopic fermionic model e.g. Bardeen-Cooper-Schrieffer (BCS) theory or involving bosonic degrees of freedom e.g. the Landau-Ginzberg framework. Although these two approaches are different from one another, both are capable of describing superconductivity with fairly good quantitative accuracy. These two descriptions are in fact equivalent near the superconducting phase transition. In a similar manner, transition of vacuum to $\rho$-supersonducting phase can be described by a Landau-Ginzburg type formalism. Here, we very briefly describe the electrodynamics of $\rho$-mesons using the Djukanovic, Schindler, Gegelia, and Scherer (DSGS) lagrangian \cite{Djukanovic:2005ag} and the condensation mechanisms in the RMF approximation. The bosonic part of the model is given by \cite{Chernodub:2010qx}
\begin{widetext}
\begin{align}
    \mathcal{L}_{\text{DSGS}} =-\frac{1}{4}F_{\mu\nu}F^{\mu\nu} - \frac{1}{4}\rho^{\dagger}_{\mu\nu}\rho^{\mu\nu} + m^2_\rho \rho^{\dagger}_{\mu}\rho^{\mu} - \frac{1}{4}\rho^{(0)\mu\nu}\rho^{(0)}_{\mu\nu} +\frac{m^2_\rho}{2}\rho^{(0)\mu}\rho^{(0)}_{\mu} +\frac{e}{2g_S}F^{\mu\nu}\rho^{(0)}_{\mu\nu}
\end{align}
\end{widetext}
where $A_\mu$ is the photon field, $\rho_\mu = (\rho^{(1)}_{\mu} - i\rho^{(2)}_{\mu})/\sqrt{2}$ and $\rho^{(0)}_{\mu} = \rho^{(3)}_{\mu}$ are fields associated with charged and neutral vector mesons with degenerate mass $m_{\rho}$. The ground state energy turns out to be
\begin{align}
    &\epsilon(\rho_{\mu}) = ieB(\rho^{\dagger}_{1}\rho_{2} - \rho^{\dagger}_{2}\rho_{1}) + m^2_\rho \rho^{\dagger}_{\mu}\rho_{\mu} \nonumber\\
    &= \mathlarger{\sum}_{a, b = 1}^{2}\rho^{\dagger}_a\mathcal{M}_{ab}\rho_b + m^2_\rho(\rho^{\dagger}_{0}\rho_0 + \rho^{\dagger}_{3}\rho_3)
\end{align}

\begin{figure*}[hbt]
    \centering
    \begin{minipage}[b]{0.5\textwidth}
        \centering
        \includegraphics[width=\linewidth]{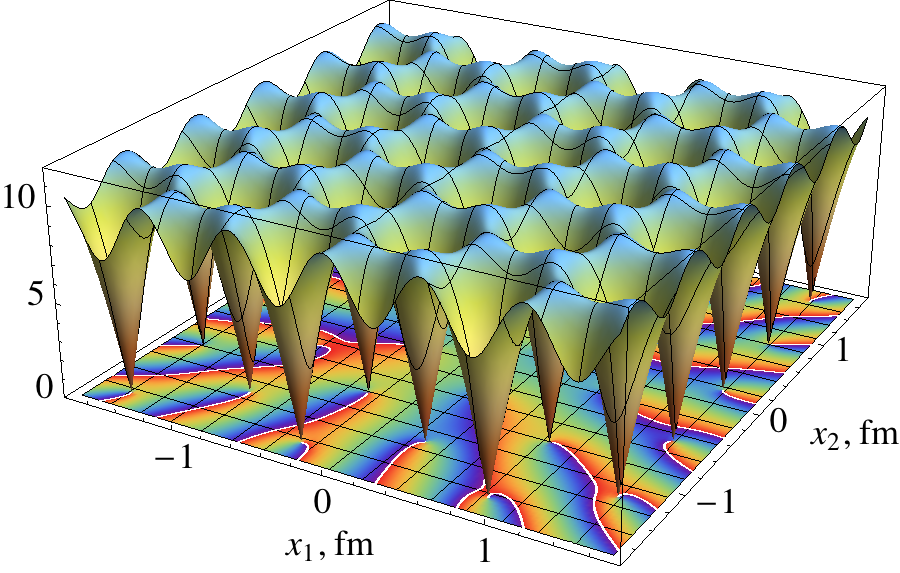}

        \vspace{2mm}
        (a)
    \end{minipage}%
    \begin{minipage}[b]{0.5\textwidth}
        \centering
        \includegraphics[width=\linewidth]{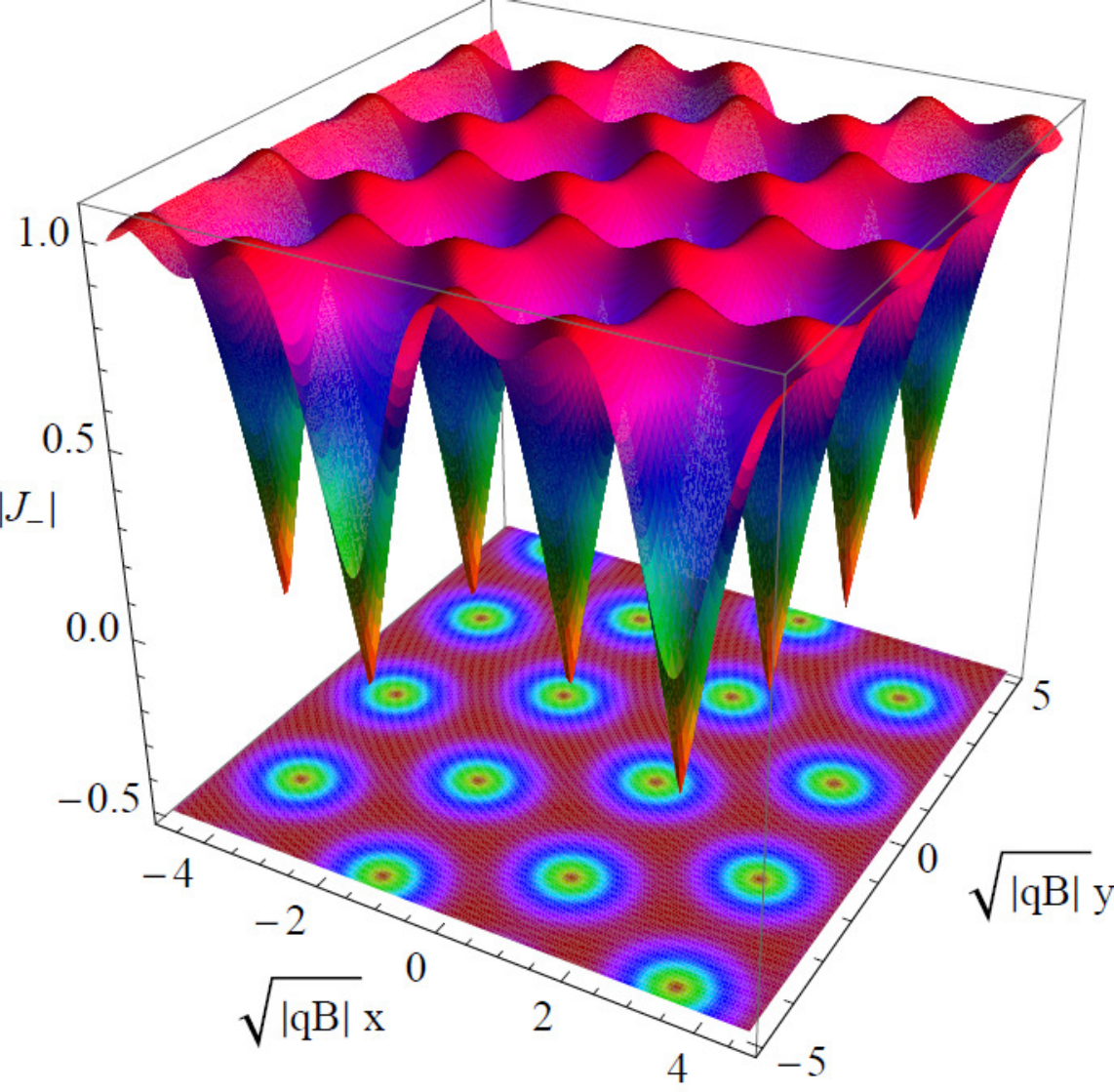}

        \vspace{2mm}
        (b)
    \end{minipage}
    \caption{(a) Abrikosov lattice structure for the $\rho$-meson condensates in the ground state at $B = 1.01\,B_C$. The plots show the behavior of $|\rho|_0$ (in MeV), while the projection onto the $(x_1,x_2)$ plane is a density plot of the corresponding condensate phases. Taken from \citep{Chernodub:2012bj}. (b) Triangular vortex lattice in the $x$--$y$ plane obtained in the holographic model, with the contour plot shown below. The condensate vanishes at the core of each vortex. Taken from \citep{Cai:2013pda}.}
    \label{fig:rho_abrikosov_N_3}
\end{figure*}

It turns out that the minimum of the energy density of the system corresponds to $\rho^{(0)}_\mu$, $\rho_0 = 0 = \rho_3$. In terms of the mass matrix $\mathcal{M}$ of the charged $\rho$ mesons
\begin{align}
\mathcal{M} &=
\begin{pmatrix}
m_{\rho}^{2} & i e B \\
- i e B & m_{\rho}^{2}
\end{pmatrix}.
\end{align}
with eigenvalues $\mu^2_{\pm} = m^2_\rho \pm eB$ and mass eigenstates $\rho_\pm = (\rho_1 \pm i\rho_2)/\sqrt{2}$, and recognizing the fact that the ground state corresponds to $\rho_{+} = 0$, and hence $\rho_1 = -i\rho_2 = \rho$ and $\rho_{-} = \sqrt{2}\rho \neq 0$, the ground state energy density becomes
\begin{align}
    \epsilon(\rho) = \frac{B^2}{2} + 2(m^2_\rho - eB)|\rho|^2 + 2g^2_S|\rho|^4 = \epsilon(|\rho|^2)
\end{align}
which is nothing but the Mexican-hat potential. One observes that for $B > B_C = m^2_\rho/e$, the coefficient of $|\rho|^2$ term changes sign and the mean field ground state of the system becomes
\begin{align}
|\rho|_0 =
\begin{cases}
0, & \text{for } B < B_C, \\[4pt]
\sqrt{\dfrac{e|B-B_C|}{2g_S^2}}, & \text{for } B > B_C.
\end{cases}
\label{rho_condensate_mean_field}
\end{align}
The field $\rho$ can, therefore, be thought as a order parameter describing the superconducting state. However, this result is a mean field solution of the DSGS model. Near the critical field, the actual solution is spatially inhomogeneous due to the existence of stringlike defects called $\rho$-vortices parallel to the magnetic field and forms an Abrikosov vortex lattice \cite{Chernodub:2010qx,Chernodub:2012bj}, analogous to the mixed state of a type-II superconductor. This solution of the condensate can be written as as a superposition of LLL wavefunctions given by
\begin{equation}
\rho(z) = \mathlarger{\sum}_{n\in\mathbb{Z}} C_n\exp\Big[-\frac{\pi}{2L^2_B}\big(|z|^2+\bar{z}^{\,2}\big) - \pi\nu^2 n^2 + \frac{2\pi\nu n\bar{z}}{L_B}\Big]
\end{equation}
where $L_B=\sqrt{2\pi/eB}$ is the magnetic length, $z = x_1 + ix_2$ is a complex coordinate in the plane transverse to the magnetic field. Here, $\nu$ and $C_n$ are chosen to minimize the energy functional and determine the lattice geometry. Conditions like $C_{n+N} =C_n$, determine the geometry of the Abrikosov lattice. The geometries are usually periodic on the $z$ plane., e.g. $N = 1$ leads to square lattice and $N = 2$ leads to equilateral triangular lattice, as is shown in Ref.\cite{Chernodub:2011gs} near the transition point $B \approx B_C$ for $C_1 = iC_2$ and $\nu = \sqrt[4]{e}/\sqrt{2}$, shown in Fig.\ref{fig:rho_abrikosov_N_3}. As opposed to the Landau-Ginzburg model, the coefficients $C_n$ depend on the magnetic field: $C_n = C_n(B)$. Each lattice cell in this phase contains a stringlike topological defect, essentially a superconductor vortex, in the charged $\rho$ condensates and three superfluid vortices and three
superfluid antivortices made of $\rho^{(0)}$ condensate. These superposition of the superconding and superfluid vortex lattices produces a very intricate kaleidoscope like pattern, analogous to condensation of $W^\pm$ and $Z$ bosons \cite{Chernodub:2012fi}. 
\par Ref.\cite{Braguta:2012fol} has observed the presence of the $\rho$ vortices in numerical simulations of the vacuum of the quenched two-color LQCD in strong magnetic field background forming a liquid. Ref.\cite{VanDoorsselaere:2011vmc} finds a hexagonal pattern of vortices obtained in a 2–parameter Abrikosov ansatz.
Ref.\cite{Chernodub:2010qx} shows that $\rho^\pm$ condensate is an anisotropic and inhomogeneous superconductor along the axis of the
magnetic field and in the plane transverse to the magnetic field the superconductivity is absent. Ref. \cite{Chernodub:2012bj} shows that this exotic superconductor induces an even more exotic superfluid phase of the neutral $\rho$ mesons given by
\begin{widetext}
\begin{equation}
\rho^{(0)}(x_1,x_2)
\equiv
\rho^{(0)}_1(x_1,x_2)
+i\rho^{(0)}_2(x_1,x_2)
=
2i g_s \cdot
\left(
\frac{\partial}{-\partial_\perp^2+m_0^2}
|\rho|^2
\right)(x_1,x_2),
\end{equation}
\end{widetext}
Charged $\rho$ condensates have also been studied within the context of holographic models \cite{Cai:2013pda} showing that there is a critical temperature below which the charged vector condenses via a second order phase transition and forming a vortex lattice structure in the spatial directions perpendicular to the magnetic field. Ref.\cite{Aharony:2007uu} shows in the context of Sakai-Sugimoto model that when the isospin chemical potential reaches a critical value $\mu_I \approx 1.7m_\rho$, $\rho$ meson state becomes massless and it condenses. Using $SU(2)_f$ NJL effective model with vector channel, Ref.\cite{Zhang:2018ome} shows that in the presence of global rotation, scalar channel pion condensate is suppressed and the $\rho$ condensate is enhanced. Ref.\cite{Liu:2016vuw} specifically shows that in the presence of IMC that the charged $\rho$ condensation can exist in the temperature region of $1 - 1.5\sim T_C$ ($T_C$ being the temperature of chiral phase transition) with critical magnetic field $eB_c \sim 0.15-0.3~\text{GeV}^2$, suggesting that that high temperature superconductor might be created through non-central heavy ion collisions at LHC energies.
\par Although rho meson condensate has been shown to have observable astrophysical signatures \citep{Maslov:2017xjp,Kolomeitsev:2017gli,Shivam:2019cmw,Pradhan:2025pol}, a corresponding body of work in the presence of a magnetic field still needs to be developed. Ref.~\citep{Mallick:2014faa} shows that $\rho^-$-meson condensates can appear in the core of the neutron star as the $\rho^-$ mass decreases with density and magnetic field. Moreover, the $\rho^-$ condensate can appear at the density $4.4\rho_0$, as shown in Figs.~\ref{fig:rhomeson_NS_ritam}(a) and (b).
\section{Discussion and Concluding Remarks}
\label{Discuss_Section}
Studies of SIM using effective field-theoretic models in the presence of magnetic fields have provided a very rich picture of the QCD phase diagram. However, the extensive body of work on this subject done by employing a wide variety of frameworks, makes the task of producing a comprehensive review extremely formidable. Instead, we work with a much more modest goal and summarize the leading results obtained using the NJL model and its extensions including magnetic fields.

\par Although the NJL model and its extensions provide valuable tools for exploring the intermediate-density regime inaccessible to first-principles methods, their predictions depend strongly on the chosen model variant and regularization scheme, and often disagree with lattice QCD results. A prominent example is the MC–IMC debate: while finite-temperature lattice QCD supports IMC, different NJL variants yield conflicting outcomes, with some predicting MC and others, such as the EPNJL and lattice-improved NJL models, reproducing IMC.

Model dependence also affects studies of confinement. PNJL and EPNJL models partially mimic the confinement–deconfinement transition through the Polyakov loop operator $\Phi$ and $\overline{\Phi}$, but the associated potentials and $\Phi$ and $\overline{\Phi}$-dependent couplings are largely ad hoc, causing predictions to vary with their parametrization. In contrast, nonlocal NJL models incorporate confinement more naturally.

Another major limitation is the strong regularization dependence of the NJL framework, which can generate unphysical oscillations mistaken for de Haas–van Alphen oscillations. Although schemes such as MFIR-MSS reduce these artifacts, the non-renormalizable nature of the NJL model makes regulator dependence unavoidable. Nonlocal PNJL models improve this behavior, but introduce additional uncertainty through the choice of form factors.
\par Such model sensitivity propagates over to the phase structure of magnetized dense quark matter, making it an area of serious debate. While phases e.g. MCFL and magnetized 2SC phases are well established in asymptotically large and intermediate dense regimes, respectively, their stability, the true ground state of magnetized QCD matter and the resolution of chromomagnetic instabilities are far from settled. Moreover, despite a considerable amount of progress in studying the properties of magnetized $\pi$-condensates, $\rho$-meson condensation under strong magnetic fields still remains a very controversial proposition. These theoretical uncertainties can lead to significant ambiguities in astrophysical predictions.
\par Taken together, these considerations indicate that no single variation of the NJL-type model currently provides a fully reliable description of magnetized SIM. Progress, therefore, requires a complementary approach in which different effective models are used to capture different aspects of the physics, and their predictions are systematically compared. Whenever possible, these results should also be confronted with LQCD calculations and astrophysical observations of compact stars.

Within such a framework, features that remain stable across different models and agree with external constraints can be considered more robust and less sensitive to model-specific assumptions, leading to a more coherent, though still model-dependent, understanding of magnetized QCD matter.

\vspace{6pt} 
 
\def\bibfont{\small}
\bibliography{Reference}{}

\end{document}